\documentclass[twocolumn,aps,prd,epsf,showpacs]{revtex4-1}
\pdfoutput=1
\usepackage[pdftex]{graphicx}
\usepackage{color}
\usepackage{colortbl}
\usepackage{xcolor}
\usepackage{amsmath}
\usepackage{amsfonts}
\usepackage{amssymb}
\usepackage{dsfont}
\usepackage[]{units}
\usepackage{braket}
\usepackage[utf8]{inputenc}
\usepackage[T1]{fontenc}
\usepackage{url}
\usepackage[english]{babel}
\usepackage{bm}
\usepackage{url}
\usepackage{verbatim}
\usepackage[]{cleveref}
\usepackage{bm}
\usepackage{slashed}
\usepackage{array,multirow}

\usepackage{algorithm}
\usepackage[noend]{algpseudocode}
\algnewcommand\algorithmicforeach{\textbf{for each}}
\algdef{S}[FOR]{ForEach}[1]{\algorithmicforeach\ #1\ \algorithmicdo}

\newcommand\Tstrut{\rule{0pt}{2.9ex}}         
\newcommand\Bstrut{\rule[-1.2ex]{0pt}{0pt}}   
\newcommand\TBstrut{\Tstrut\Bstrut}           

\newcommand{\be}{\begin{equation}}
\newcommand{\ee}{\end{equation}}
\newcommand{\bea}{\begin{eqnarray}}
\newcommand{\eea}{\end{eqnarray}}
\newcommand{\non}{\nonumber}
\newcommand{\bei}{\begin{itemize}}

\newcommand{\eei}{\end{itemize}}
\newcommand{\ben}{\begin{enumerate}}
\newcommand{\een}{\end{enumerate}}

\newcommand{\gtapprox}{\raisebox{-0.5ex}{$\,\stackrel{>}{\scriptstyle\sim}\,$}}
\newcommand{\ltapprox}{\raisebox{-0.5ex}{$\,\stackrel{<}{\scriptstyle\sim}\,$}}

\DeclareMathOperator{\diag}{diag}

\graphicspath{{figures/}}

\begin{document}

\title{Bottomonium resonances with $I = 0$ from lattice QCD correlation functions with static and light quarks}

\author{$^{(1)}$Pedro Bicudo}
\email{bicudo@tecnico.ulisboa.pt}

\author{$^{(1)}$Marco Cardoso}
\email{marco.cardoso@tecnico.ulisboa.pt}

\author{$^{(1)}$Nuno Cardoso}
\email{nuno.cardoso@tecnico.ulisboa.pt}

\author{$^{(2)}$Marc Wagner}
\email{mwagner@itp.uni-frankfurt.de}

\affiliation{\vspace{0.1cm}$^{(1)}$CeFEMA, Dep.\ F\'{\i}sica, Instituto Superior T\'ecnico, Universidade de Lisboa, Av.\ Rovisco Pais, 1049-001 Lisboa, Portugal}

\affiliation{\vspace{0.1cm}$^{(2)}$Johann Wolfgang Goethe-Universit\"at Frankfurt am Main, Institut f\"ur Theoretische Physik, Max-von-Laue-Stra{\ss}e 1, D-60438 Frankfurt am Main, Germany}

\begin{abstract}
We discuss, how to study $I = 0$ quarkonium resonances decaying into pairs of heavy-light mesons using static potentials from lattice QCD. These static potentials can be obtained from a set of correlation functions containing both static and light quarks. As a proof of concept we focus on bottomonium with relative orbital angular momentum $L_{\bar{Q} Q} = 0$ of the $\bar{b} b$ pair corresponding to $J^{P C} = 0^{- +}$ and $J^{P C} = 1^{- -}$. We use static potentials from an existing lattice QCD string breaking study and compute phase shifts and $\mbox{T}$ matrix poles for the lightest heavy-light meson-meson decay channel. We discuss our results in the context of corresponding experimental results, in particular for $\Upsilon (10860)$ and $\Upsilon (11020)$.
\end{abstract}

\pacs{12.38.Gc, 13.75.Lb, 14.40.Rt, 14.65.Fy.}

\maketitle


\section{\label{sec:intro}Introduction}

A long standing problem in QCD is to understand exotic hadrons, i.e.\ hadrons which have a structure more complicated than a mesonic quark-antiquark pair or a baryonic triplet of quarks \cite{Jaffe:1976ig}. However, the problem of identifying or predicting exotic hadrons, say tetraquarks, pentaquarks, hexaquarks, hybrids or glueballs, turned out to be much harder than initially expected (see e.g.\ Ref.\ \cite{Bicudo:2015bra}). For example the observed tetraquarks $Z_b$ and $Z_c$ are resonances high in the spectrum, not only difficult to observe, but also very technical to address in hadronic models and extremely difficult to compute from first principles e.g.\ with lattice QCD.

One approach to study hadrons composed of heavy quarks and antiquarks as well as of gluons and possibly light quarks and antiquarks, which is based on lattice QCD, is the Born-Oppenheimer approximation \cite{Born:1927}. It was succesfully applied to investigate both non-exotic and exotic quarkonium (see e.g.\ Refs.\ \cite{Juge:1999ie,Braaten:2014qka,Berwein:2015vca,Brambilla:2017uyf,Karbstein:2018mzo,Capitani:2018rox}) as well as tetraquarks with two heavy antiquarks and two light quarks. In the latter case, in a first step potentials of two static antiquarks in the presence of two light quarks are computed using state of the art lattice QCD techniques (see e.g.\ Refs.\ \cite{Detmold:2007wk,Wagner:2010ad,Bali:2010xa,Wagner:2011ev,Brown:2012tm,Bicudo:2015kna}). Then, in a second step, the dynamics of the two heavy quarks is described by a quantum mechanical Hamiltonian with the aforementioned static potentials. This requires heavy quark masses much larger than the scale of QCD, which is the case e.g.\ for $\bar{b}$ quarks. A $\bar{b} \bar{b} u d$ tetraquark bound state with quantum numbers $I(J^P) = 0(1^+)$, first predicted by model calculations \cite{Ader:1981db,Ballot:1983iv,Heller:1986bt,Carlson:1987hh,Lipkin:1986dw,Brink:1998as,Gelman:2002wf,Vijande:2003ki,Janc:2004qn,Cohen:2006jg,Vijande:2007ix}, was recently confirmed within this lattice QCD/Born-Oppenheimer approach \cite{Bicudo:2012qt,Brown:2012tm,Bicudo:2015vta,Bicudo:2015kna,Bicudo:2016ooe} (for lattice QCD work on the same $\bar{b} \bar{b} u d$ tetraquark using Non Relativistic QCD instead of static quarks see Refs.\ \cite{Francis:2016hui,Francis:2018jyb,Junnarkar:2018twb,Leskovec:2019ioa}). Very recently also $B B$ scattering was studied using similar techniques and a $\bar{b} \bar{b} u d$ tetraquark resonance with $I(J^P) = 0(1^-)$ was predicted \cite{Bicudo:2017szl}.

In this work we continue to use lattice QCD potentials and the Born-Oppenheimer approximation and focus on quarkonium bound states and resonances, which might be exotic, i.e.\ states containing a heavy quark and a heavy antiquark and possibly an additional light quark-antiquark pair. To study for example the experimentally observed $Z_b$ tetraquark resonances, it is necessary to extend the techniques introduced in \cite{Bicudo:2017szl} from a single channel to a coupled channel Schr\"odinger equation. One has to consider a confined quarkonium channel $\bar{Q} Q$ and at least one scattering channel $\bar{M} M$ with two heavy-light mesons $M = \bar{Q} q$ and $\bar{M} = \bar{q} Q$. In the least complicated case of a single scattering channel the potential in the Schr\"odinger equation is a $2 \times 2$ matrix of the form
\begin{eqnarray}
\label{eq:matrixpotential} V(r) = \left(\begin{array}{cc}
V_{\bar{Q} Q}(r) & V_{\textrm{mix}}(r) \\
V_{\textrm{mix}}(r) & V_{\bar{M} M}(r)
\end{array}\right) 
\end{eqnarray}
as we will derive in detail in section~\ref{sec:potentials}. It is important to note that the off-diagonal terms $V_{\textrm{mix}}(r)$ couple the quarkonium channel and the meson-meson channel. The consequence is that quarkonium bound states only exist below the $\bar{M} M$ threshold, whereas above this threshold all quarkonium states are resonances. Once we have set up an appropriate coupled channel Schr\"odinger equation we proceed as in Ref.\ \cite{Bicudo:2017szl} and compute phase shifts and $\mbox{T}$ matrix poles, where the latter provide masses of bound states below threshold as well as resonance masses and decay widths above threshold.

The main advantage of our approach is that it is in principle straightforward to consider several decay channels, since this is done in the framework of quantum mechanics in the form of a coupled channel Schr\"odiger equation. Each decay channel, however, requires the lattice QCD computation of static potentials with specific quantum numbers. Even though this might be time consuming and challenging, these static potentials are not only important for the approach we propose in this paper, but they will be of interest for theoretical hadron physics in general (as an example see the detailed discussion of static potentials for the $\bar{b} \bar{b} u d$ case in Ref.\ \cite{Bicudo:2015kna}). As mentioned above, it would be very interesting to study the experimentally observed $Z_b$ tetraquark resonances with $I = 1$. The lattice QCD computation of the potential matrix (\ref{eq:matrixpotential}) for $I = 1$ is, however, very difficult (see e.g.\ Refs.\ \cite{Peters:2017hon,Prelovsek:2019yae}). Therefore, we decided to first explore the simpler $I = 0$ case, where corresponding high-quality lattice QCD potentials are provided by the string breaking computation of Ref.\ \cite{Bali:2005fu}. This allows us to study the established quarkonium states $\eta_b(1S)$, $\Upsilon(1S)$, $\Upsilon(2S)$, $\Upsilon(3S)$ and $\Upsilon(4S)$, possibly also $\Upsilon(10860)$ and $\Upsilon(11020)$ and to predict additional quarkonium resonances not yet observed experimentally.

This paper is organized as follows. In section \ref{sec:potentials} we detail the theoretical basics of our approach. We start by discussing quantum numbers of quarkonium and of pairs of heavy-light mesons for $I = 0$. Then we show, how to set up a corresponding coupled channel Schr\"odiger equation in a consistent way, and explain, how the potential matrix is related to static potentials from QCD, which can be computed with lattice QCD. We also discuss the boundary conditions of the wave function, which are appropriate for a coupled channel scattering problem. Moreover, we specialize the coupled channel Schr\"odiger equation for the specific case of relative orbital angular momentum $L_{\bar{Q} Q} = 0$ for the two heavy quarks, which will be the starting point for all numerical results presented later in the paper. In section \ref{sec:Bali} we extract the static potentials we need, i.e.\ the elements of the matrix (\ref{eq:matrixpotential}), from lattice QCD data from Ref.\ \cite{Bali:2005fu}. In section \ref{sec:emergent} we discuss numerical methods to solve the coupled channel Schr\"odiger equation. We use these methods in section \ref{sec:results} to predict quarkonium bound states and resonances for $I = 0$ and compare to existing experimental results. Finally, in section \ref{sec:conclu} we conclude and present an outlook.


\section{\label{sec:potentials}Theoretical basics of studying quarkonium resonances using lattice QCD potentials}

In this section we present the theoretical basics of our approach to study quarkonium resonances for isospin $I = 0$ and various $J^{P C}$ in the Born-Oppenheimer approximation with lattice QCD potentials. We start by analyzing quantum numbers of quarkonium systems and of corresponding decay channels of two heavy-light mesons. Then we derive a coupled channel Schr\"odinger equation containing quarkonium and two-meson channels. We also discuss, how the potentials appearing in the Schr\"odinger equation are related to static potentials, which can be computed using lattice QCD. Finally we formulate the boundary conditions for meson-meson scattering and specialize the Schr\"odinger equation to a specific sector by performing a partial wave decomposition.

Notice that we ignore decays of quarkonium to a lighter quarkonium and a light $I = 0$ meson, e.g.\ a $\sigma$ or an $\eta$ meson. Such a decay is suppressed by the OZI rule \cite{Okubo:1963fa,Zweig:1981pd,Iizuka:1966fk}, when compared to the decay to a pair of heavy-light mesons. This is consistent with experimental observations, where the dominant hadronic decay is the decay to a pair of heavy-light mesons. It might be possible to also study these OZI suppressed decays using lattice QCD potentials and the Born-Oppenheimer approximation, but this seems even more technical and difficult than the decays addressed here and, thus, we leave them for future research.


\subsection{\label{SEC466}Quantum numbers of $\bar{Q} Q$ (quarkonium) and of $\bar{M} M$ (two heavy-light mesons)}

We consider systems with a heavy quark-antiquark pair $\bar{Q} Q$ and either no light quarks (i.e.\ quarkonium) or another light quark-antiquark pair $\bar{q} q$ with isospin $I = 0$ (two heavy-light mesons $M = \bar{Q} q$ and $\bar{M} = \bar{q} Q$ for large $\bar{Q} Q$ separation). The quantum numbers of these systems are denoted in the following way:
\begin{itemize}
\item $J^{P C}$: total angular momentum, parity and charge conjugation of the $\bar{Q} Q$ or the $\bar{Q} Q \bar{q} q$ system.

\item $S_Q^{P C}$: spin of $\bar{Q} Q$ and corresponding parity and charge conjugation.

\item $\widetilde{J}^{P C}$: total angular momentum excluding the heavy spins of $Q$ and $\bar{Q}$ and corresponding parity and charge conjugation (for quarkonium, i.e.\ $\bar{Q} Q$ without light quarks, $\widetilde{J}^{P C}$ coincides with the relative orbital angular momentum $L^{P C}_{\bar{Q} Q}$ of the two heavy quarks).
\end{itemize}

For the heavy quarks $Q$ and $\bar{Q}$ we use the following approximations:
\begin{itemize}
\item[(1)] \textit{Heavy quark spins are conserved quantities.}
\\ Consequently, the energy levels of the $\bar{Q} Q$ and the $\bar{Q} Q \bar{q} q$ systems as well as their decays and and resonance parameters do not depend on the spins of the heavy quarks $Q$ and $\bar{Q}$, i.e.\ are independent of $S_Q^{P C}$.

\item[(2)] \textit{Two of the four components of the Dirac spinors of the heavy quarks $Q$ and $\bar{Q}$ vanish.}
\\ Thus, one can write $Q = P_- Q$ and $\bar{Q} = \bar{Q} P_+$, where $P_\pm = (1 \pm \gamma_0) / 2$ are the projectors to the large and small components of the non-relativistic limit.
\end{itemize}
These approximations become exact for static quarks and should still yield reasonably accurate results for $b$ quarks, possibly even for $c$ quarks. For example the ground state pseudoscalar meson and the ground state vector meson have the same mass in the static limit, while for $b$ quarks $m_{\Upsilon(1S)} - m_{\eta_b(1S)} \approx 62 \, \textrm{MeV} $ and for $c$ quarks $m_{J/\Psi(1S)} - m_{\eta_c(1S)} \approx 113 \, \textrm{MeV}$. These mass differences can be considered as crude estimates of the systematic error associated with mass predictions within our approach. The estimate for $b$ quarks is supported by our numerical results for the masses of $\Upsilon(1S)$, $\Upsilon(2S)$, $\Upsilon(3S)$ and $\Upsilon(4S)$, which differ from corresponding experimental results by around $50 \, \textrm{MeV}$ or less.

Since $J^{P C}$ as well as $S_Q^{P C}$ are conserved, $\widetilde{J}^{P C}$ is also conserved. For each value of $\widetilde{J}^{P C}$ the corresponding coupled channel Schr\"odinger equation is different (see section~\ref{SEC734} for $\widetilde{J}^{P C} = 0^{+ +}$). Thus $\widetilde{J}^{P C}$ is of central importance throughout this work, similar as $J^{P C}$ for systems without heavy quarks, while both $J^{P C}$ and $S_Q^{P C}$ are less relevant.


\subsubsection{$\bar{Q} Q$ (quarkonium bound states and resonances)}

As discussed above, for quarkonium, i.e.\ a $\bar{Q} Q$ pair without light quarks, $\widetilde{J}^{P C}$ coincides with the relative orbital angular momentum $L^{P C}_{\bar{Q} Q}$ of the two quarks. Thus, possible values are $\widetilde{J}^{P C} = 0^{+ +} , 1^{- -} , 2^{+ +} \ldots$

The coupling of the two heavy spinors can be written according to 
\begin{eqnarray}
\label{EQN633} \textrm{spin}_{\bar{Q} Q} = \bar{Q} \Gamma_Q Q = (\bar{Q} P_+) \Gamma_Q (P_- Q) ,
\end{eqnarray} 
where $\Gamma_Q$ is a $4 \times 4$ matrix. It is easy to show, that there are only four linearly independent choices for $\Gamma_Q$ such that the right hand side of Eq.\ (\ref{EQN633}) does not vanish: $\Gamma_Q = P_+ \gamma_5$ corresponding to $S_Q^{P C} = 0^{- +}$ and $\Gamma_Q = P_+ \gamma_j$ ($j = 1,2,3$) corresponding to $S_Q^{P C} = 1^{- -}$ (see also Table~\ref{TAB017}, where the coupling of two heavy spinors is summarized). Within the approximations discussed at the beginning of section~\ref{SEC466} quarkonium energy levels are independent of the heavy spins and, thus, independent of $\Gamma_Q$. 

\begin{table}[htb]
\begin{center}
$\bar{Q} Q$ (two heavy spinors)
\vspace{0.1cm}
\\
\begin{tabular}{c|c}
\hline
\TBstrut
$S_Q^{P C}$ & $\Gamma_Q$ in $\bar{Q} \Gamma_Q Q$ \\
\hline
\Tstrut
$0^{- +}$ & $P_+ \gamma_5$ \\
\Bstrut
$1^{- -}$ & $P_+ \gamma_j$ \\
\hline
\end{tabular}
\vspace{0.3cm}
\\
$\bar{Q} q$ and $\bar{q} Q$ (one heavy and one light spinor)
\vspace{0.1cm}
\\
\begin{tabular}{c|c|c}
\hline
\TBstrut
$S^P$ & $\Gamma$ in $\bar{Q} \Gamma q$ & $\Gamma$ in $\bar{q} \Gamma Q$ \\
\hline
\Tstrut
$0^-$ & $P_+ \gamma_5$ & $\gamma_5 P_-$ \\
$0^+$ & $P_+$ & $P_-$ \\
$1^-$ & $P_+ \gamma_j$ & $\gamma_j P_-$ \\
\Bstrut
$1^+$ & $P_+ \gamma_j \gamma_5$ & $\gamma_j \gamma_5 P_-$ \\
\hline
\end{tabular}
\vspace{0.3cm}
\\
$\bar{q} q$ (two light spinors)
\vspace{0.1cm}
\\
\begin{tabular}{c|c}
\hline
\TBstrut
$S_q^{P C}$ & $\Gamma_q$ in $\bar{q} \Gamma_q q$ \\
\hline
\Tstrut
$0^{- +}$ & $P_+ \gamma_5$, $ P_- \gamma_5$ \\
$0^{+ +}$ & $1$ \\
$0^{+ -}$ & $\gamma_0$ \\
$1^{- -}$ & $P_+ \gamma_j$, $P_- \gamma_j$ \\
$1^{+ +}$ & $\gamma_j \gamma_5$ \\
\Bstrut
$1^{+ -}$ & $\gamma_0 \gamma_j \gamma_5$ \\
\hline
\end{tabular}
\caption{\label{TAB017}Possibilities to couple two spinors: quantum numbers and $\gamma$ matrices.}
\end{center}
\end{table}

$S_Q^{P C}$  and $\widetilde{J}^{P C} = L^{P C}_{\bar{Q} Q}$ can be coupled in the usual way to definite total angular momentum $J^{P C}$. For $\widetilde{J} = L_{\bar{Q} Q} = 0, 1, 2$ all possibilities are listed in Table~\ref{TAB004}.

\begin{table}[htb]
\begin{center}
\begin{tabular}{cc|c||c}
\hline
\TBstrut
$S_Q^{P C}$ & $\Gamma_Q$ & $ \widetilde{J}^{P C}=L^{P C}_{\bar{Q} Q}$ & $J^{P C}$ \\
\hline
\Tstrut
$0^{- +}$ & $P_+ \gamma_5$ & $0^{+ +}$ & $0^{- +}$ \\
          &                & $1^{- -}$ & $1^{+ -}$ \\
          &                & $2^{+ +}$ & $2^{- +}$ \\
\Bstrut
          &                & $\ldots $ & $\ldots $ \\
\hline
\Tstrut
$1^{- -}$ & $P_+ \gamma_j$ & $0^{+ +}$ & $1^{- -}$ \\
          &                & $1^{- -}$ & $0^{+ +} \ / \ 1^{+ +} \ / \ 2^{+ +}$ \\
          &                & $2^{+ +}$ & $1^{- -} \ / \ 2^{- -} \ / \ 3^{- -}$ \\
\Bstrut
          &                & $\ldots $ & $\ldots $ \\
\hline
\end{tabular}
\caption{\label{TAB004}$\bar{Q} Q$ (quarkonium): possibilities to couple the separately conserved $S_Q^{P C}$ and $\widetilde{J}^{P C}$ to definite $J^{P C}$.}
\end{center}
\end{table}


\subsubsection{$\bar{M} M$ (two heavy-light mesons, the decay channels of quarkonium resonances)}

\begin{table}[b]
\begin{center}
\begin{tabular}{c|c|ccc}
\hline
\TBstrut
$ \widetilde{J}^{P C}$ & $L^{P C}_{\bar{M} M}$ & $S_q^{P C}$ & $\Gamma_q$ & type of $M$ and $\bar{M}$ \\
\hline
\Tstrut
$0^{+ +}$ & $0^{+ +}$ & $0^{+ +}$ & $1$ & one $P = -$ and one $P = +$ meson \\
          & $1^{- -}$ & $1^{- -}$ & $P_+ \gamma_j$ & two $P = -$ mesons \\
\Bstrut
          &           &           & $P_- \gamma_j$ & two $P = +$ mesons \\
\hline
\Tstrut
$1^{- -}$ & $0^{+ +}$ & $1^{- -}$ & $P_+ \gamma_j$ & two $P = -$ mesons \\
          &           &           & $P_- \gamma_j$ & two $P = +$ mesons \\
          & $1^{- -}$ & $0^{+ +}$ & $1$ & one $P = -$ and one $P = +$ meson \\
          &           & $1^{+ +}$ & $\gamma_j \gamma_5$ & one $P = -$ and one $P = +$ meson \\
          & $2^{+ +}$ & $1^{- -}$ & $P_+ \gamma_j$ & two $P = -$ mesons \\
\Bstrut
          &           &           & $P_- \gamma_j$ & two $P = +$ mesons \\
\hline
\Tstrut
$2^{+ +}$ & $1^{- -}$ & $1^{- -}$ & $P_+ \gamma_j$ & two $P = -$ mesons \\
          &           &           & $P_- \gamma_j$ & two $P = +$ mesons \\
          & $2^{+ +}$ & $0^{+ +}$ & $1$ & one $P = -$ and one $P = +$ meson \\
          &           & $1^{+ +}$ & $\gamma_j \gamma_5$ & one $P = -$ and one $P = +$ meson \\
          & $3^{- -}$ & $1^{- -}$ & $P_+ \gamma_j$ & two $P = -$ mesons \\
\Bstrut
          &           &           & $P_- \gamma_j$ & two $P = +$ mesons \\
\hline
\TBstrut
$\ldots$ & $\ldots$ & $\ldots$ & $\ldots$ \\
\hline
\end{tabular}
\caption{\label{TAB005}$\bar{M} M$ (two heavy-light mesons): possibilies to couple relative orbital angular momentum $L^{P C}_{\bar{M} M}$ and light spin $S_q^{P C}$ to given $\widetilde{J}^{P C}$.}
\end{center}
\end{table}

It is convenient to write the spin coupling of the four quarks forming the two heavy-light mesons $M$ and $\bar{M}$ as
\begin{eqnarray}
\nonumber & & \textrm{spin}_{\bar{M} M} = \Gamma_{Q,A B} \Gamma_{q,C D} \underbrace{\Big((\bar{Q} P_+)_A q_D\Big)}_{\equiv M} \underbrace{\Big(\bar{q}_C (P_- Q)_B\Big)}_{\equiv \bar{M}} \\
\label{EQN030} & &
\end{eqnarray}
($A$, $B$, $C$ and $D$ are spin indices), i.e.\ such that the heavy spins are coupled with $\Gamma_Q$ and the light spins are coupled with $\Gamma_q$. To study quarkonium resonances, which can decay into two heavy-light mesons with a coupled channel Schr\"odinger equation, the quarkonium channel and the two-meson decay channels must have identical $S_Q^{P C}$ and $\widetilde{J}^{P C}$ (and identical corresponding $z$ components of $S_Q$ and $\widetilde{J}$). This implies that the heavy quark spins of $M$ and $\bar{M}$ have to be coupled with the same $\Gamma_Q$ as in the quarkonium case (\ref{EQN633}). There are, however, several possibilities to couple the relative orbital angular momentum of the two mesons $L^{P C}_{\bar{M} M}$ and the light spin $S_q^{P C}$ to given $\widetilde{J}^{P C}$. The algebra is straightforward, when using Table~\ref{TAB017}. For $\widetilde{J} = 0, 1, 2$ all possibilities are listed in Table~\ref{TAB005} together with the corresponding $\Gamma_q$.

A heavy-light meson $M$ or $\bar{M}$ with the heavy and the light quark in an S wave can have total angular momentum $J_M = 0$ or $J_M = 1$ and parity $P = -$ or $P = +$. The negative parity mesons have similar mass and the positive parity mesons have similar mass, i.e.\ $m_{J_M^P = 0^-} \approx m_{J_M^P = 1^-}$ and $m_{J_M^P = 0^+} \approx m_{J_M^P = 1^+}$. However, the positive parity mesons are roughly $400 \, \textrm{MeV} \ldots 500 \, \textrm{MeV}$ heavier (cf.\ e.g.\ \cite{Jansen:2008ht,Michael:2010aa}). Since including lighter decay channels in a coupled channel Schr\"odinger equation seems more important than including heavier decay channels, it is a necessary step to analyze, which types of heavy-light mesons correspond to the spin couplings listed in Table~\ref{TAB005}, in particular, which $\Gamma_q$ corresponds to two $P = -$ mesons. This can be done using the Fierz identity,
\begin{widetext}
\begin{eqnarray}
\label{EQN001} \Big(\bar{\psi}_1 \Gamma_A \psi_2\Big) \Big(\bar{\psi}_3 \Gamma_B \psi_4\Big) 
 = \sum_{C,D} \alpha_{A B C D} \Big(\bar{\psi}_1 \Gamma_C \psi_4\Big) \Big(\bar{\psi}_3 \Gamma_D \psi_2\Big) \quad , \quad \alpha_{A B C D} = \frac{1}{16} \textrm{Tr}\Big(\Gamma_C \Gamma_A \Gamma_D \Gamma_B\Big)
\end{eqnarray}
with $\bar{\psi}_1 \rightarrow \bar{Q}$, $\psi_2 \rightarrow Q$, $\bar{\psi}_3 \rightarrow \bar{q}$, $\psi_4 \rightarrow q$. Since the resulting energy levels and resonance parameters will not depend on the heavy spins, we consider from now on exclusively the technically simpler $S_Q^{P C} = 0^{- +}$, i.e.\ $\Gamma_A = \Gamma_Q = P_+ \gamma_5$. For $\Gamma_B = \Gamma_q$ there are four possibilities as can be seen from Table~\ref{TAB005}:
\begin{itemize}
\item $\Gamma_q = P_+ \gamma_j$:
\begin{eqnarray}
\label{EQN031} \textrm{spin}_{\bar{M} M} = +\bigg(\underbrace{\Big((\bar{Q} P_+) \gamma_5 q\Big)}_{\equiv M , \, P = -} \underbrace{\Big(\bar{q} \gamma_j (P_- Q)\Big)}_{\equiv \bar{M} , \, P = -} + (\gamma_5 \leftrightarrow \gamma_j)\bigg) - \sum_{k,l} \epsilon_{j k l} \underbrace{\Big((\bar{Q} P_+) \gamma_k q\Big)}_{\equiv M , \, P = -} \underbrace{\Big(\bar{q} \gamma_l (P_- Q)\Big)}_{\equiv \bar{M} , \, P = -} ,
\end{eqnarray}
i.e.\ a linear combination of two negative parity heavy-light mesons, which can be read off by comparing to Table~\ref{TAB017}. Notice that this is the lightest decay channel and, thus, of central importance. Later, when setting up a coupled channel Schr\"odinger equation, we will include this channel, but neglect the following three channels, which are heavier.

\item $\Gamma_q = 1$:
\begin{eqnarray}
\label{EQN031b} \textrm{spin}_{\bar{M} M} = +\frac{1}{2} \bigg(\underbrace{\Big((\bar{Q} P_+) \gamma_5 q\Big)}_{\equiv M , \, P = -} \underbrace{\Big(\bar{q} (P_- Q)\Big)}_{\equiv \bar{M} , \, P = +} + (\gamma_5 \leftrightarrow 1)\bigg) + \frac{1}{2} \sum_j \bigg(\underbrace{\Big((\bar{Q} P_+) \gamma_j q\Big)}_{\equiv M , \, P = -} \underbrace{\Big(\bar{q} \gamma_j \gamma_5 (P_- Q)\Big)}_{\equiv \bar{M} , \, P = +} - (\gamma_j \leftrightarrow \gamma_j \gamma_5)\bigg) ,
\end{eqnarray}
i.e.\ a linear combination of a negative and a positive parity heavy-light meson.

\item $\Gamma_q = \gamma_j \gamma_5$:
\begin{eqnarray}
\nonumber & & \textrm{spin}_{\bar{M} M} = +\frac{1}{2} \bigg(\underbrace{\Big((\bar{Q} P_+) \gamma_5 q\Big)}_{\equiv M , \, P = -} \underbrace{\Big(\bar{q} \gamma_j \gamma_5 (P_- Q)\Big)}_{\equiv \bar{M} , \, P = +} + (\gamma_5 \leftrightarrow \gamma_j \gamma_5)\bigg) + \frac{1}{2} \bigg(\underbrace{\Big((\bar{Q} P_+) q\Big)}_{\equiv M , \, P = +} \underbrace{\Big(\bar{q} \gamma_j (P_- Q)\Big)}_{\equiv \bar{M} , \, P = -} - (1 \leftrightarrow \gamma_j)\bigg) \\
\label{EQN031c} & & \hspace{0.7cm} - \frac{1}{2} \sum_{k,l} \epsilon_{j k l} \bigg(\underbrace{\Big((\bar{Q} P_+) \gamma_k q\Big)}_{\equiv M , \, P = -} \underbrace{\Big(\bar{q} \gamma_l \gamma_5 (P_- Q)\Big)}_{\equiv \bar{M} , \, P = +} + (\gamma_j \leftrightarrow \gamma_j \gamma_5)\bigg) ,
\end{eqnarray}
i.e.\ a linear combination of a negative and a positive parity heavy-light meson.

\item $\Gamma_q = P_- \gamma_j$:
\begin{eqnarray}
\label{EQN031d} \textrm{spin}_{\bar{M} M} = +\bigg(\underbrace{\Big((\bar{Q} P_+) q\Big)}_{\equiv M , \, P = +} \underbrace{\Big(\bar{q} \gamma_j \gamma_5 (P_- Q)\Big)}_{\equiv \bar{M} , \, P = +} - (1 \leftrightarrow \gamma_j \gamma_5)\bigg) - \sum_{k,l} \epsilon_{j k l} \underbrace{\Big((\bar{Q} P_+) \gamma_k \gamma_5 q\Big)}_{\equiv M , \, P = +} \underbrace{\Big(\bar{q} \gamma_l \gamma_5 (P_- Q)\Big)}_{\equiv \bar{M} , \, P = +} ,
\end{eqnarray}
i.e.\ a linear combination of two positive parity heavy-light mesons.
\end{itemize}

\end{widetext}


\subsection{\label{SEC885}$\bar{Q} Q$ and $\bar{M} M$ coupled channel Schr\"odinger equation}

Setting up the coupled channel Schr\"odinger equation is independent of the heavy spins. Therefore, as already stated in the previous subsection, we consider the technically simpler $S_Q^{P C} = 0^{- +}$, i.e.\ $\Gamma_Q = P_+ \gamma_5$.

In addition to a quarkonium channel characterized by $\widetilde{J}^{P C} = L^{P C}_{\bar{Q} Q}$, we consider the lightest two-meson decay channel, which contains two negative parity mesons $M$ and $\bar{M}$ (see Eq.\ (\ref{EQN031})). We ignore the heavier two-meson decay channels, which include one or even two positive parity mesons (see Eqs.\ (\ref{EQN031b}), (\ref{EQN031c}) and (\ref{EQN031d})), because their threshold energy is higher by more than $400 \, \textrm{MeV}$ or $800 \, \textrm{MeV}$, respectively. The corresponding light spin is $S_q^{P C} = 1^{- -}$ and $\Gamma_q = P_+ \gamma_j$ (see Table~\ref{TAB005}). Thus, the wave function of the coupled channel Schr\"odinger equation has four components, $\psi(\mathbf{r}) = (\psi_{\bar{Q} Q}(\mathbf{r}) , \vec{\psi}_{\bar{M} M}(\mathbf{r}))$. The first component $\psi_{\bar{Q} Q}(\mathbf{r})$ represents the $\bar{Q} Q$ quarkonium channel, while the remaining three components $\vec{\psi}_{\bar{M} M}(\mathbf{r})$ correspond to the three spin orientations of the $S_q = 1$ triplet of the $\bar{M} M$ two-meson channel. $\mathbf{r}$ is the relative coordinate of the two heavy quarks $\bar{Q} Q$, which is for $\vec{\psi}_{\bar{M} M}(\mathbf{r})$ equivalent to the separation of the two heavy-light mesons $\bar{M} M$.

The spin algebra is $[S_j , S_k] = i \epsilon_{j k l} S_l$. For the three components of $\vec{\psi}_{\bar{M} M}(\mathbf{r})$ we choose as generators
\begin{eqnarray}
\nonumber & & S_x = \left(\begin{array}{ccc}
 0 & 0 & 0 \\
 0 & 0 & -i \\
 0 & +i & 0
\end{array}\right) \quad , \quad 
S_y = \left(\begin{array}{ccc}
 0 & 0 & +i \\
 0 & 0 & 0 \\
 -i & 0 & 0
\end{array}\right) \quad , \\
 & & \hspace{0.7cm} S_z = \left(\begin{array}{ccc}
 0 & -i & 0 \\
 +i & 0 & 0 \\
 0 & 0 & 0
\end{array}\right) ,
\end{eqnarray}
i.e.\ the generators of rotations around the three Cartesian axes. The eigenvectors of $S_z$ are $\mathbf{v}_0 = (0,0,1)$ with eigenvalue $0$ and $\mathbf{v}_\pm = (1,\pm i,0) / \sqrt{2}$ with eigenvalues $\pm 1$.

The coupled channel Schr\"odinger equation reads
\begin{widetext}
\begin{eqnarray}
\label{EQN050} \bigg(-\frac{1}{2} \mu^{-1} \bigg(\partial_r^2 + \frac{2}{r} \partial_r - \frac{\mathbf{L}^2}{r^2}\bigg) + V(\mathbf{r}) + 2 m_M - E\bigg) \psi(\mathbf{r}) = 0 ,
\end{eqnarray}
where $\mu^{-1} = \diag(1/\mu_Q , 1/\mu_M , 1/\mu_M , 1/\mu_M)$ is a $4 \times 4$ diagonal matrix, $\mu_Q = m_Q / 2$ and $\mu_M = m_M / 2$ are the reduced heavy quark and heavy-light meson masses and $\mathbf{L} = \mathbf{r} \times \mathbf{p}$ is the orbital angular momentum operator. The potential $V(\mathbf{r})$ is also a $4 \times 4$ matrix, which can be written as
\begin{eqnarray}
\label{EQN051} V(\mathbf{r}) = \left(\begin{array}{cc}
V_{\bar{Q} Q}(r) & V_\textrm{mix}(r) \Big(1 \otimes \mathbf{e}_r\Big) \\
V_\textrm{mix}(r) \Big(\mathbf{e}_r \otimes 1\Big) & V_{\bar{M} M,\parallel}(r) \Big(\mathbf{e}_r \otimes \mathbf{e}_r\Big) + V_{\bar{M} M,\perp}(r) \Big(1 - \mathbf{e}_r \otimes \mathbf{e}_r\Big)
\end{array}\right) .
\end{eqnarray}
\end{widetext}
This particular structure is derived and discussed in the following section~\ref{SEC582}, where we treat the heavy quarks $Q$ and $\bar{Q}$ in the static limit and relate $V(\mathbf{r})$ to static potentials from QCD. In section~\ref{sec:Bali} we explain, how to compute the four functions $V_{\bar{Q} Q}(r)$, $V_{\bar{M} M,\parallel}(r)$, $V_{\bar{M} M,\perp}(r)$ and $V_\textrm{mix}(r)$ on the right hand side of Eq.\ (\ref{EQN051}) using lattice QCD. The non-zero off-diagonal elements in the first column and the first row proportional to $V_\textrm{mix}(r)$ lead to mixing of the quarkonium channel and the two-meson channels and, thus, to quarkonium resonances.

Note that in a previous paper \cite{Bicudo:2016ooe} we have used similar techniques to derive a coupled channel Schr\"odinger equation for an $I(J^P) = 0(1^+)$ $\bar{Q} \bar{Q} q q$ tetraquark system, to explore the effect of the heavy quark spins.


\subsection{\label{SEC582}Relating the potentials in the coupled channel Schr\"odinger equation to static potentials from QCD}

In this subsection we treat the heavy quarks $Q$ and $\bar{Q}$ as static quarks. This allows to relate the potential matrix $V(\mathbf{r})$ appearing in the coupled channel Schr\"odinger equation (\ref{EQN050}), i.e.\ the four potentials $V_{\bar{Q} Q}(r)$, $V_{\bar{M} M,\parallel}(r)$, $V_{\bar{M} M,\perp}(r)$ and $V_\textrm{mix}(r)$ (see Eq.\ (\ref{EQN051})) to static potentials from QCD, which can be computed using lattice QCD. Moreover, we explain, why $V(\mathbf{r})$ has the particular structure given in Eq.\ (\ref{EQN051}).

The positions of static quarks are frozen, for $Q$ and $\bar{Q}$ w.l.o.g.\ at $+\mathbf{r}/2$ and $-\mathbf{r}/2$, respectively, i.e.\ their separation is $r = |\mathbf{r}|$. Thus, rotational symmetry, parity and charge conjugation are broken. Remaining symmetry transformations are rotations around the $\bar{Q} Q$ separation axis, parity combined with charge conjugation (operator $P \circ C$) and spatial reflection along an axis perpendicular to the $\bar{Q} Q$ separation axis (corresponding operator denoted as $P_x$). States are labeled by quantum numbers $\Lambda_\eta^\epsilon$, where the $\bar{Q} Q$ spin is not included, because static spins are conserved quantities (for a detailed discussion of this notation, which is also used for homonuclear diatomic molecules and for excited flux tubes, see Refs.\ \cite{Juge:2002br,Bali:2005fu,Bicudo:2018jbb,Capitani:2018rox}):
\begin{itemize}
\item $\Lambda = 0, 1, 2, \ldots = \Sigma, \Pi, \Delta, \ldots$ is the absolute value of total angular momentum with respect to the $\bar{Q} Q$ separation axis.

\item $\eta = +, - = g, u$ is the eigenvalue with respect to the operator $P \circ C$.

\item $\epsilon = +, -$ is the eigenvalue with respect to the operator $P_x$ (sectors with $\Lambda \neq 0$, which differ only in the quantum number $\epsilon$, have degenerate spectra; therefore, it is common to list $\epsilon$ only for $\Lambda = \Sigma$ and to omit $\epsilon$ for $\Lambda = \Pi, \Delta, \ldots$).
\end{itemize}
Since the three quantum numbers $\Lambda_\eta^\epsilon$ do not include the $\bar{Q} Q$ spin, they play a similar role as $\widetilde{J}^{P C}$, when the heavy quark positions are not frozen.

\begin{figure}[b]
\includegraphics[width=0.49\textwidth]{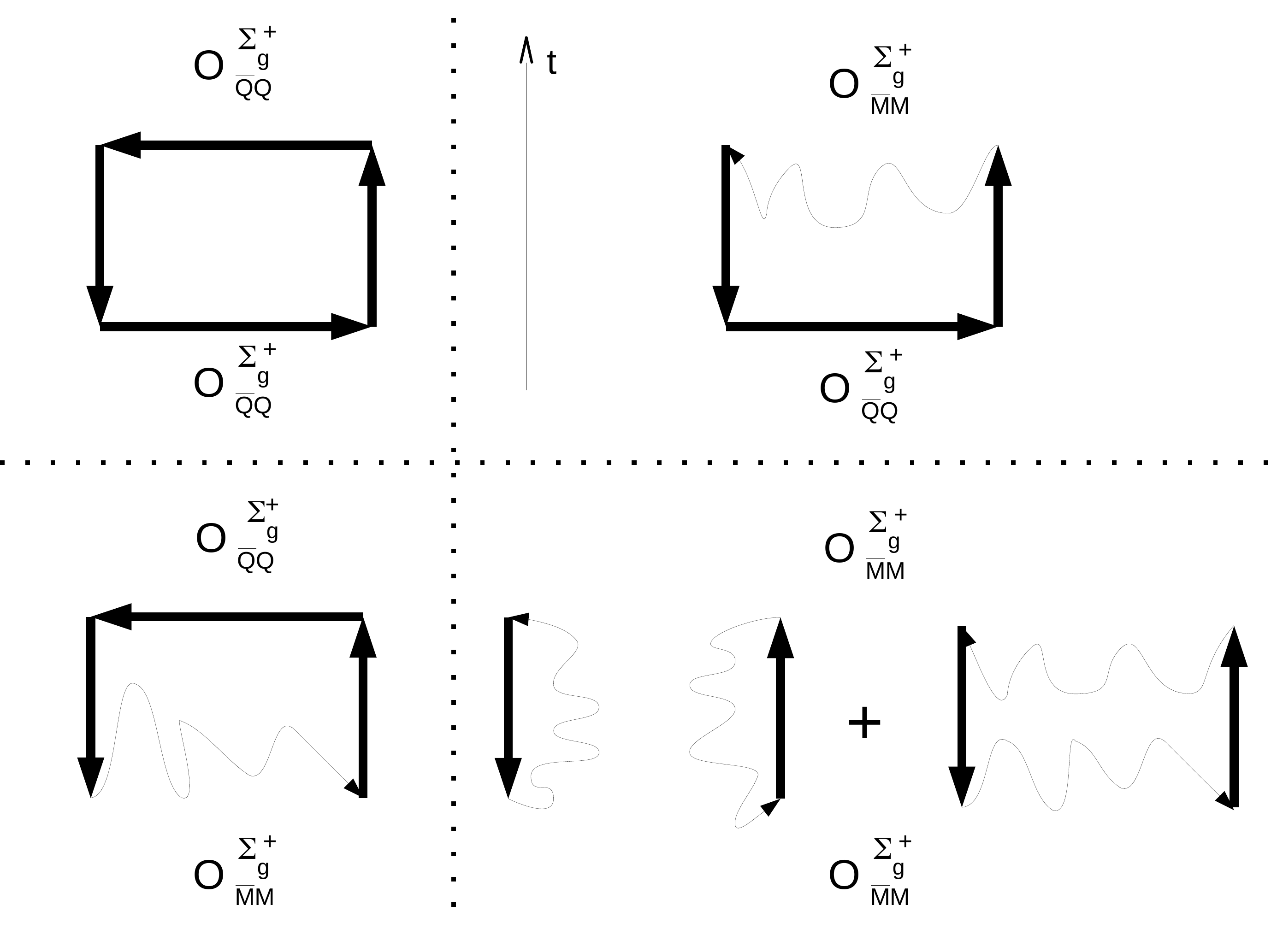}
\caption{\label{FIG_C}The correlation matrix $\langle \mathcal{O}^\dagger_j(t) \mathcal{O}_k(0) \rangle$ in diagrammatic form. The straight arrows represent parallel transporters, which appear in the creation operator $\mathcal{O}_{\bar{Q} Q}^{\Sigma_g^+}$ (see Eq.\ (\ref{EQN407})) and the static quark propagators, while the wiggly lines represent light $u$ and $d$ quark propagators.}
\end{figure}

A quarkonium system $\bar{Q} Q$ with static quarks and the gluonic flux tube in the ground state has quantum numbers $\Lambda_\eta^\epsilon = \Sigma_g^+$, i.e.\ the flux tube is invariant under rotations around and reflections parallel and perpendicular to the $\bar{Q} Q$ separation axis. For the $\bar{M} M$ channels with quark content $\bar{Q} Q \bar{q} q$ and static quarks $Q$ and $\bar{Q}$ the situation is more complicated, because of the spins of the two light quarks, which are coupled according to $\bar{q} \Gamma_q q$ (see Eq.\ (\ref{EQN030})). Since we exclusively consider channels containing two negative parity heavy-light mesons, there are only three independent possibilities, $\Gamma_q \in \{ P_+ \gamma_1 , P_+ \gamma_2 , P_+ \gamma_3 \}$ (see Table~\ref{TAB005}). The linear combination $\Gamma_q = \mathbf{e}_r P_+ \vec{\gamma} = (\mathbf{e}_r)_j P_+ \gamma_j$ corresponds to light spin $S_q = 1$ parallel to the $\bar{Q} Q$ separation axis (here and in the following $\mathbf{e}_r$, $\mathbf{e}_\vartheta$ and $\mathbf{e}_\varphi$ denote the orthogonal basis vectors of spherical coordinates). The quantum numbers are $\Lambda_\eta^\epsilon = \Sigma_g^+$, i.e.\ are identical to those of the quarkonium system. The remaining two combinations, $\Gamma_q = \mathbf{e}_\vartheta P_+ \vec{\gamma} = (\mathbf{e}_\vartheta)_j P_+ \gamma_j$ and $\Gamma_q = \mathbf{e}_\varphi P_+ \vec{\gamma}  = (\mathbf{e}_\varphi)_j P_+ \gamma_j$, correspond to light spin $S_q = 1$ perpendicular to the $\bar{Q} Q$ separation axis with quantum numbers $\Lambda_\eta^\epsilon = \Pi_g^+$ and $\Lambda_\eta^\epsilon = \Pi_g^-$, respectively (for definiteness we have choosen as direction of reflection for $P_x$ the $\varphi$ direction; note, however, that final results are independent of this choice).

The computation of static potentials with quantum numbers $\Lambda_\eta^\epsilon$ with lattice QCD can be done as explained in detail in Ref.\ \cite{Bali:2005fu}. For $\Lambda_\eta^\epsilon = \Sigma_g^+$ there is mixing between a quarkonium-like static potential and a two-meson-like static potential. To compute both potentials, two creation operators are needed,
\begin{widetext}
\begin{eqnarray}
\label{EQN407} & & \mathcal{O}_{\bar{Q} Q}^{\Sigma_g^+} = (\underbrace{P_+ \gamma_5}_{= \Gamma_Q})_{A B} \Big(\bar{Q}_A(-\mathbf{r}/2) U(-\mathbf{r}/2;+\mathbf{r}/2) Q_B(+\mathbf{r}/2)\Big) \\
\label{EQN408} & & \mathcal{O}_{\bar{M} M}^{\Sigma_g^+} = (\underbrace{P_+ \gamma_5}_{= \Gamma_Q})_{A B} (\underbrace{\mathbf{e}_r P_+ \vec{\gamma}}_{= \Gamma_q})_{C D} \Big(\bar{Q}_A(-\mathbf{r}/2) u_D(-\mathbf{r}/2)\Big) \Big(\bar{u}_C(\mathbf{r}/2) Q_B(+\mathbf{r}/2) + (u \rightarrow d)\Big) ,
\end{eqnarray}
where the first operator is of quarkonium type ($U(-\mathbf{r}/2;+\mathbf{r}/2)$ is a straight spatial parallel transporter connecting $-\mathbf{r}/2$ and $+\mathbf{r}/2$, typically a product of smeared spatial links), while the second operator is of two-meson type. From a normalized $2 \times 2$ correlation matrix $C_{j k}(t) = \langle \mathcal{O}^\dagger_j(t) \mathcal{O}_k(0) \rangle / (C_M(t))^2$, where $\langle \mathcal{O}^\dagger_j(t) \mathcal{O}_k(0) \rangle$ is visualized in Fig.\ \ref{FIG_C} and $C_M(t)$ denotes the correlation function of the $J^P = 0^-$ static-light meson (see e.g.\ Refs.\ \cite{Jansen:2008si,Michael:2010aa}), one can extract the energy eigenvalues $V_0^{\Sigma_g^+}(r)$ and $V_1^{\Sigma_g^+}(r)$. These eigenvalues correspond to the two lowest energy eigenstates in the $\Sigma_g^+$ sector, $| 0 ; \Sigma_g^+ \rangle$ and $| 1 ; \Sigma_g^+ \rangle$, and are normalized with respect to $2 m_M$. The $\Pi_g^+$ sector as well as the $\Pi_g^-$ sector do not include quarkonium states. Thus, for each of them a single creation operator is sufficient, differing from the operator (\ref{EQN408}) only in $\Gamma_q$,
\begin{eqnarray}
\label{EQN406} & & \mathcal{O}_{\bar{M} M}^{\Pi_g^+} = (\underbrace{P_+ \gamma_5}_{= \Gamma_Q})_{A B} (\underbrace{\mathbf{e}_\vartheta P_+ \vec{\gamma}}_{= \Gamma_q})_{C D} \Big(\bar{Q}_A(-\mathbf{r}/2) u_D(-\mathbf{r}/2)\Big) \Big(\bar{u}_C(\mathbf{r}/2) Q_B(+\mathbf{r}/2) + (u \rightarrow d)\Big) \\
\label{EQN409}  & & \mathcal{O}_{\bar{M} M}^{\Pi_g^-} = (\underbrace{P_+ \gamma_5}_{= \Gamma_Q})_{A B} (\underbrace{\mathbf{e}_\varphi P_+ \vec{\gamma}}_{= \Gamma_q})_{C D} \Big(\bar{Q}_A(-\mathbf{r}/2) u_D(-\mathbf{r}/2)\Big) \Big(\bar{u}_C(\mathbf{r}/2) Q_B(+\mathbf{r}/2) + (u \rightarrow d)\Big) .
\end{eqnarray}
From the corresponding normalized correlation functions one can extract the energy eigenvalues $V_0^{\Pi_g^+}(r)$ and $V_0^{\Pi_g^-}(r)$ of the lowest energy eigenstate in each of the sectors, $| 0 ; \Pi_g^+ \rangle$ and $| 0 ; \Pi_g^- \rangle$. Note that the spectra in these sectors are identical, i.e.\ $V_0^{\Pi_g}(r) = V_0^{\Pi_g^+}(r) = V_0^{\Pi_g^-}(r)$. The previously defined energy eigenvalues and eigenstates fulfill
\begin{equation}
\label{EQN924}
\Big( \langle 0 ; \Sigma_g^+ | , \langle 1 ; \Sigma_g^+ | , \langle 0 ; \Pi_g^+ | , \langle 0 ; \Pi_g^- | \Big)_j H \left( \begin{array}{c}| 0 ; \Sigma_g^+ \rangle \\ | 1 ; \Sigma_g^+ \rangle \\ | 0 ; \Pi_g^+ \rangle \\ | 0 ; \Pi_g^- \rangle \end{array}\right)_k = \left(\begin{array}{cccc}
V_0^{\Sigma_g^+}(r) & 0 & 0 & 0 \\
0 & V_1^{\Sigma_g^+}(r) & 0 & 0 \\
0 & 0 & V_0^{\Pi_g}(r) & 0 \\
0 & 0 & 0 & V_0^{\Pi_g}(r)
\end{array}\right)_{j k} ,
\end{equation}
where $H$ denotes the Hamiltonian.

If the gluonic parallel transporter $U$ and the light quark fields $u$ and $d$ appearing in the creation operators (\ref{EQN407}) and (\ref{EQN408}) are properly smeared, the corresponding normalized trial states $| \bar{Q} Q \rangle = \# \mathcal{O}_{\bar{Q} Q}^{\Sigma_g^+} | \Omega \rangle$ and $| \bar{M} M_\parallel \rangle = \# \mathcal{O}_{\bar{M} M}^{\Sigma_g^+} | \Omega \rangle$ ($\#$ denote appropriate normalization factors) are in good approximation linear combinations of the energy eigenstates $| 0 ; \Sigma_g^+ \rangle$ and $| 1 ; \Sigma_g^+ \rangle$,
\begin{eqnarray}
\label{EQN795} \Big( | 0 ; \Sigma_g^+ \rangle , | 1 ; \Sigma_g^+ \rangle \Big)_j = \left(\begin{array}{cc}
+\cos(\theta(r)) & +\sin(\theta(r)) \\
-\sin(\theta(r)) & +\cos(\theta(r))
\end{array}\right)_{j k} \Big( | \bar{Q} Q \rangle , | \bar{M} M_\parallel \rangle \Big)_k ,
\end{eqnarray}
where $\theta(r)$ is the mixing angle (for details see Ref.\ \cite{Bali:2005fu}, section~5.A). For separations $r$ somewhat below the string breaking distance $r_\textrm{sb} \approx 1.1 \, \textrm{fm}$ the lowest energy eigenstate is predominantly a quarkonium state, while the first excitation is a two-meson state, i.e.\ $\theta(r) \approx 0.3 \ldots 0.4$. For $r$ somewhat above $r_\textrm{sb}$ the situation is reversed, i.e.\ $\theta(r) \approx \pi/2$. For a detailed discussion, of how to compute the mixing angle from the $2 \times 2$ correlation matrix see Ref.\ \cite{Bali:2005fu}. One can obtain the analog of Eq.\ (\ref{EQN924}) for a basis including $| \bar{Q} Q \rangle$ and $| \bar{M} M_\parallel \rangle$ instead of $| 0 ; \Sigma_g^+ \rangle$ and $| 1 ; \Sigma_g^+ \rangle$ by using Eq.\ (\ref{EQN795}),
\begin{eqnarray}
\nonumber & & \Big( \langle \bar{Q} Q | , \langle \bar{M} M_\parallel | , \langle \bar{M} M_\perp^+ | , \langle \bar{M} M_\perp^- | \Big)_j H \Big( | \bar{Q} Q \rangle , | \bar{M} M_\parallel \rangle , | \bar{M} M_\perp^+ \rangle , | \bar{M} M_\perp^- \rangle \Big)_k = \\
\label{EQN925} & & \hspace{0.7cm} = \left(\begin{array}{cccc}
V_{\bar{Q} Q}(r) & V_{\textrm{mix}}(r) & 0 & 0 \\
V_{\textrm{mix}}(r) & V_{\bar{M} M,\parallel}(r) & 0 & 0 \\
0 & 0 & V_{\bar{M} M,\perp}(r) & 0 \\
0 & 0 & 0 & V_{\bar{M} M,\perp}(r)
\end{array}\right)_{j k} ,
\end{eqnarray}
where we have defined $| \bar{M} M_\perp^\pm \rangle = | 0 ; \Pi_g^\pm \rangle$ and
\begin{eqnarray}
\label{EQN276} & & V_{\bar{Q} Q}(r) = \cos^2(\theta(r)) V_0^{\Sigma_g^+}(r) + \sin^2(\theta(r)) V_1^{\Sigma_g^+}(r) 
\\
\label{EQN277} & & V_{\bar{M} M,\parallel}(r) = \sin^2(\theta(r)) V_0^{\Sigma_g^+}(r) + \cos^2(\theta(r)) V_1^{\Sigma_g^+}(r) 
\\
\label{EQN278} & & V_{\textrm{mix}}(r) = \cos(\theta(r)) \sin(\theta(r)) \Big(V_0^{\Sigma_g^+}(r) - V_1^{\Sigma_g^+}(r)\Big) 
\\
\label{EQN279} & & V_{\bar{M} M,\perp}(r) = V_0^{\Pi_g}(r) .
\end{eqnarray}

To express the $4 \times 4$ potential matrix $V(\mathbf{r})$ appearing in the Schr\"odinger equation (\ref{EQN050}) in terms of static potentials in QCD, recall that the first row and column of this matrix corresponds to a $\bar{Q} Q$ quarkonium channel, while the remaining three rows and columns correspond to the spin-1 triplet of a two-meson $\bar{M} M$ channel. Thus,
\begin{eqnarray}
\label{EQN926} V_{j k}(\mathbf{r}) = \Big( \langle \bar{Q} Q | , \langle \bar{M} M_1 | , \langle \bar{M} M_2 | , \langle \bar{M} M_3 | \Big)_j H \Big( | \bar{Q} Q \rangle , | \bar{M} M_1 \rangle , | \bar{M} M_2 \rangle , | \bar{M} M_3 \rangle \Big)_k ,
\end{eqnarray} 
where $| \bar{M} M_j \rangle$, $j=1,2,3$ are linear combinations of two-meson states $| \bar{M} M_\parallel \rangle$, $| \bar{M} M_\perp^+ \rangle$ and $| \bar{M} M_\perp^- \rangle$, which have light spin in $j$ direction. These linear combinations are given by
\begin{eqnarray}
\label{EQN927} | \bar{M} M_\parallel \rangle = (\mathbf{e}_r)_j | \bar{M} M_j \rangle \quad , \quad | \bar{M} M_\perp^+ \rangle = (\mathbf{e}_\vartheta)_j | \bar{M} M_j \rangle \quad , \quad | \bar{M} M_\perp^- \rangle = (\mathbf{e}_\varphi)_j | \bar{M} M_j \rangle .
\end{eqnarray}
Combining Eqs.\ (\ref{EQN925}), (\ref{EQN926}) and (\ref{EQN927}) leads to
\begin{eqnarray}
V(\mathbf{r}) = \left(\begin{array}{cc}
V_{\bar{Q} Q}(r) & V_\textrm{mix}(r) \Big(1 \otimes \mathbf{e}_r\Big) \\
V_\textrm{mix}(r) \Big(\mathbf{e}_r \otimes 1\Big) & V_{\bar{M} M,\parallel}(r) \Big(\mathbf{e}_r \otimes \mathbf{e}_r\Big) + V_{\bar{M} M,\perp}(r) \Big(1 - \mathbf{e}_r \otimes \mathbf{e}_r\Big)
\end{array}\right) ,
\end{eqnarray}
\end{widetext}
which is identical to Eq.\ (\ref{EQN051}). Thereby, we have derived the structure of the $4 \times 4$ potential matrix appearing in the Schr\"odinger Eq. (\ref{EQN050}). Moreover, via Eqs.\ (\ref{EQN276}) to (\ref{EQN277}) we have related the four potentials $V_{\bar{Q} Q}(r)$, $V_{\bar{M} M,\parallel}(r)$, $V_\textrm{mix}(r)$ and $V_{\bar{M} M,\perp}(r)$ to the static potentials $V_0^{\Sigma_g^+}(r)$, $V_1^{\Sigma_g^+}(r)$ and $V_0^{\Pi_g}(r)$ and the mixing angle $\theta(r)$, which can be computed using lattice QCD.


\subsection{Boundary conditions for $M\bar{M}$ scattering and partial wave decomposition}

Now we consider specific boundary conditions appropriate to describe scattering of two heavy-light mesons $M$ and $\bar{M}$, an incident plane wave and an emergent spherical wave for large $\bar{Q} Q$ separations $r$ (for previous work on similar systems see Refs.\ \cite{Bicudo:2015bra,Bicudo:2017szl}). Moreover, we do a partial wave decomposition and specialize the Schr\"odinger equation (\ref{EQN050}) to definite $\widetilde{J}^{P C} = 0^{+ +}$, i.e.\ vanishing total angular momentum excluding the heavy spins. In other words, we formulate the Schr\"odinger equation (\ref{EQN050}) specifically for quarkonium bound states and resonances with $\widetilde{J}^{P C} = L^{P C}_{\bar{Q} Q} = 0^{+ +}$ (for the relation of $\widetilde{J}^{P C}$ to the common $J^{P C}$ as e.g.\ used by the Particle Data Group \cite{Patrignani:2016xqp} see Table~\ref{TAB004}). This reduces the partial differential equation (\ref{EQN050}) to a system of two coupled ordinary differential equations in the radial coordinate $r = |\mathbf{r}|$, which is much simpler to solve numerically.


\subsubsection{\label{SEC566}Basis functions for the partial wave decomposition}

In standard textbooks on quantum mechanics scattering theory is typically discussed for a spin-0 system and a single channel, i.e.\ the corresponding Schr\"odinger equation has only one component and the partial wave decomposition is done in terms of spherical harmonics, which are eigenfunctions of $\mathbf{L}^2$ and $L_z$. For our particular problem an equivalent decomposition is technically more complicated, because the Schr\"odinger equation (\ref{EQN050}) has two channels, a $\bar{Q} Q$ quarkonium channel with spin $0$ (upper component of the wave function) and a $\bar{M} M$ two-meson channel with spin $1$ (lower three components of the wave function), i.e.\ $\psi(\mathbf{r}) = (\psi_{\bar{Q} Q}(\mathbf{r}) , \vec{\psi}_{\bar{M} M}(\mathbf{r}))$. The partial wave decomposition has to be done in terms of four-component eigenfunctions of $\widetilde{\mathbf{J}}^2$ and $\widetilde{J}_z$, which are conserved quantities and replace the non-conserved $\mathbf{L}^2$ and $L_z$. An orthonormal and complete set of eigenfunctions with respect to to the solid angle $\Omega$ and the four components of $\psi(\mathbf{r})$ is the following:
\begin{itemize}
\item For the spin-0 component of $\psi(\mathbf{r})$, i.e.\ for $\psi_{\bar{Q} Q}(\mathbf{r})$, $\widetilde{J} = L_{\bar{Q} Q}$. Thus, eigenfunctions of $\widetilde{\mathbf{J}}^2$ and $\widetilde{J}_z$, which are non-zero in the upper component, are
\begin{eqnarray}
\label{EQN860} Z_{\bar{Q} Q,\widetilde{J},\widetilde{J}_z}(\Omega) = (Y_{\widetilde{J},\widetilde{J}_z}(\Omega) , \vec{0}) ,
\end{eqnarray}
i.e.\ proportional to spherical harmonics.

\item Eigenfunctions of $\widetilde{\mathbf{J}}^2$ and $\widetilde{J}_z$, which are non-zero in the lower three components, can be constructed by Clebsch-Gordan coupling of spherical harmonics $Y_{L,L_z}(\Omega)$ (representing relative orbital angular momentum $L_{\bar{M} M}$; here and in the following we use $L \equiv L_{\bar{M} M}$ in lower indices, e.g.\ of the functions $Y$ and $Z$) and the three spin-1 components of the light quarks. They are given by
\begin{eqnarray}
Z_{\bar{M} M,L \rightarrow \widetilde{J},\widetilde{J}_z}(\Omega) = (0 , \mathbf{Z}_{L \rightarrow \widetilde{J},\widetilde{J}_z}(\Omega)) ,
\end{eqnarray}
where $L_{\bar{M} M} = 1$ for $\widetilde{J} = 0$ and $\widetilde{J}-1 \leq L_{\bar{M} M} \leq \widetilde{J}+1$ for $\widetilde{J} \geq 1$. $\mathbf{Z}_{L \rightarrow \widetilde{J},\widetilde{J}_z}(\Omega)$ is defined as follows:
\begin{itemize}
\item $\widetilde{J} = 0$, $\widetilde{J}_z = 0$:
\begin{eqnarray}
\mathbf{Z}_{1 \rightarrow 0,0}(\Omega) = \bigg(\frac{1}{4 \pi}\bigg)^{1/2} \mathbf{e}_r .
\end{eqnarray}

\item $\widetilde{J} = 1$, $\widetilde{J}_z = -1,0,-1$ (three possibilities for each $\widetilde{J}_z$):
\begin{eqnarray}
 & & \mathbf{Z}_{0 \rightarrow 1,j}(\Omega) = \bigg(\frac{1}{4 \pi}\bigg)^{1/2} \mathbf{e}_j \\
 & & \mathbf{Z}_{1 \rightarrow 1,j}(\Omega) = \bigg(\frac{3}{8 \pi}\bigg)^{1/2} \epsilon_{j k l} \frac{r_k}{r} \mathbf{e}_l \\
 & & \mathbf{Z}_{2 \rightarrow 1,j}(\Omega) = \bigg(\frac{18}{16 \pi}\bigg)^{1/2} \bigg(\frac{r_j}{r} \mathbf{e}_r - \frac{1}{3} \mathbf{e}_j\bigg)
\end{eqnarray}
($j = x,y,z$ replaces the index $\widetilde{J}_z = -1,0,+1$ in the usual way, i.e.\ $z \equiv 0$ and $\mp (x \pm i y) \equiv \pm 1$).

\item $\widetilde{J} \geq 2$: \\ The corresponding functions $\mathbf{Z}_{L \rightarrow \widetilde{J},\widetilde{J}_z}(\Omega)$ can be constructed in a straightforward way. Since these functions are not needed explicitly in this work, we do not provide equations. 
\end{itemize}
Any three-component function $\mathbf{G}(\mathbf{r})$ can be written as an expansion in $\mathbf{Z}_{L \rightarrow \widetilde{J},\widetilde{J}_z}(\Omega)$,
\begin{eqnarray}
\nonumber & & \mathbf{G}(\mathbf{r}) = g_{1 \rightarrow 0,0}(r) \mathbf{Z}_{1 \rightarrow 0,0}(\Omega) \\
\nonumber & & \hspace{0.7cm} + \sum_{\widetilde{J} = 1}^\infty \sum_{\widetilde{J}_z = -\widetilde{J}}^{+\widetilde{J}} \sum_{L = \widetilde{J}-1,\widetilde{J},\widetilde{J}+1} g_{L \rightarrow \widetilde{J},\widetilde{J}_z}(r) \mathbf{Z}_{L \rightarrow \widetilde{J},\widetilde{J}_z}(\Omega) , \\
\label{EQN233} & &
\end{eqnarray}
where the expansion coefficients $g_{L \rightarrow \widetilde{J},\widetilde{J}_z}(r)$ are functions of $r = |\mathbf{r}|$. Note that all $\mathbf{Z}_{L \rightarrow \widetilde{J},\widetilde{J}_z}(\Omega)$ have parity $(-1)^{L+1}$. Thus for parity even functions $\mathbf{G}(\mathbf{r})$ all coefficients $g_{L \rightarrow \widetilde{J},\widetilde{J}_z}(r)$ with even $L$ are zero, while for parity odd functions $\mathbf{G}(\mathbf{r})$ all coefficients $g_{L \rightarrow \widetilde{J},\widetilde{J}_z}(r)$ with odd $L$ are zero.
\end{itemize}


\subsubsection{\label{SEC569}Boundary conditions for $M\bar{M}$ scattering for $r \rightarrow \infty$}

In section~\ref{sec:Bali}, where we determine the potentials $V_{\bar{Q} Q}(r)$, $V_{\bar{M} M,\parallel}(r)$, $V_{\bar{M} M,\perp}(r)$ and $V_\textrm{mix}(r)$ by parameterizing lattice QCD results from Ref.\ \cite{Bali:2005fu}, we find $V_{\bar{M} M,\parallel}(r) \rightarrow 0$, $V_{\bar{M} M,\perp}(r) \rightarrow 0$ and $V_{\textrm{mix}}(r) \rightarrow 0$ for $\bar{Q} Q$ separations $r \rightarrow \infty$ (see Eqs.\ (\ref{EQN566}), (\ref{eq:fitVBB}) and (\ref{eq:fitMix})). This is expected, because the potentials are normalized by an additive constant in such a way that a value of $0$ corresponds to the $\bar{M} M$ threshold. Consequently, the $4 \times 4$ potential matrix (\ref{EQN051}) reduces to
\begin{eqnarray}
V(\mathbf{r}) = \left(\begin{array}{cccc}
V_{\bar{Q} Q}(r) & 0 & 0 & 0 \\
0 & 0 & 0 & 0 \\
0 & 0 & 0 & 0 \\
0 & 0 & 0 & 0
\end{array}\right) ,
\end{eqnarray}
i.e.\ the Schr\"odinger Eq. (\ref{EQN051}) decouples into four independent partial differential equations, one for each of the four components of the wave function $\psi(\mathbf{r}) = (\psi_{\bar{Q} Q}(\mathbf{r}) , \vec{\psi}_{\bar{M} M}(\mathbf{r}))$. 

The potential in the quarkonium equation, $V_{\bar{Q} Q}(r)$, is linear for $r \rightarrow \infty$, i.e.\ confining (see Eq.\ (\ref{eq:fitVQQ})). Thus the boundary conditions for $\psi_{\bar{Q} Q}(\mathbf{r})$ is
\begin{eqnarray}
\label{EQN570} \psi_{\bar{Q} Q}(\mathbf{r}) = 0 \quad \textrm{for } r \rightarrow \infty .
\end{eqnarray}

The three equations for the components of $\vec{\psi}_{\bar{M} M}(\mathbf{r})$ have a vanishing potential, i.e.\ are identical to the free Schr\"odinger equation. Thus the appropriate boundary conditions for $\vec{\psi}_{\bar{M} M}(\mathbf{r})$ for $\bar{M} M$ meson scattering at definite relative momentum $k = \sqrt{2 \mu_M E}$ are a superposition of an incident plane wave and an emergent spherical wave, where both are solutions of the free Schr\"odinger equation. The incident plane wave (for simplicity we choose a plane wave in positive $z$ direction) is $\mathbf{A} e^{+i k z}$ with a polarization vector $\mathbf{A}$ describing the light spin 1 of the colliding mesons. The emergent spherical wave can be expanded in terms of spherical Hankel functions of the first kind $h_L^{(1)}(k r)$ and the angular basis functions $\mathbf{Z}_{L \rightarrow \widetilde{J},\widetilde{J}_z}(\Omega)$ discussed in section~\ref{SEC566}. Thus,
\begin{eqnarray}
\nonumber & & \vec{\psi}_{\bar{M} M}(\mathbf{r}) = \mathbf{A} e^{+i k z} - \sqrt{4 \pi} A_z t_{1 \rightarrow 0,0} h_1^{(1)}(k r) \mathbf{Z}_{1 \rightarrow 0,0}(\Omega) \\
\nonumber & & \hspace{0.7cm} + \sum_{\widetilde{J} = 1}^\infty \sum_{\widetilde{J}_z = -\widetilde{J}}^{+\widetilde{J}}  \sum_{L = \widetilde{J}-1,\widetilde{J},\widetilde{J}+1} t_{L \rightarrow \widetilde{J},\widetilde{J}} h_L^{(1)}(k r) \mathbf{Z}_{L \rightarrow \widetilde{J},\widetilde{J}_z}(\Omega) \\
\label{EQN682} & & \hspace{0.7cm} \textrm{for } r \rightarrow \infty ,
\end{eqnarray}
where $t_{L \rightarrow \widetilde{J},\widetilde{J}} / k$ are the scattering amplitudes. We have included the prefactor $-\sqrt{4 \pi}$ in front of $t_{1 \rightarrow 0,0}$, because using probability conservation one can show
\begin{eqnarray}
\label{EQN672} \Big|1 + 2 i t_{1 \rightarrow 0,0}\Big| = 1 .
\end{eqnarray}
This in turn allows to define the corresponding scattering phase $\delta_{1 \rightarrow 0,0}$ as
\begin{eqnarray}
\label{EQN429} e^{2 i \delta_{1 \rightarrow 0,0}} = 1 + 2 i t_{1 \rightarrow 0,0} ,
\end{eqnarray}
which closely resembles textbook conventions for standard scattering of spinless particles. Eq.\ (\ref{EQN672}) is equivalent to
\begin{eqnarray}
\label{eq:optical} \textrm{Im}(t_{1 \rightarrow 0,0})=|t_{1 \rightarrow 0,0}|^2 ,
\end{eqnarray}
which is the optical theorem.


\subsubsection{Partial wave decomposition}

The partial wave decomposition of $\psi_{\bar{Q} Q}(\mathbf{r})$ is an ordinary expansion in spherical harmonics,
\begin{eqnarray}
\nonumber & & \psi_{\bar{Q} Q}(\mathbf{r}) = \sqrt{4 \pi} i A_z \frac{u_{0,0}(r)}{k r} Y_{0,0}(\Omega) \\
\label{EQN601} & & \hspace{0.7cm} + \sum_{\widetilde{J} = 1}^\infty \sum_{\widetilde{J}_z = -\widetilde{J}}^{+\widetilde{J}} \frac{u_{\widetilde{J},\widetilde{J}_z}(r)}{k r} Y_{\widetilde{J},\widetilde{J}_z}(\Omega)
\end{eqnarray}
with functions $u_{\widetilde{J},\widetilde{J}_z}(r)$ as coefficients (see Eq.\ (\ref{EQN860})). It is convenient to include the prefactor $\sqrt{4 \pi} i A_z$ in front of $u_{0,0}(r)$, to avoid unnecessary factors in the coupled channel Schr\"odinger equation for $\widetilde{J}^{P C} = 0^{+ +}$, which we will derive in section~\ref{SEC734}. The boundary conditions are
\begin{eqnarray}
\label{EQNbc1} & & u_{\widetilde{J},\widetilde{J}_z}(r) \propto r^{\widetilde{J}+1} \quad \textrm{for } r \rightarrow 0 \\
\label{EQNbc2} & & u_{\widetilde{J},\widetilde{J}_z}(r) = 0 \quad \textrm{for } r \rightarrow \infty ,
\end{eqnarray}
where the latter follows from Eq.\ (\ref{EQN570}).

We write $\vec{\psi}_{\bar{M} M}(\mathbf{r})$ as a sum of the incident wave $\mathbf{A} e^{+i k z}$ and an emergent wave $\mathbf{X}(r)$,
\begin{eqnarray}
\label{EQN683} \vec{\psi}_{\bar{M} M}(\mathbf{r}) = \mathbf{A} e^{+i k z} + \mathbf{X}(\mathbf{r}) .
\end{eqnarray}
The partial wave decomposition of $\vec{\psi}_{\bar{M} M}(\mathbf{r})$ follows Eq.\ (\ref{EQN233}),
\begin{eqnarray}
\nonumber & & \mathbf{A} e^{+i k z} = a_{1 \rightarrow 0,0}(r) \mathbf{Z}_{1 \rightarrow 0,0}(\Omega) \\
\nonumber & & \hspace{0.7cm} + \sum_{\widetilde{J} = 1}^\infty \sum_{\widetilde{J}_z = -\widetilde{J}}^{+\widetilde{J}} \sum_{L = \widetilde{J}-1,\widetilde{J},\widetilde{J}+1} a_{L \rightarrow \widetilde{J},\widetilde{J}_z}(r) \mathbf{Z}_{L \rightarrow \widetilde{J},\widetilde{J}_z}(\Omega) \\
\label{EQN622} & & \\
\nonumber & & \mathbf{X}(\mathbf{r}) = \sqrt{4 \pi} i A_z  \frac{\chi_{1 \rightarrow 0,0}(r) \mathbf{Z}_{1 \rightarrow 0,0}(\Omega)}{k r} \\
\nonumber & & \hspace{0.7cm} + \sum_{\widetilde{J} = 1}^\infty \sum_{\widetilde{J}_z = -\widetilde{J}}^{+\widetilde{J}} \sum_{L = \widetilde{J}-1,\widetilde{J},\widetilde{J}+1} \frac{\chi_{L \rightarrow \widetilde{J},\widetilde{J}_z}(r)}{k r} \mathbf{Z}_{L \rightarrow \widetilde{J},\widetilde{J}_z}(\Omega) . \\
\label{EQN684} & &
\end{eqnarray}
with functions $a_{L \rightarrow \widetilde{J},\widetilde{J}_z}(r)$ and $\chi_{L \rightarrow \widetilde{J},\widetilde{J}_z}(r)$ as coefficients. Again it is convenient to include the prefactor $\sqrt{4 \pi} i A_z$ in front of $\chi_{1 \rightarrow 0,0}(r)$. For the incident wave the coefficients $a_{L \rightarrow \widetilde{J},\widetilde{J}_z}(r)$ can be calculated in a straightforward way, e.g.
\begin{eqnarray}
\nonumber & & a_{1 \rightarrow 0,0}(r) = \int d\Omega \, (\mathbf{Z}_{1 \rightarrow 0,0}(\Omega))^\ast \mathbf{A} e^{i k z} = \sqrt{4 \pi} i A_z j_1(k r) \\
 & &
\end{eqnarray}
($j_1$ denotes a spherical Bessel function of the first kind), which is particularly relevant in the following. The boundary conditions are
\begin{eqnarray}
\label{EQNbc3} & & \chi_{L \rightarrow \widetilde{J},\widetilde{J}_z}(r) \propto r^{L+1} \quad \textrm{for } r \rightarrow 0 \\
\label{EQN545} & & \chi_{L \rightarrow \widetilde{J},\widetilde{J}_z}(r) = i t_{L \rightarrow \widetilde{J},\widetilde{J}_z} k r h_L^{(1)}(k r) \quad \textrm{for } r \rightarrow \infty ,
\end{eqnarray}
where the latter can be obtained by inserting Eq.\ (\ref{EQN684}) into Eq.\ (\ref{EQN683}) and by comparing to Eq.\ (\ref{EQN682}).


\subsubsection{\label{SEC734}Coupled channel Schr\"odinger equation for $\widetilde{J}^{P C} = 0^{+ +}$}

Now we project the Schr\"odinger equation (\ref{EQN050}) to definite $\widetilde{J}^{P C} = 0^{+ +}$ by integrating both the left hand side and the right hand side over the solid angle according to $\int d\Omega \, (Z_{\ldots}(\Omega))^\dagger$ with appropriate $Z_{\ldots}(\Omega)$, i.e.\ those with indices $\widetilde{J} = 0$ and $\widetilde{J}_z = 0$:
\begin{widetext}
\begin{eqnarray}
\nonumber & & Z_{\ldots}(\Omega) = Z_{\bar{Q} Q,0,0}(\Omega) \ : \\
\label{EQN404} & & \rightarrow \quad \bigg(-\frac{1}{2 \mu_Q} \partial_r^2 + V_{\bar{Q} Q}(r) + 2 m_M - E\bigg) u_{0,0}(r) + V_\textrm{mix}(r) \bigg(k r j_1(k r) + \chi_{1 \rightarrow 0,0}(r)\bigg) = 0 \\
\nonumber & & Z_{\ldots}(\Omega) = Z_{\bar{M} M,1 \rightarrow 0,0}(\Omega) \ : \\
\label{EQN405} & & \rightarrow \quad V_\textrm{mix}(r) u_{0,0}(r) + V_{\bar{M} M,\parallel}(r) k r j_1(k r) + \bigg(-\frac{1}{2 \mu_M} \bigg(\partial_r^2 - \frac{2}{r^2}\bigg) + V_{\bar{M} M,\parallel}(r) + 2 m_M - E\bigg) \chi_{1 \rightarrow 0,0}(r) = 0 .
\end{eqnarray}
These two equations can also be expressed in matrix form,
\begin{eqnarray}
\nonumber & & \left(-\frac{1}{2} \left(\begin{array}{cc} 1/\mu_Q & 0 \\ 0 & 1/\mu_M \end{array}\right) \partial_r^2 + \frac{1}{2 r^2} \left(\begin{array}{cc} 0 & 0 \\ 0 & 2/\mu_M \end{array}\right) + V_0(r) + 2 m_M - E\right)
\left(\begin{array}{c} u_{0,0}(r) \\ \chi_{1 \rightarrow 0,0}(r) \end{array}\right) =
-\left(\begin{array}{c} V_\textrm{mix}(r) \\ V_{\bar{M} M,\parallel}(r) \end{array}\right) k r j_1(k r) \quad , \\
\label{eq:coupledpot} & & \hspace{0.7cm} V_0(r) = \left(\begin{array}{cc} V_{\bar{Q} Q}(r) & V_\textrm{mix}(r) \\ V_\textrm{mix}(r) & V_{\bar{M} M,\parallel}(r) \end{array}\right) .
\end{eqnarray}
\end{widetext}
Eqs.\ (\ref{EQN404}) and (\ref{EQN405}) or equivalently Eq.\ (\ref{eq:coupledpot}) are corner stone equations of this work. In section~\ref{sec:results} we will solve them to obtain numerical results for bottomonium bound states and resonances.

Specializing the coupled channel Schr\"odinger equation (\ref{EQN050}) to $\widetilde{J}^{P C} = 1^{- -}$ or higher $\widetilde{J}$ can be done in the same way. We leave that for future publications.


\section{\label{sec:Bali}Utilizing the lattice QCD static potentials from Ref.\ \cite{Bali:2005fu}}

To determine the potentials $V_{\bar{Q} Q}(r)$, $V_{\bar{M} M,\parallel}(r)$, $V_{\bar{M} M,\perp}(r)$ and $V_{\textrm{mix}}(r)$, one has to compute the following correlation functions with lattice QCD, as discussed in section~\ref{SEC582}:
\begin{itemize}
\item A normalized $2 \times 2$ correlation matrix using the creation operators (\ref{EQN407}) and (\ref{EQN408}).

\item A normalized correlation function using either the creation operator (\ref{EQN406}) or the creation operator (\ref{EQN409}).
\end{itemize}
Such computations are quite challenging and technicalities are discussed in detail e.g.\ in \cite{Bali:2005fu,Bulava:2019iut}. In the future we plan to perform such computations. Here we follow a different strategy and reuse the existing lattice QCD results for static potentials from Ref.\ \cite{Bali:2005fu}, to determine $V_{\bar{Q} Q}(r)$, $V_{\bar{M} M,\parallel}(r)$, $V_{\bar{M} M,\perp}(r)$ and $V_{\textrm{mix}}(r)$ within certain approximations.

In Ref.\ \cite{Bali:2005fu} a $2 \times 2$ correlation matrix was computed at light $u$ and $d$ quark mass corresponding to a pion mass $m_\pi \approx 654 \, \textrm{MeV}$ and lattice spacing $a \approx 1 / (2.37 \, \textrm{GeV}) \approx 0.083 \, \textrm{fm}$ using the creation operators
\begin{widetext}
\begin{eqnarray}
\mathcal{O}_{\bar{Q} Q}^{\textrm{\cite{Bali:2005fu}}} &=& \Big(\bar{Q}(-\mathbf{r}/2) \mathbf{e}_r P_+ \vec{\gamma} U(-\mathbf{r}/2;+\mathbf{r}/2) Q(+\mathbf{r}/2)\Big) 
 \non \\
\mathcal{O}_{\bar{M} M}^{\textrm{\cite{Bali:2005fu}}} &=& \Big(\bar{Q}(-\mathbf{r}/2) P_+ \gamma_5 u(-\mathbf{r}/2)\Big) \Big(\bar{u}(+\mathbf{r}/2) \gamma_5 P_- Q(+\mathbf{r}/2)\Big)  + (u \rightarrow d)
\end{eqnarray}
(see Eqs.\ (11) and (15) in Ref.\ \cite{Bali:2005fu}; for convenience we have expressed these operators in the same notation used in previous sections of this work and we have inserted projectors $P_+$ using $Q = P_- Q$ and $\bar{Q} = \bar{Q} P_+$). Using the Fierz identity (\ref{EQN001}) these operators can be rewritten according to
\begin{eqnarray}
\label{EQN410_} & & \mathcal{O}_{\bar{Q} Q}^{\textrm{\cite{Bali:2005fu}}} = (\underbrace{\mathbf{e}_r P_+ \vec{\gamma}}_{= \Gamma_Q})_{A B} \Big(\bar{Q}_A(-\mathbf{r}/2) U(-\mathbf{r}/2;+\mathbf{r}/2) Q_B(+\mathbf{r}/2)\Big) \ \ = \ \ \mathcal{O}_{\bar{Q} Q}^{\Sigma_g^+}[\Gamma_Q = \mathbf{e}_r P_+ \vec{\gamma}] \\
\nonumber & & \mathcal{O}_{\bar{M} M}^{\textrm{\cite{Bali:2005fu}}} = \frac{1}{2} \bigg((\underbrace{P_+ \gamma_5}_{= \Gamma_Q})_{A B} (\underbrace{P_+ \gamma_5}_{= \Gamma_q})_{C D} - \sum_j (\underbrace{P_+ \gamma_j}_{= \Gamma_Q})_{A B} (\underbrace{P_+ \gamma_j}_{= \Gamma_q})_{C D}\bigg) \Big(\bar{Q}_A(-\mathbf{r}/2) u_D(-\mathbf{r}/2)\Big) \Big(\bar{u}_C(\mathbf{r}/2) Q_B(+\mathbf{r}/2)\Big) \\
\label{EQN410} & & \hspace{0.7cm} + (u \rightarrow d) = \frac{1}{2} \Big(\mathcal{O}_{\bar{M} M}^{\Sigma_u^-}[\Gamma_Q = P_+ \gamma_5] - \mathcal{O}_{\bar{M} M}^{\Pi_g^+}[\Gamma_Q = \mathbf{e}_\vartheta P_+ \vec{\gamma}] - \mathcal{O}_{\bar{M} M}^{\Pi_g^-}[\Gamma_Q = \mathbf{e}_\varphi P_+ \vec{\gamma}] - \mathcal{O}_{\bar{M} M}^{\Sigma_g^+}[\Gamma_Q = \mathbf{e}_r P_+ \vec{\gamma}]\Big) .
\end{eqnarray}
Four of the five operators appearing on the right hand sides are defined in Eqs.\ (\ref{EQN407}) to (\ref{EQN409}), where $\Gamma_Q$ has to be replaced as indicated in square brackets. $\mathcal{O}_{\bar{M} M}^{\Sigma_u^-}$ is an operator probing the $\Sigma_u^-$ sector, defined via
\begin{eqnarray}
\mathcal{O}_{\bar{M} M}^{\Sigma_u^-} = (\Gamma_Q)_{A B} (\underbrace{P_+ \gamma_5}_{= \Gamma_q})_{C D} \Big(\bar{Q}_A(-\mathbf{r}/2) u_D(-\mathbf{r}/2)\Big) \Big(\bar{u}_C(\mathbf{r}/2) Q_B(+\mathbf{r}/2)\Big) + (u \rightarrow d) .
\end{eqnarray}
Note that the two-meson creation operator $\mathcal{O}_{\bar{M} M}^{\textrm{\cite{Bali:2005fu}}}$ used in Ref.\ \cite{Bali:2005fu} does not only probe the $\Sigma_g^+$ sector, i.e.\ the sector of the ordinary static potential, but also the three sectors $\Sigma_u^-$, $\Pi_g^+$ and $\Pi_g^-$.

To parameterize the three independent elements of the normalized $2 \times 2$ correlation matrix for large temporal separations $t$, we follow the arguments of Ref.\ \cite{Bali:2005fu},
\begin{eqnarray}
\label{EQN377} & & C_{\bar{Q} Q,\bar{Q} Q}(t) = \frac{1}{(C_M(t))^2} \Big\langle \mathcal{O}_{\bar{Q} Q}^{\textrm{\cite{Bali:2005fu}} \, \dagger}(t) \mathcal{O}_{\bar{Q} Q}^{\textrm{\cite{Bali:2005fu}}}(0) \Big\rangle = \Big(a_{\bar{Q} Q}^{\Sigma_g^+}(r)\Big)^2 \Big(\cos^2(\theta(r)) e^{-V_0^{\Sigma_g^+}(r) t} + \sin^2(\theta(r)) e^{-V_1^{\Sigma_g^+}(r) t}\Big) \\
\nonumber & & C_{\bar{M} M,\bar{M} M}(t) = \frac{1}{(C_M(t))^2} \Big\langle \mathcal{O}_{\bar{M} M}^{\textrm{\cite{Bali:2005fu}} \, \dagger}(t) \mathcal{O}_{\bar{M} M}^{\textrm{\cite{Bali:2005fu}}}(0) \Big\rangle = \frac{1}{4} \Big(a_{\bar{M} M}^{\Sigma_u^-}(r)\Big)^2 e^{-V_0^{\Sigma_u^-}(r) t} + \frac{1}{2} \Big(a_{\bar{M} M}^{\Pi_g}(r)\Big)^2 e^{-V_0^{\Pi_g}(r) t} \\
 & & \hspace{0.7cm} + \frac{1}{4} \Big(a_{\bar{M} M}^{\Sigma_g^+}(r)\Big)^2 \Big(\sin^2(\theta(r)) e^{-V_0^{\Sigma_g^+}(r) t} + \cos^2(\theta(r)) e^{-V_1^{\Sigma_g^+}(r) t}\Big) \\
\label{EQN378} & & C_{\bar{Q} Q,\bar{M} M}(t) = \frac{1}{(C_M(t))^2} \Big\langle \mathcal{O}_{\bar{Q} Q}^{\textrm{\cite{Bali:2005fu}} \, \dagger}(t) \mathcal{O}_{\bar{M} M}^{\textrm{\cite{Bali:2005fu}}}(0) \Big\rangle = \frac{1}{2} a_{\bar{Q} Q}^{\Sigma_g^+}(r) a_{\bar{M} M}^{\Sigma_g^+}(r) \cos(\theta(r)) \sin(\theta(r)) \Big(e^{-V_0^{\Sigma_g^+}(r) t} - e^{-V_1^{\Sigma_g^+}(r) t}\Big) .
\end{eqnarray}
\end{widetext}
It is assumed that for large $t$ two energy eigenstates are sufficient to describe the contributions from the $\Sigma_g^+$ sector (which are a mixtures of a $\bar{Q} Q$ quarkonium state and a $\bar{M} M$ two-meson state), while for each of the sectors $\Sigma_u^-$, $\Pi_g^+$ and $\Pi_g^-$ only a single energy eigenstate is needed (a $\bar{M} M$ two-meson state). $V_0^{\Sigma_g^+}(r)$, $V_1^{\Sigma_g^+}(r)$, $V_0^{\Sigma_u^-}(r)$ and $V_0^{\Pi_g}(r)$ are the corresponding potentials and the coefficients $a_j^{\Lambda_\eta^{(\epsilon)}}$ are proportional to the overlaps of the energy eigenstates and the trial states generated by the creation operators $\mathcal{O}_{\bar{Q} Q}^{\Sigma_g^+}$, $\mathcal{O}_{\bar{M} M}^{\Sigma_u^-}$, $\mathcal{O}_{\bar{M} M}^{\Pi_g^+}$, $\mathcal{O}_{\bar{M} M}^{\Pi_g^-}$ and $\mathcal{O}_{\bar{M} M}^{\Sigma_g^+}$ (note that $a_{\bar{M} M}^{\Pi_g} = a_{\bar{M} M}^{\Pi_g^+} = a_{\bar{M} M}^{\Pi_g^-}$, as discussed in section~\ref{SEC582}). Eqs.\ (\ref{EQN377}) to (\ref{EQN378}) represent a corrected version of Eqs.\ (68) to (70) in Ref.\ \cite{Bali:2005fu}, where it is now taken into account that $\mathcal{O}_{\bar{M} M}^{\textrm{\cite{Bali:2005fu}}}$ probes several $\Lambda_\eta^\epsilon$ sectors.

\begin{figure}[htb]
\includegraphics[width=1.0\linewidth]{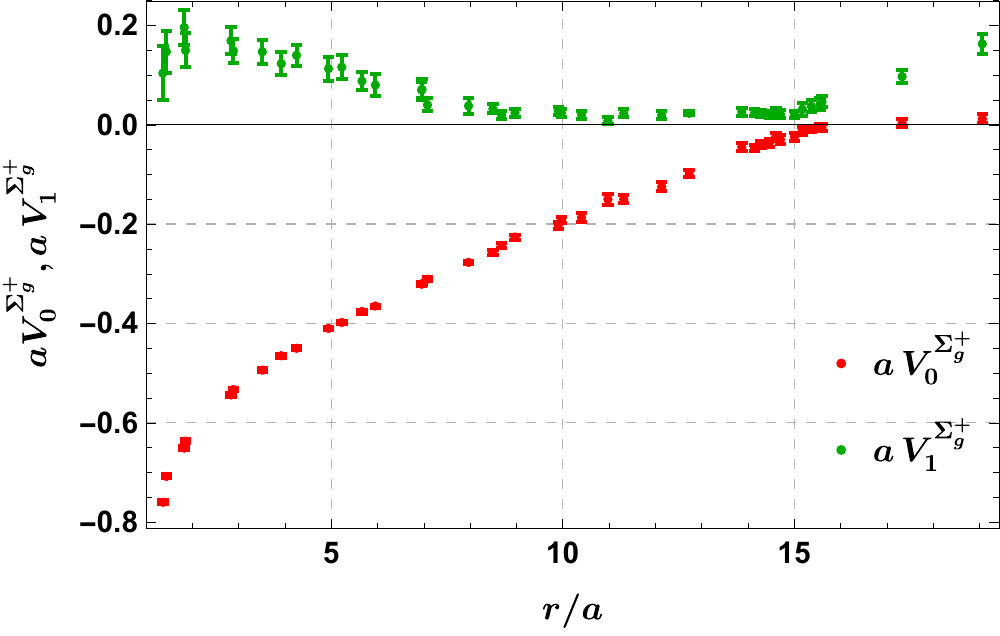} \\
\includegraphics[width=1.0\linewidth]{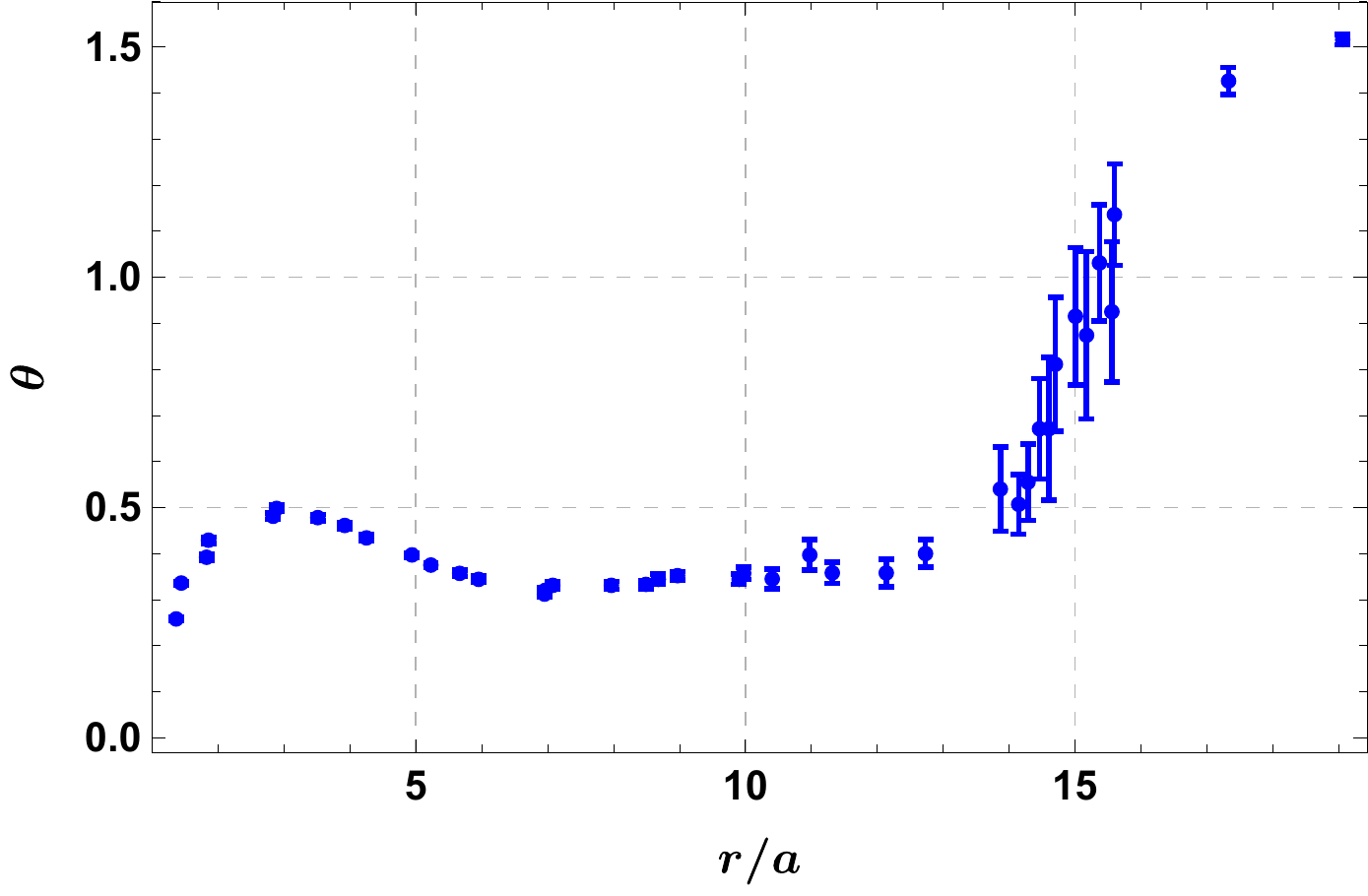}
\caption{\label{fig:fit_E1_E2_theta}(Color online.) Potentials $V_0^{\Sigma_g^+}(r)$ and $V_1^{\Sigma_g^+}(r)$ (upper plot) and mixing angle $\theta(r)$ (lower plot) as functions of the $\bar{Q} Q$ separation $r$ (lattice spacing $a \approx 1 / (2.37 \, \textrm{GeV})$).}
\end{figure}

The correlation matrix data of Ref.\ \cite{Bali:2005fu} is not publicly available, but a parametrization of this data is given (see.\ Eqs.\ (68) to (70) and Table~I in Ref.\ \cite{Bali:2005fu}). This parametrization allows us to resample the correlators from Ref.\ \cite{Bali:2005fu} and to perform $\chi^2$ minimizing fits of our parametrization (\ref{EQN377}) to (\ref{EQN378}) to the resampled data. For stable fits the number of parameters is, however, too large. Thus, we assume $V_0^{\Sigma_u^-}(r) = V_0^{\Pi_g}(r) = 0$ (i.e.\ the ground state energy in the sectors, where no string-like state is present, is around two times the static-light meson mass) and $a_{\bar{M} M}^{\Sigma_u^-}(r) = a_{\bar{M} M}^{\Pi_g}(r) = a_{\bar{M} M}^{\Sigma_g^+}(r)$ (i.e.\ all two-meson creation operators create similar overlaps \footnote{For $r$ larger than the size of a static-light meson, i.e.\ $r \gtapprox 0.5 \, \textrm{fm} \ldots 1.0 \, \textrm{fm}$, this assumption is exactly fulfilled, because the overlaps $a_{\bar{M} M}^{\Lambda_\eta^{(\epsilon)}}(r)$ are then equal to the squares of the overlaps of a $P = -$ static-light meson with the corresponding single-meson creation operator, which do not depend on the static spin orientation.}). For each $\bar{Q} Q$ separation $r$ we perform a $\chi^2$ minimizing fit with the remaining five parameters $V_0^{\Sigma_g^+}(r)$, $V_1^{\Sigma_g^+}(r)$, $\theta(r)$, $a_{\bar{Q} Q}^{\Sigma_g^+}(r)$ and $a_{\bar{M} M}^{\Sigma_g^+}(r)$ with $\chi^2 / \textrm{dof}$ indicating good fits. The corresponding results are collected in Table~\ref{TAB_data}, columns 2 to 6. Moreover, the potentials $V_0^{\Sigma_g^+}(r)$, $V_1^{\Sigma_g^+}(r)$ and the mixing angle $\theta(r)$ are shown as functions of $r$ in Fig.\ \ref{fig:fit_E1_E2_theta}.

\begin{table*}[htb]
\begin{center}
\begin{tabular}{c|ccccc|ccc}
\hline
$r/a$ & $V_0^{\Sigma_g^+}(r) a$ & $V_1^{\Sigma_g^+}(r) a$ & $\theta(r)$ & $a_{\bar{Q} Q}^{\Sigma_g^+}(r)$ & $a_{\bar{M} M}^{\Sigma_g^+}(r)$ & $V_{\bar{Q} Q}(r) a$ & $V_{\bar{M} M,\parallel}(r) a$ & $V_{\textrm{mix}}(r) a$ \\
\hline
$\phantom{0}1.365$ & $-0.760(04)$ & $+0.104(54)$ & $0.258(004)$ & $1.038(11)$ & $0.493(6)$ & $-0.704(05)$ & $+0.048(51)$ & $-0.213(12)$ \\
$\phantom{0}1.442$ & $-0.708(04)$ & $+0.147(42)$ & $0.336(004)$ & $1.057(11)$ & $0.498(5)$ & $-0.615(05)$ & $+0.054(39)$ & $-0.267(12)$ \\
$\phantom{0}1.826$ & $-0.651(04)$ & $+0.196(35)$ & $0.392(006)$ & $1.072(11)$ & $0.502(4)$ & $-0.528(06)$ & $+0.073(31)$ & $-0.299(12)$ \\
$\phantom{0}1.855$ & $-0.637(04)$ & $+0.150(34)$ & $0.429(007)$ & $1.088(11)$ & $0.498(4)$ & $-0.500(07)$ & $+0.014(29)$ & $-0.298(13)$ \\
$\phantom{0}2.836$ & $-0.544(04)$ & $+0.169(27)$ & $0.481(007)$ & $1.103(12)$ & $0.500(3)$ & $-0.391(06)$ & $+0.016(22)$ & $-0.292(11)$ \\
\hline
$\phantom{0}2.889$ & $-0.533(04)$ & $+0.149(24)$ & $0.498(007)$ & $1.113(12)$ & $0.498(4)$ & $-0.378(06)$ & $-0.007(20)$ & $-0.286(10)$ \\
$\phantom{0}3.513$ & $-0.494(04)$ & $+0.147(24)$ & $0.478(007)$ & $1.095(12)$ & $0.499(3)$ & $-0.358(06)$ & $+0.011(20)$ & $-0.262(10)$ \\
$\phantom{0}3.922$ & $-0.465(04)$ & $+0.123(23)$ & $0.461(007)$ & $1.075(12)$ & $0.498(3)$ & $-0.349(05)$ & $+0.007(20)$ & $-0.234(09)$ \\
$\phantom{0}4.252$ & $-0.450(04)$ & $+0.140(21)$ & $0.434(007)$ & $1.051(13)$ & $0.500(3)$ & $-0.346(05)$ & $+0.035(19)$ & $-0.225(08)$ \\
$\phantom{0}4.942$ & $-0.410(03)$ & $+0.113(24)$ & $0.397(005)$ & $1.026(10)$ & $0.499(2)$ & $-0.332(03)$ & $+0.035(22)$ & $-0.186(08)$ \\
\hline
$\phantom{0}5.229$ & $-0.398(03)$ & $+0.116(24)$ & $0.375(005)$ & $1.013(10)$ & $0.500(3)$ & $-0.329(03)$ & $+0.047(22)$ & $-0.175(07)$ \\
$\phantom{0}5.666$ & $-0.376(04)$ & $+0.088(18)$ & $0.357(007)$ & $1.001(13)$ & $0.497(3)$ & $-0.319(04)$ & $+0.031(17)$ & $-0.152(06)$ \\
$\phantom{0}5.954$ & $-0.365(03)$ & $+0.080(22)$ & $0.344(007)$ & $0.979(10)$ & $0.497(2)$ & $-0.314(03)$ & $+0.029(21)$ & $-0.141(06)$ \\
$\phantom{0}6.953$ & $-0.321(04)$ & $+0.070(17)$ & $0.312(007)$ & $0.953(13)$ & $0.497(3)$ & $-0.284(04)$ & $+0.033(16)$ & $-0.114(05)$ \\
$\phantom{0}6.962$ & $-0.320(03)$ & $+0.071(21)$ & $0.320(007)$ & $0.953(09)$ & $0.497(3)$ & $-0.281(03)$ & $+0.032(19)$ & $-0.116(06)$ \\
\hline
$\phantom{0}7.079$ & $-0.311(04)$ & $+0.040(13)$ & $0.331(007)$ & $0.966(11)$ & $0.491(2)$ & $-0.274(04)$ & $+0.003(12)$ & $-0.108(04)$ \\
$\phantom{0}7.967$ & $-0.277(03)$ & $+0.038(16)$ & $0.331(009)$ & $0.937(09)$ & $0.492(4)$ & $-0.244(03)$ & $+0.005(16)$ & $-0.097(05)$ \\
$\phantom{0}8.492$ & $-0.257(05)$ & $+0.032(09)$ & $0.333(009)$ & $0.925(15)$ & $0.491(1)$ & $-0.226(04)$ & $+0.001(08)$ & $-0.089(03)$ \\
$\phantom{0}8.680$ & $-0.243(05)$ & $+0.020(08)$ & $0.344(011)$ & $0.943(16)$ & $0.489(1)$ & $-0.213(04)$ & $-0.010(08)$ & $-0.083(03)$ \\
$\phantom{0}8.971$ & $-0.226(05)$ & $+0.024(08)$ & $0.352(009)$ & $0.947(14)$ & $0.490(1)$ & $-0.197(04)$ & $-0.006(08)$ & $-0.081(03)$ \\
\hline
$\phantom{0}9.905$ & $-0.202(07)$ & $+0.028(08)$ & $0.344(011)$ & $0.889(18)$ & $0.491(1)$ & $-0.176(06)$ & $+0.002(08)$ & $-0.073(04)$ \\
$\phantom{0}9.974$ & $-0.192(06)$ & $+0.024(08)$ & $0.357(013)$ & $0.908(17)$ & $0.490(1)$ & $-0.166(05)$ & $-0.003(08)$ & $-0.071(04)$ \\
          $10.408$ & $-0.187(09)$ & $+0.020(08)$ & $0.345(021)$ & $0.864(26)$ & $0.489(1)$ & $-0.163(08)$ & $-0.004(08)$ & $-0.066(05)$ \\
          $10.977$ & $-0.150(11)$ & $+0.008(07)$ & $0.397(033)$ & $0.898(30)$ & $0.489(1)$ & $-0.126(10)$ & $-0.015(08)$ & $-0.056(06)$ \\
          $11.319$ & $-0.150(08)$ & $+0.023(08)$ & $0.358(023)$ & $0.849(20)$ & $0.490(1)$ & $-0.129(07)$ & $+0.002(09)$ & $-0.057(04)$ \\
\hline
          $12.138$ & $-0.125(09)$ & $+0.019(08)$ & $0.358(030)$ & $0.819(23)$ & $0.490(1)$ & $-0.107(08)$ & $+0.001(09)$ & $-0.047(05)$ \\
          $12.733$ & $-0.099(07)$ & $+0.023(04)$ & $0.400(030)$ & $0.821(15)$ & $0.490(1)$ & $-0.080(06)$ & $+0.004(05)$ & $-0.044(04)$ \\
          $13.869$ & $-0.045(08)$ & $+0.026(08)$ & $0.540(091)$ & $0.828(14)$ & $0.491(1)$ & $-0.026(05)$ & $+0.007(11)$ & $-0.031(06)$ \\
          $14.147$ & $-0.047(06)$ & $+0.024(07)$ & $0.507(065)$ & $0.796(09)$ & $0.491(1)$ & $-0.030(04)$ & $+0.007(09)$ & $-0.030(04)$ \\
          $14.288$ & $-0.040(07)$ & $+0.022(07)$ & $0.555(083)$ & $0.798(12)$ & $0.490(2)$ & $-0.023(04)$ & $+0.005(10)$ & $-0.028(05)$ \\
\hline
          $14.463$ & $-0.037(08)$ & $+0.020(07)$ & $0.671(109)$ & $0.797(12)$ & $0.489(2)$ & $-0.015(04)$ & $-0.002(11)$ & $-0.028(05)$ \\
          $14.605$ & $-0.024(08)$ & $+0.024(10)$ & $0.671(155)$ & $0.804(12)$ & $0.490(1)$ & $-0.006(04)$ & $+0.005(14)$ & $-0.023(05)$ \\
          $14.704$ & $-0.030(09)$ & $+0.021(08)$ & $0.811(145)$ & $0.798(11)$ & $0.489(2)$ & $-0.003(04)$ & $-0.006(14)$ & $-0.025(05)$ \\
          $15.008$ & $-0.024(08)$ & $+0.021(07)$ & $0.915(149)$ & $0.787(11)$ & $0.489(1)$ & $+0.004(03)$ & $-0.007(12)$ & $-0.021(04)$ \\
          $15.176$ & $-0.012(08)$ & $+0.031(13)$ & $0.874(182)$ & $0.789(12)$ & $0.491(1)$ & $+0.014(05)$ & $+0.006(16)$ & $-0.021(06)$ \\
\hline
          $15.372$ & $-0.009(06)$ & $+0.037(12)$ & $1.031(126)$ & $0.794(14)$ & $0.490(1)$ & $+0.025(07)$ & $+0.003(11)$ & $-0.020(06)$ \\
          $15.561$ & $-0.005(06)$ & $+0.041(12)$ & $0.925(152)$ & $0.773(14)$ & $0.491(1)$ & $+0.024(06)$ & $+0.012(13)$ & $-0.022(07)$ \\
          $15.600$ & $-0.006(07)$ & $+0.046(11)$ & $1.136(110)$ & $0.795(15)$ & $0.490(1)$ & $+0.036(08)$ & $+0.003(10)$ & $-0.020(06)$ \\
          $17.331$ & $+0.004(08)$ & $+0.097(13)$ & $1.426(029)$ & $0.756(21)$ & $0.491(1)$ & $+0.095(13)$ & $+0.006(08)$ & $-0.013(03)$ \\
          $19.063$ & $+0.012(09)$ & $+0.163(20)$ & $1.516(011)$ & $0.732(29)$ & $0.491(1)$ & $+0.162(20)$ & $+0.013(09)$ & $-0.008(02)$ \\
\hline
\end{tabular}
\caption{\label{TAB_data}Columns 2 to 6: potentials $V_0^{\Sigma_g^+}(r) a$ and $V_1^{\Sigma_g^+}(r) a$, mixing angle $\theta(r)$, overlaps $a_{\bar{Q} Q}^{\Sigma_g^+}(r)$ and $a_{\bar{M} M}^{\Sigma_g^+}(r)$ from $\chi^2$ minimizing fits to the resampled correlation functions of Ref.\ \cite{Bali:2005fu} (lattice spacing $a \approx 1 / (2.37 \, \textrm{GeV})$). Columns 7 to 9: potentials $V_{\bar{Q} Q}(r) a$, $V_{\bar{M} M,\parallel}(r) a$ and $V_{\textrm{mix}}(r) a$ obtained via Eqs.\ (\ref{EQN276}) to (\ref{EQN278}).}
\end{center}
\end{table*}

\begin{figure}[htb]
\includegraphics[width=0.49\textwidth]{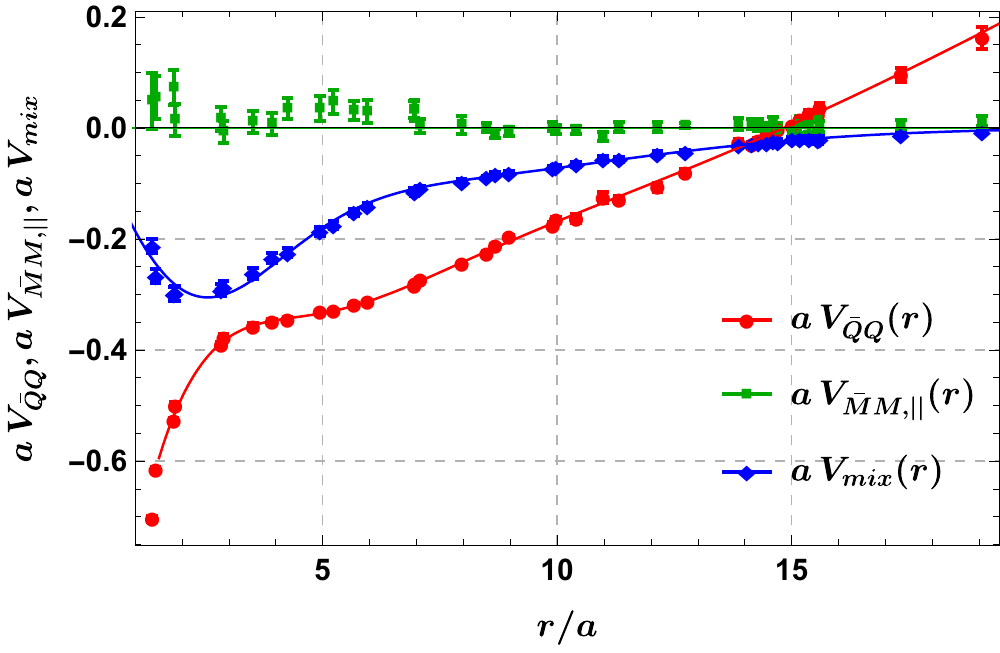}
\caption{\label{fig:fit_V} (Color online.) Potentials $V_{\bar{Q} Q}(r)$, $V_{\bar{M} M,\parallel}(r)$ and $V_{\textrm{mix}}(r)$ as functions of the $\bar{Q} Q$ separation $r$ (lattice spacing $a \approx 1 / (2.37 \, \textrm{GeV})$). The curves are the parametrizations (\ref{eq:fitVQQ}) to (\ref{eq:fitMix}).}
\end{figure}

Inserting the fit results for $V_0^{\Sigma_g^+}(r)$, $V_1^{\Sigma_g^+}(r)$ and $\theta(r)$ into Eqs.\ (\ref{EQN276}) to (\ref{EQN278}) leads to the potentials $V_{\bar{Q} Q}(r)$, $V_{\bar{M} M,\parallel}(r)$ and $V_{\textrm{mix}}(r)$, which appear in the coupled channel Schr\"odinger equation (Eqs.\ (\ref{EQN050}) and (\ref{EQN051}) or specifically for $\widetilde{J}^{P C} = 0^{+ +}$ Eq.\ (\ref{eq:coupledpot})). These potentials are also collected in Table~\ref{TAB_data}, columns 7 to 9, and shown in Fig.\ \ref{fig:fit_V}.

There are several interesting aspects concerning the potentials $V_0^{\Sigma_g^+}(r)$, $V_1^{\Sigma_g^+}(r)$ (see Figure~\ref{fig:fit_E1_E2_theta}, upper plot), which correspond to energy eigenvalues, and the potentials $V_{\bar{Q} Q}(r)$, $V_{\bar{M} M,\parallel}(r)$, $V_{\textrm{mix}}(r)$ (see Figure~\ref{fig:fit_V}), which represent the interaction in the quarkonium and the two-meson channel and the mixing between these channels:
\begin{itemize}
\item $V_0^{\Sigma_g^+}(r)$ for separations $r \ltapprox 1 \, \textrm{fm}$ is a standard quantity computed in lattice QCD and commonly referred to as ``the static potential''. It has a negative curvature for small $r$ and is almost linear for larger $r$. Quite often it is parameterized via $V_0^{\Sigma_g^+}(r) = \textrm{const} - \alpha/r + \sigma r$ (see e.g.\ Ref.\ \cite{Karbstein:2018mzo}). With this parameterization it is straightforward to determine the common hadronic scale $r_0$ defined via $r_0^2 |(d/dr) V_0^{\Sigma_g^+}(r)|_{r = r_0}| = 1.65$ in Ref.\ \cite{Sommer:1993ce}. Fitting the parameterization in the range $3 a \leq r \leq 12 a$ yields $r_0 = 6.004(41) \, a$, which is in excellent agreement with the original result quoted in Ref.\ \cite{Bali:2005fu}), $r_0 = 6.009(53) \, a$.


\item $V_{\bar{Q} Q}(r)$ is the potential between a static quark and a static antiquark. In contrast to $V_0^{\Sigma_g^+}(r)$ this potential is linear also for separations $r \gtapprox 1 \, \textrm{fm}$, i.e.\ separations larger than the string breaking distance. At smaller $r$ there is a sizable $r$-dependent mixing of the two lowest energy eigenstates (see Fig.\ \ref{fig:fit_E1_E2_theta}, lower plot), and $V_{\bar{Q} Q}(r)$ is an $r$-dependent linear combination of the potentials $V_0^{\Sigma_g^+}(r)$ and $V_1^{\Sigma_g^+}(r)$ (see Eq.\ (\ref{EQN276})). Thus one should not expect a simple parameterization similar to the previously mentioned $\textrm{const} - \alpha/r + \sigma r$. The mixing manifests itself by a clearly visible bump around $r \approx 3 a$ and needs to be taken into account, when parameterizing $V_{\bar{Q} Q}(r)$.

\item $V_{\bar{M} M,\parallel}(r)$ is the potential between two static-light mesons. For larger $r$ the residual strong force between the two mesons is expected to vanish and the potential should approach two times the static-light meson mass. This expectation is in excellent agreement to what we observe for $r \gtapprox 8 a$. For smaller $r$ the statistical errors are larger and the potential might either be constant or slightly repulsive. Note that $V_1^{\Sigma_g^+}(r)$ is somewhat larger than $V_{\bar{M} M,\parallel}(r)$. Again, this is a consequence of the non-vanishing mixing angle.
\end{itemize}

To solve the Schr\"odinger equation (\ref{EQN050}), which we do in the following for $\widetilde{J}^{P C} = 0^{+ +}$ (see Eq.\ (\ref{eq:coupledpot})), it is necessary to have a continuous parameterization of the potentials $V_{\bar{Q} Q}(r)$, $V_{\bar{M} M,\parallel}(r)$ and $V_{\textrm{mix}}(r)$. We found the parametrization
\begin{eqnarray}
\nonumber & & V_{\bar{Q} Q}(r) = E_0 - \frac{\alpha}{r} + \sigma r + \sum_{j=1}^2 c_{\bar{Q} Q,j} r \exp\bigg(-\frac{r^2}{2 \lambda_{\bar{Q} Q,j}^2}\bigg) \\
\label{eq:fitVQQ} & & \\
\label{eq:fitVBB} & & V_{\bar{M} M,\parallel}(r) = 0 \\
\label{eq:fitMix} & & V_{\textrm{mix}}(r) = \sum_{j=1}^2 c_{\textrm{mix},j} r \exp\bigg(-\frac{r^2}{2 \lambda_{\textrm{mix},j}^2}\bigg)
\end{eqnarray}
most suitable.
The $11$ parameters, $E_0$, $\alpha$, $\sigma$, $c_{\bar{Q} Q,j}$, $\lambda_{\bar{Q} Q,j}$, $c_{\textrm{mix},j}$ and $\lambda_{\textrm{mix},j}$ (where $j = 1,2$) can be determined by $\chi^2$ minimizing fits to the data points collected in Table~\ref{TAB_data} in a stable way and the corresponding $\chi^2 / \textrm{dof}$ indicate reasonable fits. It is known that discretization errors for data points corresponding to small separations $r/a \ltapprox 2$ are sizable. To reduce these discretization errors the authors of Ref.\ \cite{Bali:2005fu} used a tree-level improvement technique \cite{Sommer:1993ce}. Since data points are only available for a single lattice spacing, it is not obvious, for which separations $r$ discretization errors can be neglected. Thus, we performed several fits and computations, where data points with $r < r_\textrm{min}$ are excluded with $r_\textrm{min} / a \in \{ 1.365 , 1.442 , 1.826 , 1.855 , 2.836 \}$. Choosing $r_\textrm{min} / a \geq 1.442$ (i.e.\ excluding at least the smallest separation $r/a = 1.365$) leads to results, which are quite stable, while the results for $r_\textrm{min} / a = 1.365$ are somewhat different. All results presented in the following correspond to $r_\textrm{min}/a = 1.442$. Results for the other $r_\textrm{min}$ values are used to estimate systematic errors (see the discussion in section~\ref{SEC807}).
 
We also tested several other parameterizations, e.g.\ with more or less than two terms in the sums of Eqs.\ (\ref{eq:fitVQQ}) and (\ref{eq:fitMix}) or with exponentials with arguments linear in $r$ instead of quadratic, but the resulting $\chi^2 / \textrm{dof}$ were either larger or the fits less stable. Results for the parameterization (\ref{eq:fitVQQ}) and (\ref{eq:fitMix}) are collected in Table~\ref{tab:fitsGevFm} and the fitted parameterizations are shown in Fig.\ \ref{fig:fit_V}.

Moreover, Eq.\ (\ref{EQN279}) and our above assumption $V_0^{\Pi_g}(r) = 0$ lead to
\begin{eqnarray}
\label{EQN566} V_{\bar{M} M,\perp}(r) = 0 .
\end{eqnarray}

\begin{table*}[htb]
\begin{center}
\begin{tabular}{c|c|c|c|c}
\hline
\TBstrut
potential & parameter & in units of $a$ & in units of GeV & $\chi^2 / \textrm{dof}$ \\ 
\hline
\Tstrut
$V_{\bar{Q} Q}(r)$ & $E_0$                      & $-0.675(114) \, a^{-1}$ & $-1.599(269) \, \textrm{GeV}\phantom{1.^{-1}}$ & $0.91$ \\
                   & $\alpha$                   & $+0.320(94) \phantom{0 \, a^{-1}}$ & $+0.320(94) \phantom{1.0 \, \textrm{GeV}^{-1}}$ &  \\
                   & $\sigma$                   & $+0.045(006) \, a^{-2}$ & $+0.253(035) \, \textrm{GeV}^{2\phantom{-}}\phantom{1.}$ &  \\
                   & $c_{\bar{Q} Q,1}$          & $+0.147(157) \, a^{-2}$ & $+0.826(882) \, \textrm{GeV}^{2\phantom{-}}\phantom{1.}$ &  \\
                   & $\lambda_{\bar{Q} Q,1}$    & $+2.285(112) \, a\phantom{^{-1}}$ & $+0.964(47) \, \textrm{GeV}^{-1}\phantom{1.0}$ &  \\
                   & $c_{\bar{Q} Q,2}$          & $+0.031(179) \, a^{-2}$ & $+0.174(1.004) \, \textrm{GeV}^{2\phantom{-}}$ &  \\
\Bstrut
                   & $\lambda_{\bar{Q} Q,2}$    & $+6.311(1.008) \, a\phantom{\ }$ & $+2.663(425) \, \textrm{GeV}^{-1}\phantom{1.}$ &  \\
\hline
\TBstrut
$V_{\bar{M} M,\parallel}(r)$ & -- & & & $1.14$ \\
\hline
\Tstrut
$V_{\textrm{mix}}(r)$ & $c_{\textrm{mix},1}$       & $-0.176(6) \, a^{-2}\phantom{00}$ & $-0.988(32) \, \textrm{GeV}^{2\phantom{-}}\phantom{1.0}$ & $0.79$ \\
                      & $\lambda_{\textrm{mix},1}$ & $+2.327(42) \, a\phantom{0^{-1}}$ & $+0.982(18) \, \textrm{GeV}^{-1}\phantom{1.0}$ &  \\
                      & $c_{\textrm{mix},2}$       & $-0.025(1) \, a^{-2}\phantom{00}$ & $-0.142(7) \, \textrm{GeV}^{2\phantom{-}}\phantom{1.00}$ &  \\
\Bstrut
                      & $\lambda_{\textrm{mix},2}$ & $+6.319(108) \, a\phantom{^{-1}}$ & $+2.666(46) \, \textrm{GeV}^{-1}\phantom{1.0}$ &  \\
\hline
\end{tabular}
\end{center}
\caption{\label{tab:fitsGevFm}The parameters of the potential parametrizations (\ref{eq:fitVQQ}) to (\ref{eq:fitMix}) in units of the lattice spacing $a \approx 1 / (2.37 \, \textrm{GeV})$ as well as in GeV.}
\end{table*}




\section{\label{sec:emergent}Numerical methods to study quarkonium resonances}


\subsection{$\mbox{S}$ and $\mbox{T}$ matrix poles in the complex energy plane and their relation to quarkonium resonances}

The quantity $t_{1 \rightarrow 0,0}$ appearing in the $r \rightarrow \infty$ boundary condition (\ref{EQN545}) of the radial wave function $\chi_{1 \rightarrow 0,0}(r)$ of the coupled channel Schr\"odinger equation (\ref{eq:coupledpot}) is an eigenvalue of the $\mbox{T}$ matrix. From $t_{1 \rightarrow 0,0}$ we can read off the corresponding $\mbox{S}$ matrix eigenvalue,
\begin{equation}
\label{eq:Smatrix} s_{1 \rightarrow 0,0} = 1 + 2 i t_{1 \rightarrow 0,0} = e^{2 i \delta_{1 \rightarrow 0,0}} .
\end{equation}
Moreover, both the $\mbox{S}$ matrix and the $\mbox{T}$ matrix are analytical in the complex plane. They are well-defined for complex energies $E$. The poles of the $\mbox{S}$ and the $\mbox{T}$ matrix, i.e.\ the poles of $t_{L \rightarrow \widetilde{J},\widetilde{J}_z}$, which are in the second Riemann sheet of the complex energy plane with a negative imaginary part, correspond to quarkonium resonances. For a pole at complex energy $E$ the resonance energy and the decay width are
\begin{eqnarray}
\label{EQN521} m = \textrm{Re}(E) \quad , \quad \Gamma = -2 \textrm{Im}(E) .
\end{eqnarray}
For more details see e.g.\ our recent work \cite{Bicudo:2017szl}.


\subsection{Numerical methods to determine $t_{1 \rightarrow 0,0}$ and to find poles}

Since we restrict our numerical calculations in this work to the sector $\widetilde{J}^{P C} = 0^{+ +}$, i.e.\ the Schr\"odinger equation (\ref{eq:coupledpot}), it is convenient to use the simplified notation $u(r) \equiv u_{0,0}(r)$ and $\chi(r) \equiv \chi_{1 \rightarrow 0,0}(r)$.

The boundary conditions of the solutions $(u(r) , \chi(r))$ of the Schr\"odinger equation (\ref{eq:coupledpot}) can be read of from Eqs.\ (\ref{EQNbc1}), (\ref{EQNbc2}), (\ref{EQNbc3}) and (\ref{EQN545}) and are
\begin{eqnarray}
 & & u(r) \propto r \quad \textrm{for } r \rightarrow 0 \\
\label{EQNbc02} & & u(r) = 0 \quad \textrm{for } r \rightarrow \infty \\
 & & \chi(r) \propto r^2 \quad \textrm{for } r \rightarrow 0 \\
\label{EQNbc04} & & \chi(r) = i t_{1 \rightarrow 0,0} k r h_1^{(1)}(k r) \quad \textrm{for } r \rightarrow \infty .
\end{eqnarray}
Note that the boundary condition (\ref{EQNbc04}) depends on $t_{1 \rightarrow 0,0}$. For a given value of the energy $E$ this boundary condition is only fulfilled for a specific corresponding value of $t_{1 \rightarrow 0,0}$. In other words the boundary condition (\ref{EQNbc04}) fixes $t_{1 \rightarrow 0,0}$ as a function of $E$. Thus, our numerical goals in the following are to compute $t_{1 \rightarrow 0,0}$ for given values of the energy $E$ and to find the poles of $t_{1 \rightarrow 0,0}$ in the complex energy plane.

To this end we replace $r \rightarrow \infty$ in Eqs.\ (\ref{EQNbc02}) and (\ref{EQNbc04}) by $r \geq R$, where $R$ is finite, but sufficiently large, such that the boundary conditions are still valid. We discretize the interval $[0,R]$ using a uniform 1-dimensional lattice with $N+1$ sites and spacing $d = R / N$, i.e.\ $r \rightarrow r_n = n d$. Moreover, $u(r) \rightarrow u_n = u(r_n)$ and $\chi(r) \rightarrow \chi_n = \chi(r_n)$ with boundary conditions $u_0 = 0$, $\chi_0 = 0$, $u_N = 0$ and
\begin{eqnarray}
\nonumber & & \chi_N = i t_{1 \rightarrow 0,0} h_1^{(1)}(k R) k R = \\
\label{EQN952} & & \hspace{0.7cm} = \chi_{N-1} \frac{h_1^{(1)}(k R) R}{h_1^{(1)}(k (R-d)) (R-d)} .
\end{eqnarray}
%
%
%
%
%
%
%
The second derivative is discretized according to
\begin{eqnarray}
\partial_r^2 \rightarrow (\Delta_r^2)_{n,n'} = \frac{\delta_{n+1,n'} - \delta_{n,n'} + \delta_{n-1,n'}}{d^2} .
\end{eqnarray}
The coupled channel Schr\"odinger equation (\ref{eq:coupledpot}) is then a system of $2 (N-1)$ linear equations,
%
%
%
%
%
%
\begin{widetext}
\begin{eqnarray}
\nonumber & & \hspace{-0.7cm} \sum_{n'=0}^N \bigg(-\frac{1}{2} \bigg(\begin{array}{cc} 1/\mu_Q & 0 \\ 0 & 1/\mu_M \end{array}\bigg) (\Delta_r^2)_{n,n'} + \bigg(\frac{1}{2 r_n^2} \bigg(\begin{array}{cc} 0 & 0 \\ 0 & 2/\mu_M \end{array}\bigg) + V_0(r_n) + 2 m_M - E\bigg) \delta_{n,n'}\bigg)
\left(\begin{array}{c} u_{n'} \\ \chi_{n'} \end{array}\right) = \\
 & & = -\left(\begin{array}{c} V_{\textrm{mix}}(r_n) \\ V_{\bar{M} M,\parallel}(r_n) \end{array}\right) k r_n j_1(k r_n) \quad , \quad n = 1,\ldots,N-1 ,
\end{eqnarray}
for the $2 (N-1)$ unknowns $u_1 ,\ldots, u_{N-1}$ and $\chi_1 ,\ldots, \chi_{N-1}$ representing the radial wave functions. Inserting the boundary conditions leads to
%
%
%
%
%
%
%
\begin{eqnarray}
\nonumber & & \hspace{-0.7cm} \sum_{n'=1}^{N-1} \bigg(-\frac{1}{2} \bigg(\begin{array}{cc} 1/\mu_Q & 0 \\ 0 & 1/\mu_M \end{array}\bigg) (\Delta_r^2)_{n,n'} + \bigg(\frac{1}{2 r_n^2} \bigg(\begin{array}{cc} 0 & 0 \\ 0 & 2/\mu_M \end{array}\bigg) + V_0(r_n) + 2 m_M - E\bigg) \delta_{n,n'} \\
\nonumber & & \hspace{0.7cm} -\frac{1}{2 \mu_M d^2} \frac{h_1^{(1)}(k R) R}{h_1^{(1)}(k (R-d)) (R-d)} \bigg(\begin{array}{cc} 0 & 0 \\ 0 & 1 \end{array}\bigg) \delta_{n,N-1} \delta_{n,n'}\bigg)
\left(\begin{array}{c} u_{n'} \\ \chi_{n'} \end{array}\right) = \\
\label{EQN754} & & = -\left(\begin{array}{c} V_{\textrm{mix}}(r_n) \\ V_{\bar{M} M,\parallel}(r_n) \end{array}\right) k r_n j_1(k r_n) \quad , \quad n = 1,\ldots,N-1 ,
\end{eqnarray}
\end{widetext}
which is of the form $M(E) \mathbf{x} = \mathbf{b}$. $\mathbf{x}$ is a vector with $2 (N-1)$ components, which are $u_1 ,\ldots, u_{N-1}$ and $\chi_1 ,\ldots, \chi_{N-1}$. $M(E)$ is a square matrix with $2 (N-1) \times 2 (N-1)$ entries and $\mathbf{b}$ is a vector with $2 (N-1)$ components, which can be read off from the left hand side and the right hand side of Eq.\ (\ref{EQN754}), respectively.

To determine $t_{1 \rightarrow 0,0}$ for a given value of the energy $E$ is now straightforward. We solve the linear system (\ref{EQN754}) and insert $\chi_{N-1}$ in Eq.\ (\ref{EQN952}).

In case we are just interested to find the positions of poles of $t_{1 \rightarrow 0,0}$ in the complex energy plane, we use a more efficient method. From $\mathbf{x} = M^{-1}(E) \mathbf{b}$, where all components of $\mathbf{b}$ are finite, it is obvious that $M^{-1}(E)$ must have at least one infinite eigenvalue, if $|t_{1 \rightarrow 0,0}| \rightarrow \infty$ or equivalently $|\chi_{N-1}| \rightarrow \infty$. Thus, $M(E)$ must have at least one zero mode, which implies $\det(M(E)) = 0$. Thus, we determine poles of $t_{1 \rightarrow 0,0}$ by finding the roots of $\det(M(E))$. To this end we apply the Newton-Raphson method for analytic functions of a single complex variable. Note that $\det(M(E))$ is typically a large number beyond machine double precision. Thus, we rescale $M(E)$ by an appropriate factor, after a $LU$ decomposition, but before multiplying the diagonal elements to obtain the determinant. The roots of the determinant are, of course independent, of such a rescaling, i.e.\ the final results for the positions of the poles of $t_{1 \rightarrow 0,0}$ in the complex energy plane are unaffected.

Instead of rewriting the coupled channel Schrödinger equation (\ref{eq:coupledpot}) as a large system of linear equations (\ref{EQN754}), one can also solve it and determine $t_{1 \rightarrow 0,0}$ using Runge-Kutta methods. We cross checked and verified our numerical results by implementing a 4th order Runge-Kutta solver.


\section{\label{sec:results}Numerical results for bottomonium}


\subsection{\label{SEC489}Choice of parameters and error analysis}

In the following we focus on $b$ quarks and bottomonium, where the heavy quark approximations discussed in section~\ref{SEC466} are more accurate. For $m_M$, which is the energy reference of our system, we use the spin-averaged mass of the $B$ meson and the $B^*$ meson, i.e.\ $m_M = (m_B + 3 m_{B^*}) / 4 = 5.313 \, \textrm{GeV}$ \cite{Patrignani:2016xqp}. $\mu_Q = m_Q/2$ in the kinetic term of the coupled channel Schr\"odinger equation is the reduced mass of the $b$ quark. In the quark model perspective the $B$ meson is composed of a $b$ quark/antiquark and a light antiquark/quark $u$ or $d$. Thus the $B$ meson mass is heavier than the heavy quark mass, where the difference is of the order of the light constituent quark mass $m_l$, i.e.\ $m_M = m_Q + m_l$. Since results are only weakly dependent on $m_Q$ (see e.g.\ previous work following a similar approach \cite{Bicudo:2015kna,Karbstein:2018mzo}), we use for simplicity $m_Q = 4.977 \, \textrm{GeV}$ from quark models \cite{Godfrey:1985xj}.

For the upper boundary of the $r$ axis we use $R = 15.0 / \textrm{GeV} \approx 2.96 \, \textrm{fm} \approx 35.6 \, a$, where $V_{\bar{Q} Q}(r)$ is quite large and both $V_{\bar{M} M,\parallel}(r)$ and $V_{\textrm{mix}}(r)$ are essentially vanishing (see e.g.\ Figure~\ref{fig:fit_V}). For the 1-dimensional lattice discretizing the interval $[0,R]$ we use $N = 600$ sites corresponding to the spacing $d = R/N = 0.025 / \textrm{GeV} \approx 0.005 \, \textrm{fm}$. We verified the independence of our results from these parameters for $R \geq 15.0 / \textrm{GeV}$ and $N \geq 600$ by performing identical computations with several different $R$ and $N$.

We propagate the uncertainties provided in Ref.\ \cite{Bali:2005fu}, TABLE~I by resampling. We generate 1000 statistically independent samples and repeat all computations on each of the samples. This is a computer time consuming task, for which we employ GPUs utilizing the algebra package of CUDA \cite{CUDA1} and the cuSOLVER library \cite{CUDA2}. For $t_{1 \rightarrow 0,0}$, the corresponding phase shift $\delta_{1 \rightarrow 0,0}$, energy levels and positions of poles of $t_{1 \rightarrow 0,0}$ in the complex energy plane we quote asymmetric errors, which are defined via the 16th and 84th percentile of the 1000 samples, respectively.


\subsection{The bottomonium spectrum from single channel Schr\"odinger equations}

In this subsection we are interested in a qualitative understanding, how results from single channel Schr\"odinger equations compare to experimental results. Thus we just use the mean values of the parameters of the potentials $V_{\bar{Q} Q}(r)$, $V_{\bar{M} M,\parallel}(r)$ and $V_{\textrm{mix}}(r)$ (Eqs.\ (\ref{eq:fitVQQ}) to (\ref{eq:fitMix})), but ignore their uncertainties. It will also be interesting to compare these single channel results to more realistic results from the coupled channel Schr\"odinger equation (\ref{eq:coupledpot}), which will be discussed in sections \ref{SEC806} and \ref{SEC807}.

In a first step we compute the quarkonium spectrum, setting $V_{\textrm{mix}}(r) = 0$, i.e.\ decoupling the quarkonium channel from the meson-meson channel. The corresponding Schr\"odinger equation is then the upper component of Eq.\ (\ref{eq:coupledpot}),
\begin{eqnarray}
\label{eq:schro1} \left(-\frac{1}{2 \mu_Q} \partial_r^2 + V_{\bar{Q} Q}(r) + 2 m_M - E\right) u_{0,0}(r) = 0 .
\end{eqnarray}

There is an infinite number of bound states, because the potential $V_{\bar{Q} Q}(r)$ is confining. The energies of the lightest states are listed in Table~\ref{tab:energyconfined}. Three states are below the $B^{(*)} \bar{B}^{(*)}$ threshold at $2 m_M$, which is marked by a horizontal line, all other states are above. In particular the mass of the lowest state ($n=1$) is significantly larger than the corresponding masses from experiment (compare columns ``from $V_{\bar{Q} Q}(r)$'' and ``from experiment'' of Table~\ref{tab:energyconfined}; for a full summary of the experimental results on bottomonium see Table~\ref{tab:bottomonium}). This sizable discrepancy is expected, because the mixing angle is non-zero, where the wave function is large, i.e.\ $0.3 \ltapprox \theta(r) \ltapprox 0.5$ for separations $r \ltapprox 1.1 \, \textrm{fm}$ (see Fig.\ \ref{fig:fit_E1_E2_theta}). Consequently, the potential $V_{\bar{Q} Q}(r)$ is a mixture of the ground state potential and the first exited potential, which leads to unphysically large bottomonium masses.

\begin{table}[htb]
\begin{ruledtabular}
\begin{tabular}{c|c|c|cc}
\Tstrut
    & from $V_{\bar{Q} Q}(r)$ & from $V_0^{\Sigma_g^+}(r)$ & \multicolumn{2}{c}{from experiment} \\
\Bstrut
$n$ & $E$ [GeV]               & $E$ [GeV]                  & name & $m$ [GeV] \\
\hline
\Tstrut
$1$ & $\phantom{0}9.766$ & $\phantom{0}9.497$ & $\eta_b(1S)$     & $\phantom{0}9.399$ \\
    &                    &                    & $\Upsilon_b(1S)$ & $\phantom{0}9.460$ \\
$2$ & $10.129$           & $\phantom{0}9.998$ & $\Upsilon_b(2S)$ & $10.023$           \\
\Bstrut
$3$ & $10.436$           & $10.330$           & $\Upsilon_b(3S)$ & $10.355$           \\
\cline{2-2}
\TBstrut
$4$ & $10.696$           & $10.591$           & $\Upsilon_b(4S)$ & $10.579$           \\
\cline{3-5}
\Tstrut
$5$ & $10.941$           &          & \\
$6$ & $11.176$           &          & \\
$7$ & $11.403$           &          & \\
$8$ & $11.622$           &          & \\
\Bstrut
$\ldots$ & $\ldots$      &          &
\end{tabular}
\end{ruledtabular}
\caption{\label{tab:energyconfined}Masses for $\widetilde{J}^{P C} = 0^{+ +}$ bottomonium from single channel Schr\"odinger equations with potentials $V_{\bar{Q} Q}(r)$ and $V_0^{\Sigma_g^+}(r)$, i.e.\ Eqs.\ (\ref{eq:schro1}) and (\ref{eq:schro2}), and corresponding experimental results. Relevant $B^{(*)} \bar{B}^{(*)}$ thresholds are marked by horizontal lines.} 
\end{table}


\begin{table}[htb]
\begin{ruledtabular}
\begin{tabular}{cc|cc|c}
\TBstrut
name  & $I^G (J^{PC})$ & $m$ [GeV] & $\Gamma$ [MeV] & $ \widetilde{J}^{P C}$ \\ 
\hline
\Tstrut
$\eta_b (1S)$     & $0^+(0^{+-})$ & $\phantom{0}9.3990(23)$           & $10(5)$  & $0^{++}$\\ 
$\Upsilon_b (1S)$ & $0^-(1^{--})$ & $\phantom{0}9.4603(3)\phantom{0}$ & $54.0(1.3) \times 10^{-3}$ & $0^{++}$\\ 
$\chi_{b0} (1P)$  & $0^+(0^{++})$ & $\phantom{0}9.8594(7)\phantom{0}$ & -- & $1^{--}$\\ 
$\chi_{b1} (1P)$  & $0^+(1^{++})$ & $\phantom{0}9.8928(6)\phantom{0}$ & -- & $1^{--}$\\ 
$h_{b} (1P)$      & $\ ?^?(1^{+-})$ & $\phantom{0}9.8993(8)\phantom{0}$ & -- & $1^{--}$\\ 
$\chi_{b2} (1P)$  & $0^+(2^{++})$ & $\phantom{0}9.9122(6)\phantom{0}$ & -- & $1^{--}$\\ 
$\Upsilon (2S)$   & $0^-(1^{--})$ & $10.0233(3)\phantom{0}$           & $32.0(2.6) \times 10^{-3}$  & $0^{++}$\\ 
$\Upsilon (1D)$   & $0^-(2^{--})$ & $10.1637(14)$                     & -- & $2^{++}$\\ 
$\chi_{b0} (2P)$  & $0^+(0^{++})$ & $10.2325(9)$\phantom{0}           & -- & $1^{--}$\\ 
$\chi_{b1} (2P)$  & $0^+(1^{++})$ & $10.2555(8)$\phantom{0}           & -- & $1^{--}$\\ 
$\chi_{b2} (2P)$  & $0^+(1^{++})$ & $10.2687(7)$\phantom{0}           & -- & $1^{--}$\\ 
$\Upsilon (3S)$   & $0^-(1^{--})$ & $10.3552(5)$\phantom{0}           & $20.3(1.9) \times 10^{-3}$  & $0^{++}$\\ 
\Bstrut
$\chi_{b1} (3P)$  & $0^+(1^{++})$ & $10.5121(23)$                     & -- & $1^{--}$\\ 
\hline
\TBstrut
$\Upsilon (4S)$ & $0^-(1^{--})$   & $10.5794(12)$                     & $20.5(2.5)$   & $0^{++}$\\
\hline 
\Tstrut
$\Upsilon (10860)$ & $0^-(1^{--})$ & $10.8899(32)$                    & $51(7)\phantom{0}$  & $0^{++}$\\ 
\Bstrut
$\Upsilon (11020)$ & $0^-(1^{--})$ & $10.9929(10)$                    & $49(15)$  & $0^{++}$\\ 
\end{tabular}
\end{ruledtabular}
\caption{\label{tab:bottomonium}Bottomonium states with isopspin $I = 0$ according to the Review of Particle Physics \cite{Patrignani:2016xqp}. We also list the quantum numbers $\widetilde{J}^{PC}$ conserved in the limit of infinite $b$ quark mass ($\widetilde{J} = 0,1,2$ corresponds to $S,P,D$ in the meson name; the quantum numbers $J^{PC} = 1^{--}$ of $\Upsilon (10860)$ and $\Upsilon (11020)$ are consistent with $\widetilde{J}^{PC} = 0^{++}$, which is the sector we focus on in this work). $B^{(*)} \bar{B}^{(*)}$ thresholds are marked by horizontal lines.}
\end{table}

More realistic estimates for these masses are obtained, when replacing $V_{\bar{Q} Q}(r)$ in Eq.\ (\ref{eq:schro1}) by the ground state potential $V_0^{\Sigma_g^+}(r)$, i.e.\ by solving the Schr\"odinger equation
\begin{eqnarray}
\label{eq:schro2} \left(-\frac{1}{2 \mu_Q} \partial_r^2 + V_0^{\Sigma_g^+}(r) + 2 m_M - E\right) u(r) = 0 .
\end{eqnarray}
This is a standard approach appearing frequently in the literature (for recent work see e.g.\ \cite{Karbstein:2018mzo}). The resulting bottomonium masses are now smaller and compare much better to experimental results (see Table~\ref{tab:energyconfined}).

From Table~\ref{tab:bottomonium} one can also read off that the decay widths of bottomonium states are indeed quite small below the $B \bar{B}$ threshold, except for $\eta_b(1S)$, which diagrammatically couples to two gluons, whereas the $\Upsilon$ states couple to three gluons \cite{Braaten:2000cm,Bodwin:2001pt,Maltoni:2004hv}. Thus, our approach to neglect the OZI suppressed decays of excited bottomonium states to lighter bottomonium and a light $I = 0$ meson should be a good approximation.


\subsection{\label{SEC806}$t_{1 \rightarrow 0,0}$ and the phase shift $\delta_{1 \rightarrow 0,0}$ for real energies}

We proceed by computing the scattering amplitude $t_{1 \rightarrow 0,0}$ and the phase shift $\delta_{1 \rightarrow 0,0}$ for real energies $E$ above the $B^{(*)} \bar{B}^{(*)}$ threshold at $10.627 \, \textrm{GeV}$. In contrast to the previous subsection, we now include the meson-meson channel, i.e.\ we consider the coupled channel Schr\"odinger equation (\ref{eq:coupledpot}). The available experimental data for bottomonium goes up to $\approx 11 \, \textrm{GeV}$ (see Table~\ref{tab:bottomonium}). Two resonances consistent with $\widetilde{J}^{PC} = 0^{++}$ were already observed, $\Upsilon (10860)$ and $\Upsilon (11020)$. We perform our computations up to $11.6 \, \textrm{GeV}$.

\begin{figure}[htb]
\includegraphics[width=\columnwidth]{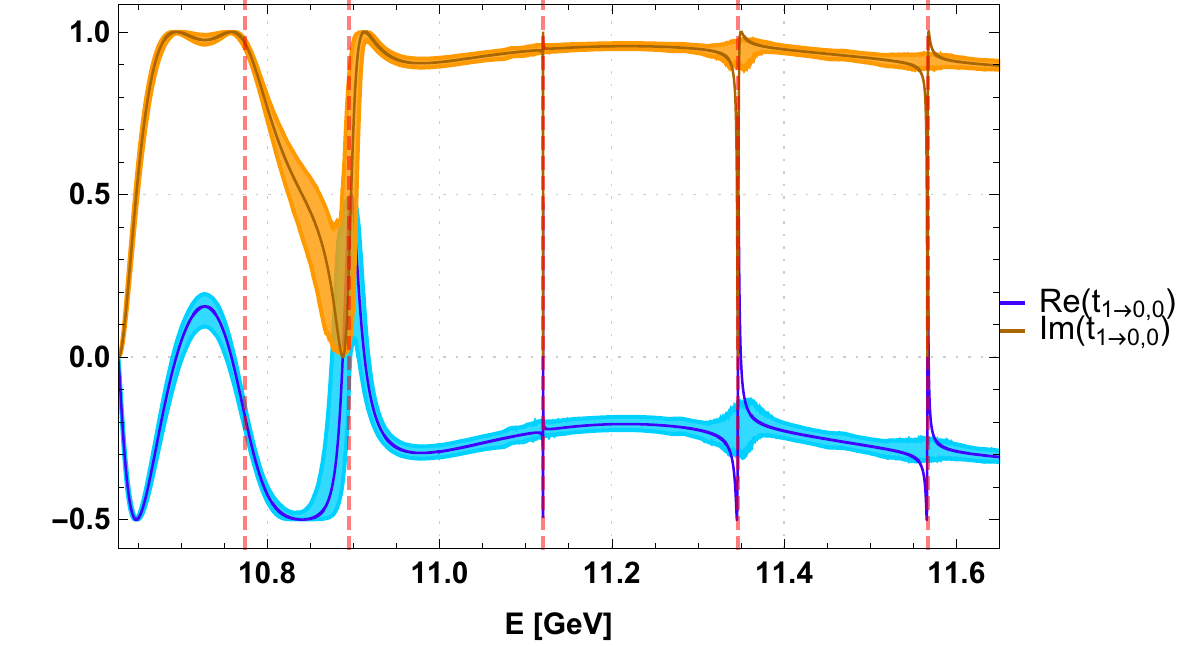}
\caption{\label{fig:realimagT}(Color online.) $\textrm{Re}(t_{1 \rightarrow 0,0})$ and $\textrm{Im}(t_{1 \rightarrow 0,0})$ as functions of the energy $E$. The real parts of the positions of the poles of $t_{1 \rightarrow 0,0}$ are indicated by red dashed lines.}
\end{figure}

In Fig.\ \ref{fig:realimagT} we show the real and the imaginary part of $t_{1 \rightarrow 0,0}$ as functions of the energy $E$. If there would be ``simple resonances'', $\textrm{Im}(t_{1 \rightarrow 0,0})$ would exhibit clear peaks, top to bottom, since according to Eq.\ (\ref{eq:optical}) it is identical to $|t_{1 \rightarrow 0,0}|^2$, i.e.\ proportional to the absolute square of the partial wave scattering amplitude. In such a case we could determine the decay widths from the peaks at half height. However, the system we investigate is more complicated, e.g.\ the resonances seem to mutually impact each other and there is a large background.

\begin{figure}[htb]
\includegraphics[width=\columnwidth]{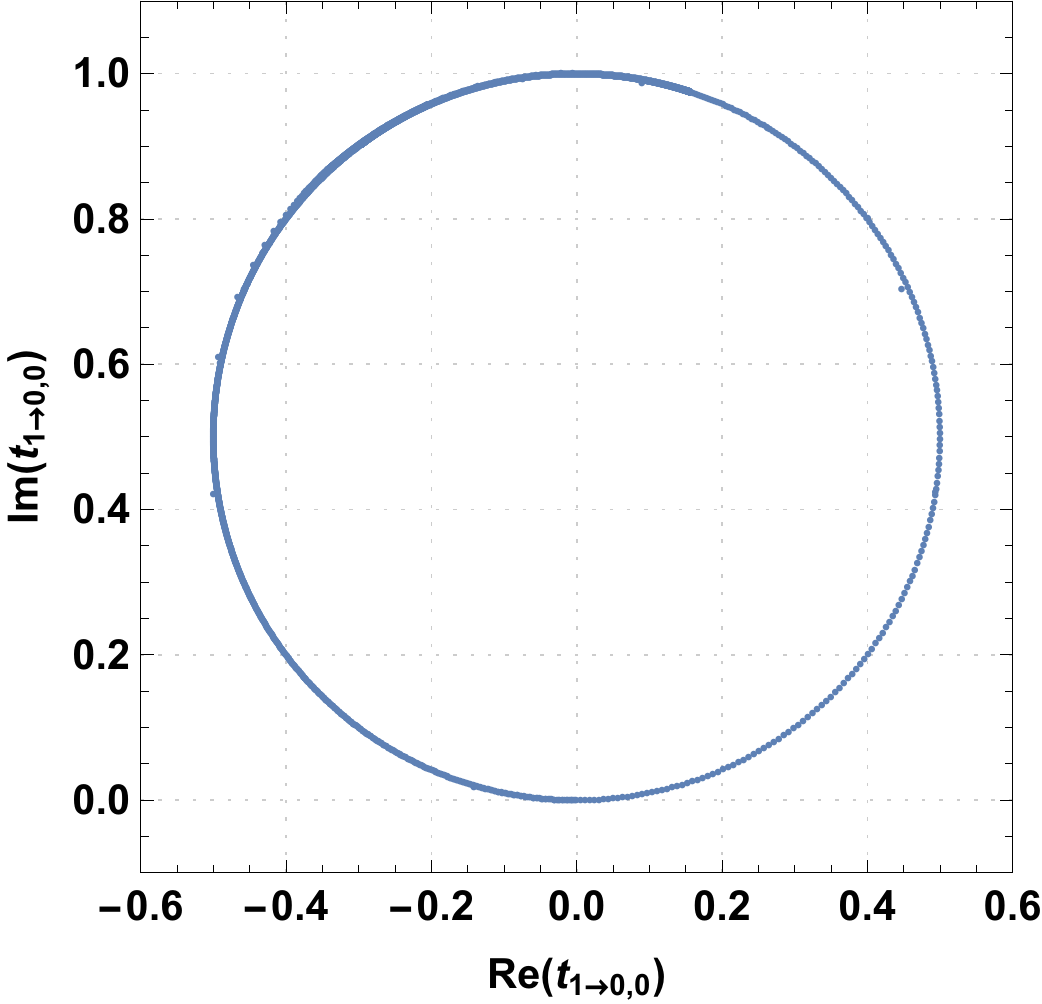}
\caption{\label{fig:argand}(Color online.) $t_{1 \rightarrow 0,0}$ in the complex plane for energies in the range $10.62 \, \textrm{GeV} \leq E \leq 11.6 \, \textrm{GeV}$, where the energy step is $1 \, \textrm{MeV}$.}
\end{figure}

Notice Eq.\ (\ref{eq:optical}) is equivalent to
\begin{eqnarray}
\nonumber & & \Big(\textrm{Re}(t_{1 \rightarrow 0,0})\Big)^2 + \Big(\textrm{Im}(t_{1 \rightarrow 0,0}) - 1/2\Big)^2 = \Big(1/2\Big)^2 , \\
 & &
\end{eqnarray}
the equation of circle in the complex plane centered at $i/2$ with radius $1/2$. In Fig.\ \ref{fig:argand} we show the corresponding Argand diagram. We get the expected circle, which is a good test that we are complying with the probability conservation expressed in the optical theorem as well as of the numerical precision of our results.

\begin{figure}[htb]
\includegraphics[width=\columnwidth]{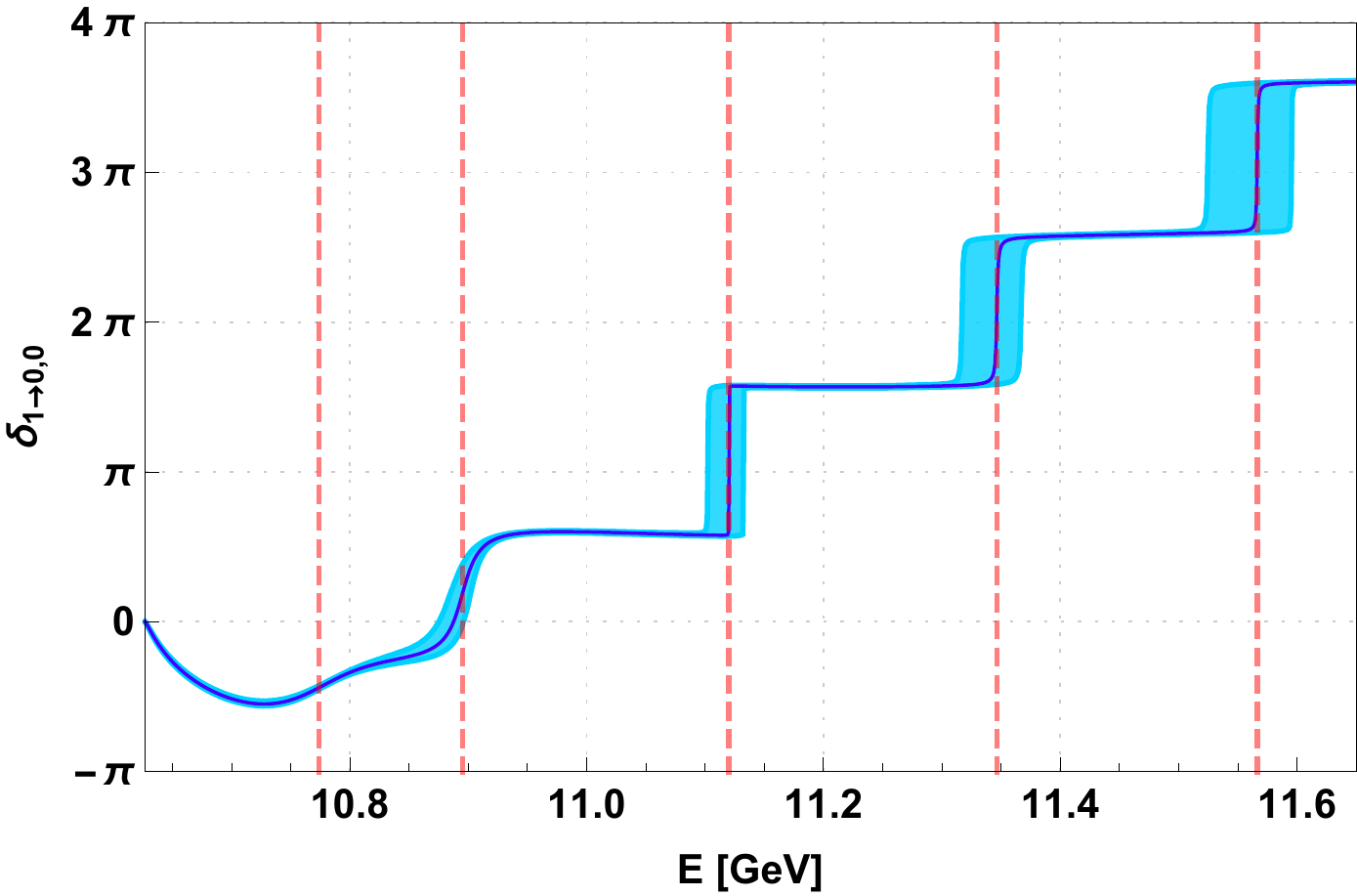}
\caption{\label{fig:phaseshift}(Color online.) Phase shift $\delta_{1 \rightarrow 0,0}$ as a function of the energy $E$. The real parts of the positions of the poles of $t_{1 \rightarrow 0,0}$ are indicated by red dashed lines.}
\end{figure}

In Fig.\ \ref{fig:phaseshift} we show the phase shift $\delta_{1 \rightarrow 0,0}$, which can be obtained from $t_{1 \rightarrow 0,0}$ via Eq.\ (\ref{EQN429}), as function of the energy. In this plot the resonances can be identified more clearly, since each of them corresponds to a ``jump'' of the order of $\pi$. The steepness of each jump is inversely proportional to the the corresponding decay widths. There are three clear resonances close to $11.125 \, \textrm{GeV}$, $11.350 \, \textrm{GeV}$ and $11.575 \, \textrm{GeV}$. Moreover, there seems to be another wider and less clear resonance around $10.900 \, \textrm{GeV}$.

\subsection{\label{SEC807}Poles of $t_{1 \rightarrow 0,0}$ in the plane of complex energies}

To determine resonance energies and decay widths precisely, we consider the analytic continuation of our scattering problem to the complex energy plane. There we search for the poles of $t_{1 \rightarrow 0,0}$ using the Newton-Raphson method as discussed in section~\ref{SEC489}. The positions of the poles $E$ are related to the resonance masses and decay widths according to Eq.\ (\ref{EQN521}), i.e.\ $m = \textrm{Re}(E)$ and $\Gamma = -2 \textrm{Im}(E)$.

\begin{figure*}[htb]
\includegraphics[width=2.0\columnwidth]{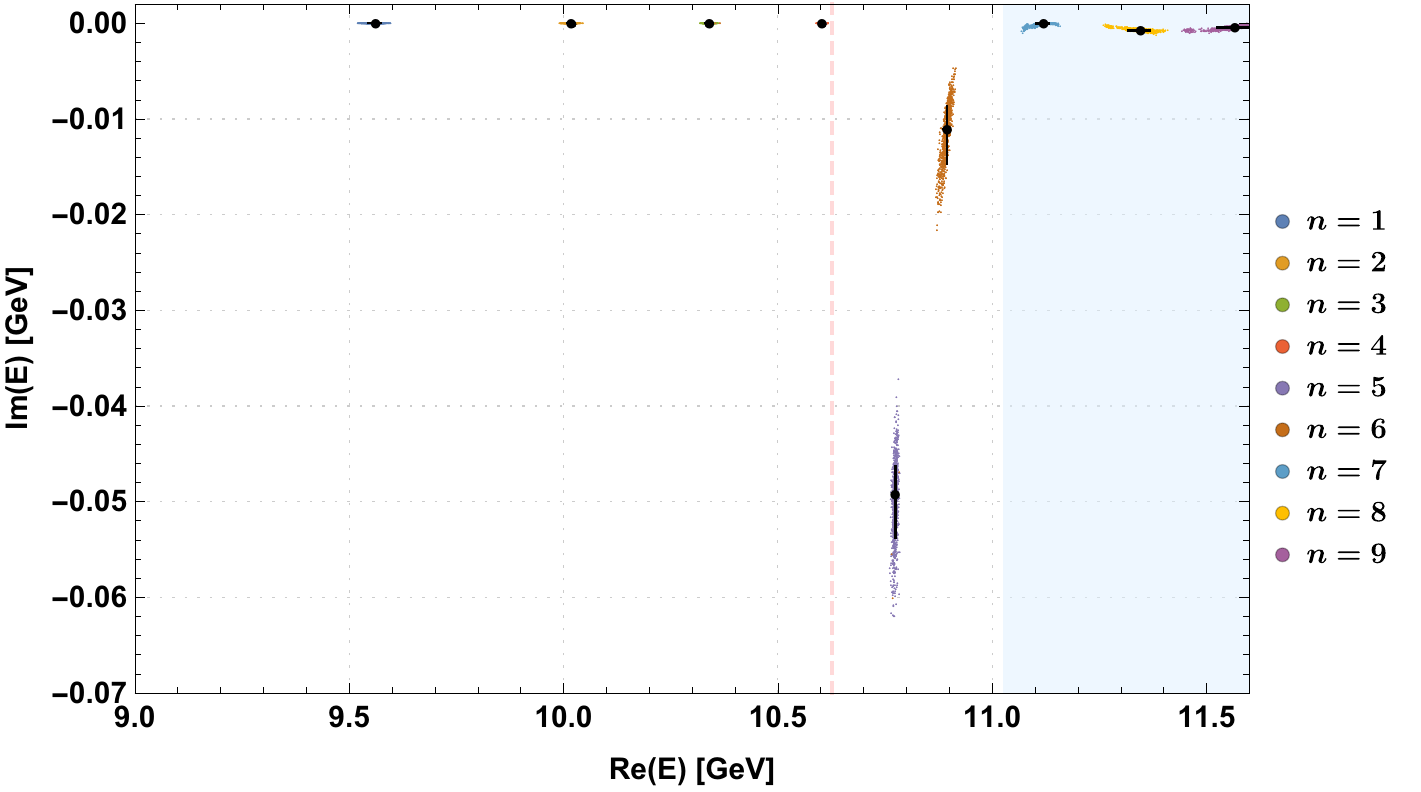}
\caption{\label{fig:errorpole}(Color online.) Positions of the poles of $t_{1 \rightarrow 0,0}$ in the complex energy plane for all bound states and resonances below $11.6 \, \textrm{GeV}$. Colored point clouds represent the 1000 resampled sets of parameters of the potentials, while black points and crosses represent the corresponding mean values and error bars. The vertical dashed line marks the spin-averaged $B^{(*)} \bar{B}^{(*)}$ threshold at $10.627 \, \textrm{GeV}$. The shaded region above $11.025 \, \textrm{GeV}$ marks the opening of the threshold of one heavy-light meson with negative parity and another with positive parity, beyond which our results should not be trusted anymore.}
\end{figure*}

In Fig.\ \ref{fig:errorpole} we show the positions of the poles of $t_{1 \rightarrow 0,0}$ in the complex energy plane for all bound states and resonances below $11.6 \, \textrm{GeV}$. For each bound state and each resonance there is a differently colored point cloud representing the 1000 resampled sets of parameters of the potentials, which we use to determine statistical errors (see section~\ref{SEC489} for details). For the bound states the poles are located on the real axis below the $B^{(*)} \bar{B}^{(*)}$ threshold. For the resonances the positions of the poles follow curved bands, where the imaginary parts range from almost vanishing values to finite values comparable to those observed in experiments (see Table~\ref{tab:bottomonium}), i.e.\ of the order of tens of MeV. There are clear gaps between the point clouds representing different bottomonium bound states and resonances, which allows a straightforward error analysis. The corresponding mean values and errors are indicated by the black circles and crosses. These results are also summarized in Table \ref{tab:polesemergent}, where we show in addition corresponding results from the single channel Schr\"odinger equation (\ref{eq:schro2}) with the ground state potential $V_0^{\Sigma_g^+}(r)$ as well as experimental results. 

\begin{table*}[htb]
\begin{ruledtabular}
\begin{tabular}{c|c|ccc|ccc}
\Tstrut
    & from $V_0^{\Sigma_g^+}(r)$ & \multicolumn{3}{c|}{from poles of $t_{1 \rightarrow 0,0}$} & \multicolumn{3}{c}{from experiment} \\
\Bstrut
$n$ & $E$ [GeV]                  & $m = \textrm{Re}(E)$ [GeV] & $\textrm{Im}(E)$ [MeV] & $\Gamma$ [MeV] & name & $m$ [GeV] & $\Gamma$ [MeV] \\
\hline
\Tstrut
$1$ & $\phantom{0}9.497_{-19}^{+13}$ &
 $\phantom{0}9.562_{-17}^{+11}$ &
 $0$ & -- & $\eta_b(1S)$ & $\phantom{0}9.399(2)$ & $10(5)\phantom{0}$ \\
    & & & & & $\Upsilon_b(1S)$ & $\phantom{0}9.460(0)$ & $\approx 0$ \\
$2$ & $\phantom{0}9.998_{-10}^{+7\phantom{0}}$ &
 $10.018_{-10}^{+8\phantom{0}}$ &
 $0$ & -- & $\Upsilon_b(2S)$ & $10.023(0)$ & $\approx 0$ \\
$3$ & $10.330_{-8\phantom{0}}^{+6\phantom{0}}$ &
 $10.340_{-9\phantom{0}}^{+7\phantom{0}}$ &
 $0$ & -- & $\Upsilon_b(3S)$ & $10.355(1)$ & $\approx 0$ \\
\Bstrut
$4$ & $10.591_{-5\phantom{0}}^{+4\phantom{0}}$ &
 $10.603_{-6\phantom{0}}^{+5\phantom{0}}$ &
 $0$ & -- & $\Upsilon_b(4S)$ & $10.579(1)$ & $21(3)\phantom{0}$ \\
\cline{2-8} \Tstrut
\Tstrut
$5$ & &
 $10.774_{-4\phantom{0}}^{+4\phantom{0}}$ &
 $-49.3_{-4.6}^{+3.0}$ &
 $98.5_{-5.9}^{+9.2}$ & & & \\
$6$ & &
 $10.895_{-10}^{+7\phantom{0}}$ &
 $-11.1_{-3.6}^{+2.4}$ &
 $22.2_{-4.9}^{+7.1}$ & $\Upsilon(10860)$ & $10.890(3)$ & $51(7)\phantom{0}$ \\
$7$ & &
 \cellcolor{lightgray}$11.120_{-18}^{+13}$ &
 \cellcolor{lightgray}$\phantom{0}$$-0.0_{-0.2}^{+0.0}$ &
 \cellcolor{lightgray}$\phantom{0}0.0_{-0.0}^{+0.4}$ & $\Upsilon(11020)$ & $10.993(1)$ & $49(15)$ \\
$8$ & &
 \cellcolor{lightgray}$11.347_{-30}^{+20}$ &
 \cellcolor{lightgray}$\phantom{0}$$-0.8_{-0.1}^{+0.2}$ &
 \cellcolor{lightgray}$\phantom{0}1.6_{-0.4}^{+0.2}$ & & & \\
\Bstrut
$9$ & &
 \cellcolor{lightgray}$11.567_{-42}^{+29}$ &
 \cellcolor{lightgray}$\phantom{0}$$-0.4_{-0.3}^{+0.2}$ &
 \cellcolor{lightgray}$\phantom{0}0.8_{-0.4}^{+0.5}$ & & & 
\end{tabular}
\end{ruledtabular}

\caption{\label{tab:polesemergent}Masses and decay widths for $\widetilde{J}^{P C} = 0^{+ +}$ bottomonium from the coupled channel Schr\"odinger equation (\ref{eq:coupledpot}) (column ``from poles of $t_{1 \rightarrow 0,0}$''). For comparison we also list corresponding single channel results (column ``from $V_0^{\Sigma_g^+}(r)$'') and experimental results (column ``from experiment''). Relevant $B^{(*)} \bar{B}^{(*)}$ thresholds are marked by horizontal lines. Errors on our theoretical results are purely statistical. Resonances with $n \geq 7$ are above the threshold of one heavy-light meson with negative parity and another with positive parity at around $11.025 \, \textrm{GeV}$ and, thus, should not be trusted anymore (indicated by gray shaded background).}
\end{table*}

We first remark that the positions of the poles of the four bound states (i.e.\ $n = 1,2,3,4$) are similar to the energies obtained from the single channel Schr\"odinger equation. This is hardly surprising, because the inclusion of a meson-meson channel is expected to have a sizable effect in particular for energies above the $B^{(*)} \bar{B}^{(*)}$ threshold. The positions of the poles of the four bound states are also reasonably close to the experimental results for the masses of $\eta_b(1S)$, $\Upsilon_b(1S)$, $\Upsilon(2S)$, $\Upsilon(3S)$ and $\Upsilon(4S)$.

We also find several resonances (i.e.\ $n \geq 5$), where the majority has not yet been observed experimentally. Our resonance mass for $n = 6$ is quite similar to the experimental result for $\Upsilon(10860)$, which might indicate that $\Upsilon(10860)$ should be interpreted as $\Upsilon(5S)$ (see also the recent paper \cite{Chen:2019uzm}, which supports this interpretation). For $\Upsilon(11020)$, on the other hand, there is no perfect match among our theoretical results. The closest resonance we find ($n = 7$) is almost $100 \, \textrm{MeV}$ heavier. An explanation could be that $\Upsilon(11020)$ is not a state with $\widetilde{J}^{P C} = 0^{+ +}$, but with higher $\widetilde{J}$. It will be interesting to explore this further in the future, by deriving and solving coupled channel Schr\"odiger equations also for $\widetilde{J} \geq 1$.

\begin{figure}[htb]
\includegraphics[width=\columnwidth]{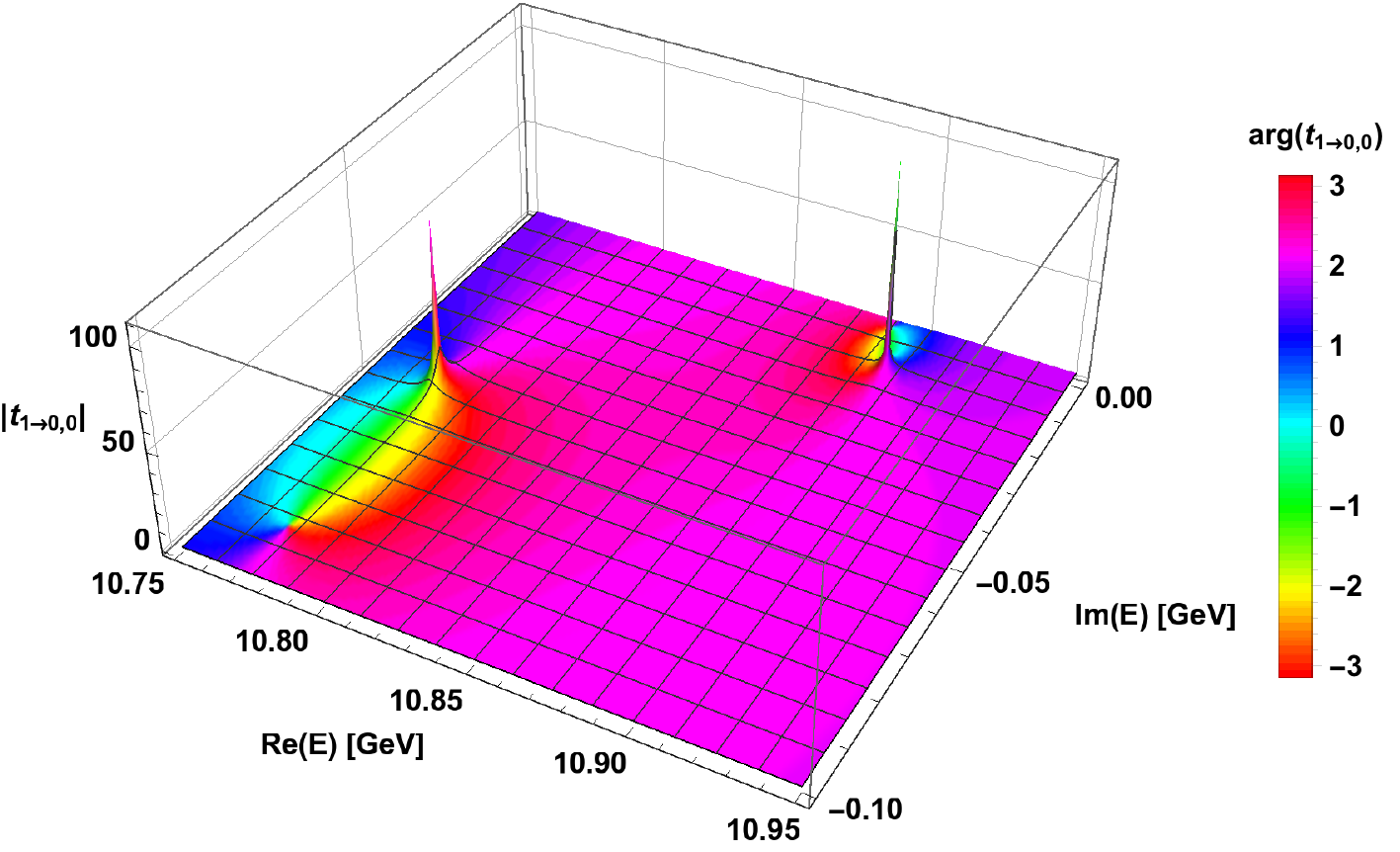}
\caption{\label{fig:pole3D}(Color online.) Visualization of the poles of $t_{1 \rightarrow 0,0}$ corresponding to the two lowest resonances ($n = 5$ and $n = 6$) in the complex energy plane.}
\end{figure}

Note that we also predict a resonance ($n = 5$) below the resonance masses of the two experimental $\widetilde{J}^{P C} = 0^{+ +}$ candidates $\Upsilon(10860)$ and $\Upsilon(11020)$, not far away from the $B^{(*)} \bar{B}^{(*)}$ threshold. This resonance and the resonance with $n = 6$, which is a candidate for $\Upsilon(10860)$, with masses close to  $10.8 \, \textrm{GeV}$ and $10.9 \, \textrm{GeV}$, respectively, are illustrated in Fig.\ \ref{fig:pole3D}, which is a 3D plot of the absolute value and the phase of $t_{1 \rightarrow 0,0}$ in the complex energy plane. As expected, the phase performs a full $2 \pi$ revolution around each of the corresponding poles. Note that a clear identification and separation of those two resonances is only possible from a pole analysis in the complex energy plane, but not from $\textrm{Re}(t_{1 \rightarrow 0,0})$, $\textrm{Im}(t_{1 \rightarrow 0,0})$ and $\delta_{1 \rightarrow 0,0}$ at real energies (see Fig.\ \ref{fig:realimagT} and Fig.\ \ref{fig:phaseshift}). For example, as discussed at the end of section~\ref{SEC806}, Fig.\ \ref{fig:phaseshift} suggests that there might be a wide resonance around $10.9 \, \textrm{GeV}$, but it is almost impossible to see from that figure that there are actually two resonances in that energy region.

\begin{figure*}[htb]
\includegraphics[width=1.0\columnwidth]{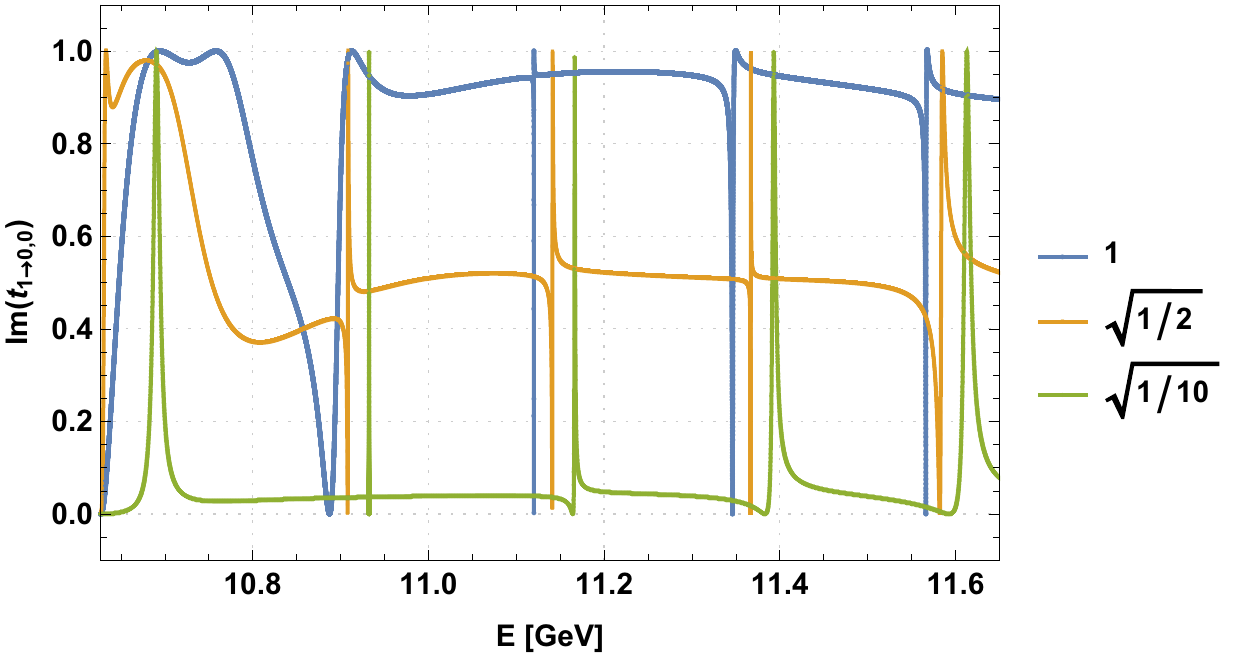}
\includegraphics[width=1.0\columnwidth]{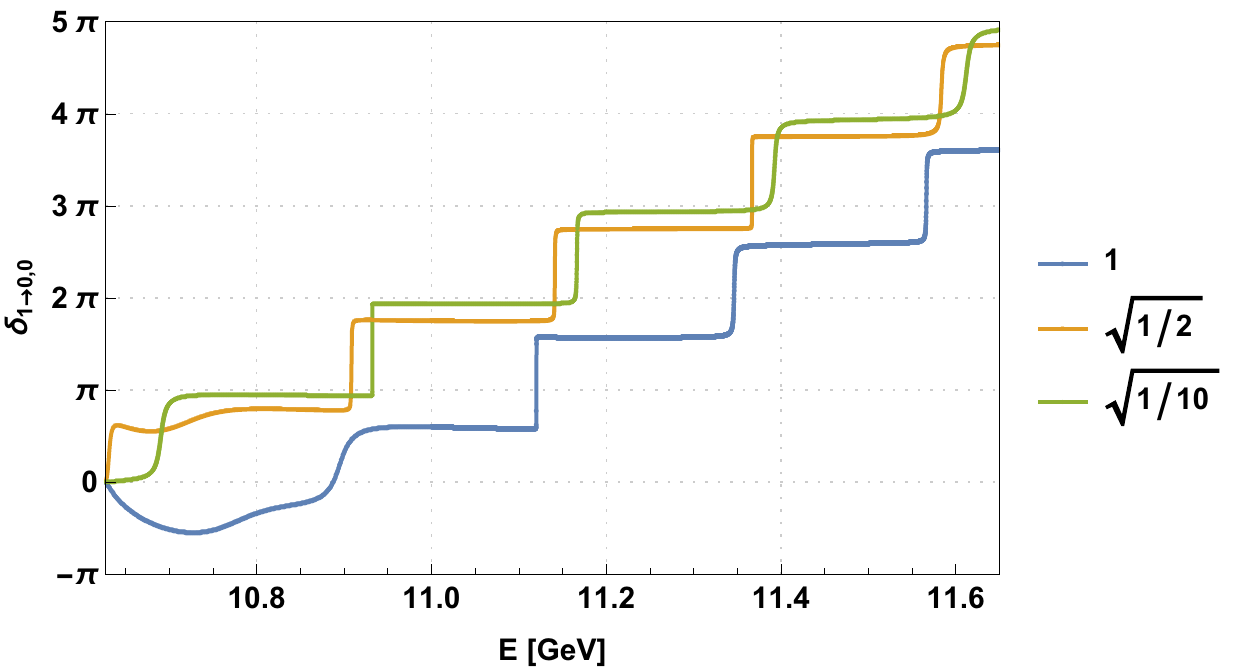}
\caption{\label{fig:magfac}(Color online.) $\textrm{Im}(t_{1 \rightarrow 0,0})$ and phase shift $\delta_{1 \rightarrow 0,0}$ as functions of the energy $E$ for artificially reduced mixing potentials $c V_\text{mix}(r)$, $c \in \{ 1/\sqrt{10} , 1/\sqrt{2} , 1 \}$.}
\end{figure*}

A closer inspection reveals that the lower of the two resonances ($n = 5$) has a fully dynamical origin. This can be seen from Fig.\ \ref{fig:magfac}, where we study both $\textrm{Im}(t_{1 \rightarrow 0,0})$ and $\delta_{1 \rightarrow 0,0}$ for mixing potentials $c V_\text{mix}(r)$ with $c \in \{ 1/\sqrt{10} , 1/\sqrt{2} , 1 \}$. In particular in the left plot, where we show $\textrm{Im}(t_{1 \rightarrow 0,0})$ as a function of real energy $E$, one can see that the peak at around $10.9 \, \textrm{GeV}$ becomes more pronounced for decreasing $c$. At the same time the very wide peak below $10.8 \, \textrm{GeV}$ also transforms into a sharp and clear peak and moves to significantly smaller energies slightly above the $B^{(*)} \bar{B}^{(*)}$ threshold. At small mixing, $c = 1 / \sqrt{10}$, however, there are only three bound states, not four as for $c = 1$. In other words, when decreasing the mixing potential, the bound state with $n = 4$ becomes a clear resonance close to the threshold, while the wide resonance with $n = 5$ disappears.

In what concerns the imaginary parts of the positions of the poles, we obtain below the $B^{(*)} \bar{B}^{(*)}$ threshold exactly zero as expected. Above the $B^{(*)} \bar{B}^{(*)}$ threshold only the lightest resonances ($n = 5$ and $n = 6$) have decay widths similar to the experimental $\widetilde{J}^{P C} = 0^{+ +}$ candidates $\Upsilon(10860)$ and $\Upsilon(11020)$. All higher resonances ($n \geq 7$) have significantly smaller widths. The reason for this is most likely that we consider only the coupling of quarkonium to the lightest meson-meson channel. To obtain larger widths, we would need to include all excited meson-meson channels up to the respective resonance masses. Clearly, this goes beyond the scope of the present work.

We expect only a weak dependence of our results on the $b$ quark mass $m_Q$ (see e.g.\ Refs.\ \cite{Bicudo:2015kna,Karbstein:2018mzo}). We use for simplicity $m_Q = 4.977 \, \textrm{GeV}$ from quark models \cite{Godfrey:1985xj}, but have tested explicitly the dependence of our results on $m_Q$. Qualitatively nothing changes, when we change the heavy quark mass, i.e.\ we obtain the same number and pattern of bound states and resonances. Even when varying the quark mass drastically by a few hundred MeV, the meson spectrum is shifted, but mass differences remain almost constant, i.e.\ the meson masses are strongly correlated with respect to such changes. On a quantitative level, a variation of the quark mass by e.g.\ $\pm 50 \, \textrm{MeV}$, which is the order of the error in the determination of the running mass of the $b$ quark \cite{Patrignani:2016xqp}, changes the meson masses masses by around $\pm 4 \textrm{MeV}$. The widths of the wide resonances with $n = 5$ and $n = 6$ are anti-correlated and change by around $\pm 10 \, \textrm{MeV}$. Note that the $b$ quark mass $m_Q$ could also be tuned to optimally reproduce the experimentally observed results collected in Table~\ref{tab:bottomonium}. We plan to do this in a future publication, where we will apply our approach not only to $\widetilde{J} = 0$, but also to $\widetilde{J} = 1$ and $\widetilde{J} = 2$.

There is also a certain uncertainty associated with the parameterization of the potentials $V_{\bar{Q} Q}(r)$, $V_{\bar{M} M,\parallel}(r)$ and $V_{\textrm{mix}}(r)$, in particular with the choice of $r_\textrm{min}$ (see the discussion below Eq.\ (\ref{eq:fitMix})). When varying $r_\textrm{min} / a$ between $1.442$ and $2.836$, bottomonium masses change by around $25 \, \textrm{MeV}$, while decay widths change by around $10 \, \textrm{MeV}$. While this is still significantly below the total systematic error we estimate below, it is clear that a more precise lattice QCD computation of the potentials at small $r$ would be very worthwhile.

We stress that the errors on our theoretical results quoted in Table \ref{tab:polesemergent} and shown in the figures are purely statistical. These results, however, were obtained by resorting to certain approximations. First of all, the coupled channel Schr\"odinger equation (\ref{EQN050}) was derived  in the static limit (see the discussion in section~\ref{SEC466}), while $b$ quarks have a large, but finite mass. Similarly, we use lattice QCD potentials $V_0^{\Sigma_g^+}(r)$ and $V_1^{\Sigma_g^+}(r)$ and a mixing angle $\theta(r)$ computed in the static limit, at unphysically heavy $u$ and $d$ quark mass corresponding to $m_\pi \approx 654 \, \textrm{MeV}$ and at a single lattice spacing $a \approx 1 / (2.37 \, \textrm{GeV}) \approx 0.083 \, \textrm{fm}$, which also introduces systematic errors. Moreover, we assumed $V_0^{\Pi_g}(r) = 0$. As already mentioned, our coupled channel Schr\"odinger equation contains only the lightest decay channel to two negative parity heavy-light mesons, while the next ones containing a negative and a positive parity meson are around $400 \, \textrm{MeV} \ldots 500 \, \textrm{MeV}$ above (see Eqs.\ (\ref{EQN031b}) and (\ref{EQN031c})). Thus we expect that the neglect of this channel has little effect on our results up to the corresponding threshold around $11.025 \, \textrm{GeV} \ldots 11.125 \, \textrm{GeV}$, i.e.\ for $n \leq 6$, while higher resonances with $n \geq 7$ might be strongly affected. Finally we separated the treatment of heavy and light degrees of freedom using the Born-Oppenheimer approximation. For the obtained masses with $n \leq 6$ we crudely estimate systematic errors to be around $50 \, \textrm{MeV}$, which is the order of the discrepancy of experimental results and our theoretical predictions for the four states $\Upsilon(1S)$, $\Upsilon(2S)$, $\Upsilon(3S)$ and $\Upsilon(4S)$.


\section{\label{sec:conclu}Conclusions}

We proposed and derived a formalism to study quarkonium bound states and resonances with $I = 0$ based on static potentials from QCD, which can be computed with lattice QCD. We applied the Born-Oppenheimer approximation by inserting these potentials in a specifically derived coupled channel Schrödinger equation for the dynamics of heavy quarks. This equation, which contains a quarkonium and a heavy-light meson meson channel, allows to predict masses of bound states and resonances as well as decay widths. Since we only consider decays to a pair of heavy-light mesons, we are neglecting the OZI suppressed decay channels, which have rather small partial decay widths, as shown by the experimental results collected in Table~\ref{tab:bottomonium}. For the resonances we apply scattering theory, which enables to compute phase shifts and eigenvalues of the $\mbox{T}$ matrix in the complex energy plane.

Within our framework we studied bottomonium states with $I = 0$ up to $11.6 \, \textrm{GeV}$ focusing on the $\widetilde{J}^{P C} = 0^{+ +}$ channel, which corresponds to $L_{\bar{Q} Q} = 0$ for the $\bar{b} b$ pair in the quarkonium channel and $L_{\bar{M} M} = 1$ for the $\bar{B}^{(\ast)} B^{(\ast)}$ pair in the meson-meson channel. Even though we resort to several approximations (see the discussion at the end of section \ref{SEC807}), we find reasonable agreement with the experimentally observed bottomonium spectrum. There are four bound states, which can clearly be identified with $\eta_b \equiv \Upsilon(1S)$, $\Upsilon(2S)$, $\Upsilon(3S)$, $\Upsilon(4S)$. We also obtain a resonance around $10.870 \, \textrm{GeV}$, which matches $\Upsilon(10860)$ rather well, suggesting that $\Upsilon(10860)$ could be interpreted as $\Upsilon(5S)$. For $\Upsilon(11020)$, on the other hand, we do not find a close-by resonance, which might be an indication that $\Upsilon(11020)$ is not an $S$ wave state. Moreover, we predict a new, dynamically generated resonance close the the $\bar{B}^{(\ast)} B^{(\ast)}$ threshold with mass $\approx 10.774 \, \textrm{GeV}$ and decay width $\approx 99 \, \textrm{MeV}$.

A straightforward next step will be to study bottomonium with $\widetilde{J} \geq 1$. The corresponding coupled channel Schr\"odiger equations will have at least a $3 \times 3$ matrix structure. For instance $\widetilde{J}^{P C} = 1^{- -}$ corresponds to $L_{\bar{Q} Q} = 1$ for the $\bar{b} b$ pair in the quarkonium channel and $L_{\bar{M} M} = 0$ or $L_{\bar{M} M} = 2$ for the $\bar{B}^{(\ast)} B^{(\ast)}$ pair in the meson-meson channels. Thus, one can study a possibly existing $X_b$ meson, the counterpart of the famous $X_c(3872)$ \cite{Choi:2003ue}.

Another direction for the future could be to include the decay channels to a negative and a positive parity heavy-light meson. This would allow to make more realistic predictions for resonances with $n \geq 7$ up to the threshold of two positive  parity heavy-light mesons at around $11.525 \, \textrm{GeV}$. The corresponding static potentials, however, have not yet been computed with lattice QCD. Moreover, our current determination of the potentials $V_{\bar{Q} Q}(r)$, $V_{\bar{M} M,\parallel}(r)$, $V_{\bar{M} M,\perp}(r)$ and $V_{\textrm{mix}}(r)$ from the lattice QCD results of Ref.\ \cite{Bali:2005fu} requires certain assumptions. Thus, we plan to perform a dedicated lattice QCD computation of all those static potentials, possibly also with $u$ and $d$ quark mass closer to the physical value.

Finally it would be worthwhile to include the effects of the heavy spins, either on the level of the coupled channel Schr\"odinger equation as in Ref.\ \cite{Bicudo:2016ooe} or even in a direct way, by computing $1/m_Q$ and $1/m_Q^2$ corrections to the static potentials using effective field theories like pNRQCD and lattice QCD (see e.g.\ Refs.\ \cite{Bali:1997am,Brambilla:2000gk,Pineda:2000sz,Brambilla:2004jw,Koma:2006si,Koma:2012bc,Brambilla:2018pyn,Brambilla:2019jfi}).


\begin{acknowledgements}

We acknowledge useful conversations with Gunnar Bali.

We thank the referee at Physical Review D for several helpful comments and suggestions, in particular concerning the dependence of our results on $r_\textrm{min}$.

P.B.\ and N.C.\ acknowledge the support of CeFEMA\mbox{} under the FCT contract for R\&D Units UID/CTM/04540/2013, and the FCT project grant CERN/FIS-COM/0029/2017. N.C.\ acknowledges the FCT contract SFRH/BPD/109443/2015. M.W.\ acknowledges funding by the Heisenberg Programme of the Deutsche Forschungsgemeinschaft (DFG, German Research Foundation) -- Projektnummer 399217702.

P.B.\ and M.W.\ are grateful to the Mainz Institute for Theoretical Physics (MITP) for its hospitality and its partial support during the completion of this work.

This work was supported in part by the Helmholtz International Center for FAIR within the framework of the LOEWE program launched by the State of Hesse.

Calculations on GPU servers of PtQCD partly supported by NVIDIA were conducted for this research. Calculations on the Goethe-HLR and on the on the FUCHS-CSC high-performance computer of the Frankfurt University were conducted for this research. We would like to thank HPC-Hessen, funded by the State Ministry of Higher Education, Research and the Arts, for programming advice.

\end{acknowledgements}


\bibliographystyle{apsrev4-1}
\bibliography{literature.bib}

\begin{thebibliography}{65}%
\makeatletter
\providecommand \@ifxundefined [1]{%
 \@ifx{#1\undefined}
}%
\providecommand \@ifnum [1]{%
 \ifnum #1\expandafter \@firstoftwo
 \else \expandafter \@secondoftwo
 \fi
}%
\providecommand \@ifx [1]{%
 \ifx #1\expandafter \@firstoftwo
 \else \expandafter \@secondoftwo
 \fi
}%
\providecommand \natexlab [1]{#1}%
\providecommand \enquote  [1]{``#1''}%
\providecommand \bibnamefont  [1]{#1}%
\providecommand \bibfnamefont [1]{#1}%
\providecommand \citenamefont [1]{#1}%
\providecommand \href@noop [0]{\@secondoftwo}%
\providecommand \href [0]{\begingroup \@sanitize@url \@href}%
\providecommand \@href[1]{\@@startlink{#1}\@@href}%
\providecommand \@@href[1]{\endgroup#1\@@endlink}%
\providecommand \@sanitize@url [0]{\catcode `\\12\catcode `\$12\catcode
  `\&12\catcode `\#12\catcode `\^12\catcode `\_12\catcode `\%12\relax}%
\providecommand \@@startlink[1]{}%
\providecommand \@@endlink[0]{}%
\providecommand \url  [0]{\begingroup\@sanitize@url \@url }%
\providecommand \@url [1]{\endgroup\@href {#1}{\urlprefix }}%
\providecommand \urlprefix  [0]{URL }%
\providecommand \Eprint [0]{\href }%
\providecommand \doibase [0]{http://dx.doi.org/}%
\providecommand \selectlanguage [0]{\@gobble}%
\providecommand \bibinfo  [0]{\@secondoftwo}%
\providecommand \bibfield  [0]{\@secondoftwo}%
\providecommand \translation [1]{[#1]}%
\providecommand \BibitemOpen [0]{}%
\providecommand \bibitemStop [0]{}%
\providecommand \bibitemNoStop [0]{.\EOS\space}%
\providecommand \EOS [0]{\spacefactor3000\relax}%
\providecommand \BibitemShut  [1]{\csname bibitem#1\endcsname}%
\let\auto@bib@innerbib\@empty
\bibitem [{\citenamefont {Jaffe}(1977)}]{Jaffe:1976ig}%
  \BibitemOpen
  \bibfield  {author} {\bibinfo {author} {\bibfnamefont {R.~L.}\ \bibnamefont
  {Jaffe}},\ }\href {\doibase 10.1103/PhysRevD.15.267} {\bibfield  {journal}
  {\bibinfo  {journal} {Phys. Rev.}\ }\textbf {\bibinfo {volume} {D15}},\
  \bibinfo {pages} {267} (\bibinfo {year} {1977})}\BibitemShut {NoStop}%
\bibitem [{\citenamefont {Bicudo}\ and\ \citenamefont
  {Cardoso}(2016)}]{Bicudo:2015bra}%
  \BibitemOpen
  \bibfield  {author} {\bibinfo {author} {\bibfnamefont {P.}~\bibnamefont
  {Bicudo}}\ and\ \bibinfo {author} {\bibfnamefont {M.}~\bibnamefont
  {Cardoso}},\ }\href {\doibase 10.1103/PhysRevD.94.094032} {\bibfield
  {journal} {\bibinfo  {journal} {Phys. Rev.}\ }\textbf {\bibinfo {volume}
  {D94}},\ \bibinfo {pages} {094032} (\bibinfo {year} {2016})},\ \Eprint
  {http://arxiv.org/abs/1509.04943} {arXiv:1509.04943 [hep-ph]} \BibitemShut
  {NoStop}%
\bibitem [{\citenamefont {Born}\ and\ \citenamefont
  {Oppenheimer}(1927)}]{Born:1927}%
  \BibitemOpen
  \bibfield  {author} {\bibinfo {author} {\bibfnamefont {M.}~\bibnamefont
  {Born}}\ and\ \bibinfo {author} {\bibfnamefont {R.}~\bibnamefont
  {Oppenheimer}},\ }\href@noop {} {\bibfield  {journal} {\bibinfo  {journal}
  {Annalen der Physik}\ }\textbf {\bibinfo {volume} {389}},\ \bibinfo {pages}
  {457} (\bibinfo {year} {1927})}\BibitemShut {NoStop}%
\bibitem [{\citenamefont {Juge}\ \emph {et~al.}(1999)\citenamefont {Juge},
  \citenamefont {Kuti},\ and\ \citenamefont {Morningstar}}]{Juge:1999ie}%
  \BibitemOpen
  \bibfield  {author} {\bibinfo {author} {\bibfnamefont {K.~J.}\ \bibnamefont
  {Juge}}, \bibinfo {author} {\bibfnamefont {J.}~\bibnamefont {Kuti}}, \ and\
  \bibinfo {author} {\bibfnamefont {C.~J.}\ \bibnamefont {Morningstar}},\
  }\href {\doibase 10.1103/PhysRevLett.82.4400} {\bibfield  {journal} {\bibinfo
   {journal} {Phys. Rev. Lett.}\ }\textbf {\bibinfo {volume} {82}},\ \bibinfo
  {pages} {4400} (\bibinfo {year} {1999})},\ \Eprint
  {http://arxiv.org/abs/hep-ph/9902336} {arXiv:hep-ph/9902336 [hep-ph]}
  \BibitemShut {NoStop}%
\bibitem [{\citenamefont {Braaten}\ \emph {et~al.}(2014)\citenamefont
  {Braaten}, \citenamefont {Langmack},\ and\ \citenamefont
  {Smith}}]{Braaten:2014qka}%
  \BibitemOpen
  \bibfield  {author} {\bibinfo {author} {\bibfnamefont {E.}~\bibnamefont
  {Braaten}}, \bibinfo {author} {\bibfnamefont {C.}~\bibnamefont {Langmack}}, \
  and\ \bibinfo {author} {\bibfnamefont {D.~H.}\ \bibnamefont {Smith}},\ }\href
  {\doibase 10.1103/PhysRevD.90.014044} {\bibfield  {journal} {\bibinfo
  {journal} {Phys. Rev.}\ }\textbf {\bibinfo {volume} {D90}},\ \bibinfo {pages}
  {014044} (\bibinfo {year} {2014})},\ \Eprint {http://arxiv.org/abs/1402.0438}
  {arXiv:1402.0438 [hep-ph]} \BibitemShut {NoStop}%
\bibitem [{\citenamefont {Berwein}\ \emph {et~al.}(2015)\citenamefont
  {Berwein}, \citenamefont {Brambilla}, \citenamefont {Tarrus~Castella},\ and\
  \citenamefont {Vairo}}]{Berwein:2015vca}%
  \BibitemOpen
  \bibfield  {author} {\bibinfo {author} {\bibfnamefont {M.}~\bibnamefont
  {Berwein}}, \bibinfo {author} {\bibfnamefont {N.}~\bibnamefont {Brambilla}},
  \bibinfo {author} {\bibfnamefont {J.}~\bibnamefont {Tarrus~Castella}}, \ and\
  \bibinfo {author} {\bibfnamefont {A.}~\bibnamefont {Vairo}},\ }\href
  {\doibase 10.1103/PhysRevD.92.114019} {\bibfield  {journal} {\bibinfo
  {journal} {Phys. Rev.}\ }\textbf {\bibinfo {volume} {D92}},\ \bibinfo {pages}
  {114019} (\bibinfo {year} {2015})},\ \Eprint
  {http://arxiv.org/abs/1510.04299} {arXiv:1510.04299 [hep-ph]} \BibitemShut
  {NoStop}%
\bibitem [{\citenamefont {Brambilla}\ \emph {et~al.}(2018)\citenamefont
  {Brambilla}, \citenamefont {Krein}, \citenamefont {Tarrus~Castella},\ and\
  \citenamefont {Vairo}}]{Brambilla:2017uyf}%
  \BibitemOpen
  \bibfield  {author} {\bibinfo {author} {\bibfnamefont {N.}~\bibnamefont
  {Brambilla}}, \bibinfo {author} {\bibfnamefont {G.}~\bibnamefont {Krein}},
  \bibinfo {author} {\bibfnamefont {J.}~\bibnamefont {Tarrus~Castella}}, \ and\
  \bibinfo {author} {\bibfnamefont {A.}~\bibnamefont {Vairo}},\ }\href
  {\doibase 10.1103/PhysRevD.97.016016} {\bibfield  {journal} {\bibinfo
  {journal} {Phys. Rev.}\ }\textbf {\bibinfo {volume} {D97}},\ \bibinfo {pages}
  {016016} (\bibinfo {year} {2018})},\ \Eprint
  {http://arxiv.org/abs/1707.09647} {arXiv:1707.09647 [hep-ph]} \BibitemShut
  {NoStop}%
\bibitem [{\citenamefont {Karbstein}\ \emph {et~al.}(2018)\citenamefont
  {Karbstein}, \citenamefont {Wagner},\ and\ \citenamefont
  {Weber}}]{Karbstein:2018mzo}%
  \BibitemOpen
  \bibfield  {author} {\bibinfo {author} {\bibfnamefont {F.}~\bibnamefont
  {Karbstein}}, \bibinfo {author} {\bibfnamefont {M.}~\bibnamefont {Wagner}}, \
  and\ \bibinfo {author} {\bibfnamefont {M.}~\bibnamefont {Weber}},\ }\href
  {\doibase 10.1103/PhysRevD.98.114506} {\bibfield  {journal} {\bibinfo
  {journal} {Phys. Rev.}\ }\textbf {\bibinfo {volume} {D98}},\ \bibinfo {pages}
  {114506} (\bibinfo {year} {2018})},\ \Eprint
  {http://arxiv.org/abs/1804.10909} {arXiv:1804.10909 [hep-ph]} \BibitemShut
  {NoStop}%
\bibitem [{\citenamefont {Capitani}\ \emph {et~al.}(2019)\citenamefont
  {Capitani}, \citenamefont {Philipsen}, \citenamefont {Reisinger},
  \citenamefont {Riehl},\ and\ \citenamefont {Wagner}}]{Capitani:2018rox}%
  \BibitemOpen
  \bibfield  {author} {\bibinfo {author} {\bibfnamefont {S.}~\bibnamefont
  {Capitani}}, \bibinfo {author} {\bibfnamefont {O.}~\bibnamefont {Philipsen}},
  \bibinfo {author} {\bibfnamefont {C.}~\bibnamefont {Reisinger}}, \bibinfo
  {author} {\bibfnamefont {C.}~\bibnamefont {Riehl}}, \ and\ \bibinfo {author}
  {\bibfnamefont {M.}~\bibnamefont {Wagner}},\ }\href {\doibase
  10.1103/PhysRevD.99.034502} {\bibfield  {journal} {\bibinfo  {journal} {Phys.
  Rev.}\ }\textbf {\bibinfo {volume} {D99}},\ \bibinfo {pages} {034502}
  (\bibinfo {year} {2019})},\ \Eprint {http://arxiv.org/abs/1811.11046}
  {arXiv:1811.11046 [hep-lat]} \BibitemShut {NoStop}%
\bibitem [{\citenamefont {Detmold}\ \emph {et~al.}(2007)\citenamefont
  {Detmold}, \citenamefont {Orginos},\ and\ \citenamefont
  {Savage}}]{Detmold:2007wk}%
  \BibitemOpen
  \bibfield  {author} {\bibinfo {author} {\bibfnamefont {W.}~\bibnamefont
  {Detmold}}, \bibinfo {author} {\bibfnamefont {K.}~\bibnamefont {Orginos}}, \
  and\ \bibinfo {author} {\bibfnamefont {M.~J.}\ \bibnamefont {Savage}},\
  }\href {\doibase 10.1103/PhysRevD.76.114503} {\bibfield  {journal} {\bibinfo
  {journal} {Phys. Rev.}\ }\textbf {\bibinfo {volume} {D76}},\ \bibinfo {pages}
  {114503} (\bibinfo {year} {2007})},\ \Eprint
  {http://arxiv.org/abs/hep-lat/0703009} {arXiv:hep-lat/0703009 [hep-lat]}
  \BibitemShut {NoStop}%
\bibitem [{\citenamefont {Wagner}(2010)}]{Wagner:2010ad}%
  \BibitemOpen
  \bibfield  {author} {\bibinfo {author} {\bibfnamefont {M.}~\bibnamefont
  {Wagner}} (\bibinfo {collaboration} {ETM}),\ }\href@noop {} {\bibfield
  {journal} {\bibinfo  {journal} {PoS}\ }\textbf {\bibinfo {volume}
  {LATTICE2010}},\ \bibinfo {pages} {162} (\bibinfo {year} {2010})},\ \Eprint
  {http://arxiv.org/abs/1008.1538} {arXiv:1008.1538 [hep-lat]} \BibitemShut
  {NoStop}%
\bibitem [{\citenamefont {Bali}\ and\ \citenamefont
  {Hetzenegger}(2010)}]{Bali:2010xa}%
  \BibitemOpen
  \bibfield  {author} {\bibinfo {author} {\bibfnamefont {G.}~\bibnamefont
  {Bali}}\ and\ \bibinfo {author} {\bibfnamefont {M.}~\bibnamefont
  {Hetzenegger}} (\bibinfo {collaboration} {QCDSF}),\ }\href@noop {} {\bibfield
   {journal} {\bibinfo  {journal} {PoS}\ }\textbf {\bibinfo {volume}
  {LATTICE2010}},\ \bibinfo {pages} {142} (\bibinfo {year} {2010})},\ \Eprint
  {http://arxiv.org/abs/1011.0571} {arXiv:1011.0571 [hep-lat]} \BibitemShut
  {NoStop}%
\bibitem [{\citenamefont {Wagner}(2011)}]{Wagner:2011ev}%
  \BibitemOpen
  \bibfield  {author} {\bibinfo {author} {\bibfnamefont {M.}~\bibnamefont
  {Wagner}} (\bibinfo {collaboration} {ETM}),\ }\href {\doibase
  10.5506/APhysPolBSupp.4.747} {\bibfield  {journal} {\bibinfo  {journal} {Acta
  Phys. Polon. Supp.}\ }\textbf {\bibinfo {volume} {4}},\ \bibinfo {pages}
  {747} (\bibinfo {year} {2011})},\ \Eprint {http://arxiv.org/abs/1103.5147}
  {arXiv:1103.5147 [hep-lat]} \BibitemShut {NoStop}%
\bibitem [{\citenamefont {Brown}\ and\ \citenamefont
  {Orginos}(2012)}]{Brown:2012tm}%
  \BibitemOpen
  \bibfield  {author} {\bibinfo {author} {\bibfnamefont {Z.~S.}\ \bibnamefont
  {Brown}}\ and\ \bibinfo {author} {\bibfnamefont {K.}~\bibnamefont
  {Orginos}},\ }\href {\doibase 10.1103/PhysRevD.86.114506} {\bibfield
  {journal} {\bibinfo  {journal} {Phys. Rev.}\ }\textbf {\bibinfo {volume}
  {D86}},\ \bibinfo {pages} {114506} (\bibinfo {year} {2012})},\ \Eprint
  {http://arxiv.org/abs/1210.1953} {arXiv:1210.1953 [hep-lat]} \BibitemShut
  {NoStop}%
\bibitem [{\citenamefont {Bicudo}\ \emph {et~al.}(2016)\citenamefont {Bicudo},
  \citenamefont {Cichy}, \citenamefont {Peters},\ and\ \citenamefont
  {Wagner}}]{Bicudo:2015kna}%
  \BibitemOpen
  \bibfield  {author} {\bibinfo {author} {\bibfnamefont {P.}~\bibnamefont
  {Bicudo}}, \bibinfo {author} {\bibfnamefont {K.}~\bibnamefont {Cichy}},
  \bibinfo {author} {\bibfnamefont {A.}~\bibnamefont {Peters}}, \ and\ \bibinfo
  {author} {\bibfnamefont {M.}~\bibnamefont {Wagner}},\ }\href {\doibase
  10.1103/PhysRevD.93.034501} {\bibfield  {journal} {\bibinfo  {journal} {Phys.
  Rev.}\ }\textbf {\bibinfo {volume} {D93}},\ \bibinfo {pages} {034501}
  (\bibinfo {year} {2016})},\ \Eprint {http://arxiv.org/abs/1510.03441}
  {arXiv:1510.03441 [hep-lat]} \BibitemShut {NoStop}%
\bibitem [{\citenamefont {Ader}\ \emph {et~al.}(1982)\citenamefont {Ader},
  \citenamefont {Richard},\ and\ \citenamefont {Taxil}}]{Ader:1981db}%
  \BibitemOpen
  \bibfield  {author} {\bibinfo {author} {\bibfnamefont {J.~P.}\ \bibnamefont
  {Ader}}, \bibinfo {author} {\bibfnamefont {J.~M.}\ \bibnamefont {Richard}}, \
  and\ \bibinfo {author} {\bibfnamefont {P.}~\bibnamefont {Taxil}},\ }\href
  {\doibase 10.1103/PhysRevD.25.2370} {\bibfield  {journal} {\bibinfo
  {journal} {Phys. Rev.}\ }\textbf {\bibinfo {volume} {D25}},\ \bibinfo {pages}
  {2370} (\bibinfo {year} {1982})}\BibitemShut {NoStop}%
\bibitem [{\citenamefont {Ballot}\ and\ \citenamefont
  {Richard}(1983)}]{Ballot:1983iv}%
  \BibitemOpen
  \bibfield  {author} {\bibinfo {author} {\bibfnamefont {J.~l.}\ \bibnamefont
  {Ballot}}\ and\ \bibinfo {author} {\bibfnamefont {J.~M.}\ \bibnamefont
  {Richard}},\ }\href {\doibase 10.1016/0370-2693(83)90991-7} {\bibfield
  {journal} {\bibinfo  {journal} {Phys. Lett.}\ }\textbf {\bibinfo {volume}
  {B123}},\ \bibinfo {pages} {449} (\bibinfo {year} {1983})}\BibitemShut
  {NoStop}%
\bibitem [{\citenamefont {Heller}\ and\ \citenamefont
  {Tjon}(1987)}]{Heller:1986bt}%
  \BibitemOpen
  \bibfield  {author} {\bibinfo {author} {\bibfnamefont {L.}~\bibnamefont
  {Heller}}\ and\ \bibinfo {author} {\bibfnamefont {J.~A.}\ \bibnamefont
  {Tjon}},\ }\href {\doibase 10.1103/PhysRevD.35.969} {\bibfield  {journal}
  {\bibinfo  {journal} {Phys. Rev.}\ }\textbf {\bibinfo {volume} {D35}},\
  \bibinfo {pages} {969} (\bibinfo {year} {1987})}\BibitemShut {NoStop}%
\bibitem [{\citenamefont {Carlson}\ \emph {et~al.}(1988)\citenamefont
  {Carlson}, \citenamefont {Heller},\ and\ \citenamefont
  {Tjon}}]{Carlson:1987hh}%
  \BibitemOpen
  \bibfield  {author} {\bibinfo {author} {\bibfnamefont {J.}~\bibnamefont
  {Carlson}}, \bibinfo {author} {\bibfnamefont {L.}~\bibnamefont {Heller}}, \
  and\ \bibinfo {author} {\bibfnamefont {J.~A.}\ \bibnamefont {Tjon}},\ }\href
  {\doibase 10.1103/PhysRevD.37.744} {\bibfield  {journal} {\bibinfo  {journal}
  {Phys. Rev.}\ }\textbf {\bibinfo {volume} {D37}},\ \bibinfo {pages} {744}
  (\bibinfo {year} {1988})}\BibitemShut {NoStop}%
\bibitem [{\citenamefont {Lipkin}(1986)}]{Lipkin:1986dw}%
  \BibitemOpen
  \bibfield  {author} {\bibinfo {author} {\bibfnamefont {H.~J.}\ \bibnamefont
  {Lipkin}},\ }\href {\doibase 10.1016/0370-2693(86)90843-9} {\bibfield
  {journal} {\bibinfo  {journal} {Phys. Lett.}\ }\textbf {\bibinfo {volume}
  {B172}},\ \bibinfo {pages} {242} (\bibinfo {year} {1986})}\BibitemShut
  {NoStop}%
\bibitem [{\citenamefont {Brink}\ and\ \citenamefont
  {Stancu}(1998)}]{Brink:1998as}%
  \BibitemOpen
  \bibfield  {author} {\bibinfo {author} {\bibfnamefont {D.~M.}\ \bibnamefont
  {Brink}}\ and\ \bibinfo {author} {\bibfnamefont {F.}~\bibnamefont {Stancu}},\
  }\href {\doibase 10.1103/PhysRevD.57.6778} {\bibfield  {journal} {\bibinfo
  {journal} {Phys. Rev.}\ }\textbf {\bibinfo {volume} {D57}},\ \bibinfo {pages}
  {6778} (\bibinfo {year} {1998})}\BibitemShut {NoStop}%
\bibitem [{\citenamefont {Gelman}\ and\ \citenamefont
  {Nussinov}(2003)}]{Gelman:2002wf}%
  \BibitemOpen
  \bibfield  {author} {\bibinfo {author} {\bibfnamefont {B.~A.}\ \bibnamefont
  {Gelman}}\ and\ \bibinfo {author} {\bibfnamefont {S.}~\bibnamefont
  {Nussinov}},\ }\href {\doibase 10.1016/S0370-2693(02)03069-1} {\bibfield
  {journal} {\bibinfo  {journal} {Phys. Lett.}\ }\textbf {\bibinfo {volume}
  {B551}},\ \bibinfo {pages} {296} (\bibinfo {year} {2003})},\ \Eprint
  {http://arxiv.org/abs/hep-ph/0209095} {arXiv:hep-ph/0209095 [hep-ph]}
  \BibitemShut {NoStop}%
\bibitem [{\citenamefont {Vijande}\ \emph {et~al.}(2004)\citenamefont
  {Vijande}, \citenamefont {Fernandez}, \citenamefont {Valcarce},\ and\
  \citenamefont {Silvestre-Brac}}]{Vijande:2003ki}%
  \BibitemOpen
  \bibfield  {author} {\bibinfo {author} {\bibfnamefont {J.}~\bibnamefont
  {Vijande}}, \bibinfo {author} {\bibfnamefont {F.}~\bibnamefont {Fernandez}},
  \bibinfo {author} {\bibfnamefont {A.}~\bibnamefont {Valcarce}}, \ and\
  \bibinfo {author} {\bibfnamefont {B.}~\bibnamefont {Silvestre-Brac}},\ }\href
  {\doibase 10.1140/epja/i2003-10128-9} {\bibfield  {journal} {\bibinfo
  {journal} {Eur. Phys. J.}\ }\textbf {\bibinfo {volume} {A19}},\ \bibinfo
  {pages} {383} (\bibinfo {year} {2004})},\ \Eprint
  {http://arxiv.org/abs/hep-ph/0310007} {arXiv:hep-ph/0310007 [hep-ph]}
  \BibitemShut {NoStop}%
\bibitem [{\citenamefont {Janc}\ and\ \citenamefont
  {Rosina}(2004)}]{Janc:2004qn}%
  \BibitemOpen
  \bibfield  {author} {\bibinfo {author} {\bibfnamefont {D.}~\bibnamefont
  {Janc}}\ and\ \bibinfo {author} {\bibfnamefont {M.}~\bibnamefont {Rosina}},\
  }\href {\doibase 10.1007/s00601-004-0068-9} {\bibfield  {journal} {\bibinfo
  {journal} {Few Body Syst.}\ }\textbf {\bibinfo {volume} {35}},\ \bibinfo
  {pages} {175} (\bibinfo {year} {2004})},\ \Eprint
  {http://arxiv.org/abs/hep-ph/0405208} {arXiv:hep-ph/0405208 [hep-ph]}
  \BibitemShut {NoStop}%
\bibitem [{\citenamefont {Cohen}\ and\ \citenamefont
  {Hohler}(2006)}]{Cohen:2006jg}%
  \BibitemOpen
  \bibfield  {author} {\bibinfo {author} {\bibfnamefont {T.~D.}\ \bibnamefont
  {Cohen}}\ and\ \bibinfo {author} {\bibfnamefont {P.~M.}\ \bibnamefont
  {Hohler}},\ }\href {\doibase 10.1103/PhysRevD.74.094003} {\bibfield
  {journal} {\bibinfo  {journal} {Phys. Rev.}\ }\textbf {\bibinfo {volume}
  {D74}},\ \bibinfo {pages} {094003} (\bibinfo {year} {2006})},\ \Eprint
  {http://arxiv.org/abs/hep-ph/0606084} {arXiv:hep-ph/0606084 [hep-ph]}
  \BibitemShut {NoStop}%
\bibitem [{\citenamefont {Vijande}\ \emph {et~al.}(2007)\citenamefont
  {Vijande}, \citenamefont {Valcarce},\ and\ \citenamefont
  {Richard}}]{Vijande:2007ix}%
  \BibitemOpen
  \bibfield  {author} {\bibinfo {author} {\bibfnamefont {J.}~\bibnamefont
  {Vijande}}, \bibinfo {author} {\bibfnamefont {A.}~\bibnamefont {Valcarce}}, \
  and\ \bibinfo {author} {\bibfnamefont {J.~M.}\ \bibnamefont {Richard}},\
  }\href {\doibase 10.1103/PhysRevD.76.114013} {\bibfield  {journal} {\bibinfo
  {journal} {Phys. Rev.}\ }\textbf {\bibinfo {volume} {D76}},\ \bibinfo {pages}
  {114013} (\bibinfo {year} {2007})},\ \Eprint {http://arxiv.org/abs/0707.3996}
  {arXiv:0707.3996 [hep-ph]} \BibitemShut {NoStop}%
\bibitem [{\citenamefont {Bicudo}\ and\ \citenamefont
  {Wagner}(2013)}]{Bicudo:2012qt}%
  \BibitemOpen
  \bibfield  {author} {\bibinfo {author} {\bibfnamefont {P.}~\bibnamefont
  {Bicudo}}\ and\ \bibinfo {author} {\bibfnamefont {M.}~\bibnamefont
  {Wagner}},\ }\href {\doibase 10.1103/PhysRevD.87.114511} {\bibfield
  {journal} {\bibinfo  {journal} {Phys. Rev.}\ }\textbf {\bibinfo {volume}
  {D87}},\ \bibinfo {pages} {114511} (\bibinfo {year} {2013})},\ \Eprint
  {http://arxiv.org/abs/1209.6274} {arXiv:1209.6274 [hep-ph]} \BibitemShut
  {NoStop}%
\bibitem [{\citenamefont {Bicudo}\ \emph {et~al.}(2015)\citenamefont {Bicudo},
  \citenamefont {Cichy}, \citenamefont {Peters}, \citenamefont {Wagenbach},\
  and\ \citenamefont {Wagner}}]{Bicudo:2015vta}%
  \BibitemOpen
  \bibfield  {author} {\bibinfo {author} {\bibfnamefont {P.}~\bibnamefont
  {Bicudo}}, \bibinfo {author} {\bibfnamefont {K.}~\bibnamefont {Cichy}},
  \bibinfo {author} {\bibfnamefont {A.}~\bibnamefont {Peters}}, \bibinfo
  {author} {\bibfnamefont {B.}~\bibnamefont {Wagenbach}}, \ and\ \bibinfo
  {author} {\bibfnamefont {M.}~\bibnamefont {Wagner}},\ }\href {\doibase
  10.1103/PhysRevD.92.014507} {\bibfield  {journal} {\bibinfo  {journal} {Phys.
  Rev.}\ }\textbf {\bibinfo {volume} {D92}},\ \bibinfo {pages} {014507}
  (\bibinfo {year} {2015})},\ \Eprint {http://arxiv.org/abs/1505.00613}
  {arXiv:1505.00613 [hep-lat]} \BibitemShut {NoStop}%
\bibitem [{\citenamefont {Bicudo}\ \emph
  {et~al.}(2017{\natexlab{a}})\citenamefont {Bicudo}, \citenamefont
  {Scheunert},\ and\ \citenamefont {Wagner}}]{Bicudo:2016ooe}%
  \BibitemOpen
  \bibfield  {author} {\bibinfo {author} {\bibfnamefont {P.}~\bibnamefont
  {Bicudo}}, \bibinfo {author} {\bibfnamefont {J.}~\bibnamefont {Scheunert}}, \
  and\ \bibinfo {author} {\bibfnamefont {M.}~\bibnamefont {Wagner}},\ }\href
  {\doibase 10.1103/PhysRevD.95.034502} {\bibfield  {journal} {\bibinfo
  {journal} {Phys. Rev.}\ }\textbf {\bibinfo {volume} {D95}},\ \bibinfo {pages}
  {034502} (\bibinfo {year} {2017}{\natexlab{a}})},\ \Eprint
  {http://arxiv.org/abs/1612.02758} {arXiv:1612.02758 [hep-lat]} \BibitemShut
  {NoStop}%
\bibitem [{\citenamefont {Francis}\ \emph {et~al.}(2017)\citenamefont
  {Francis}, \citenamefont {Hudspith}, \citenamefont {Lewis},\ and\
  \citenamefont {Maltman}}]{Francis:2016hui}%
  \BibitemOpen
  \bibfield  {author} {\bibinfo {author} {\bibfnamefont {A.}~\bibnamefont
  {Francis}}, \bibinfo {author} {\bibfnamefont {R.~J.}\ \bibnamefont
  {Hudspith}}, \bibinfo {author} {\bibfnamefont {R.}~\bibnamefont {Lewis}}, \
  and\ \bibinfo {author} {\bibfnamefont {K.}~\bibnamefont {Maltman}},\ }\href
  {\doibase 10.1103/PhysRevLett.118.142001} {\bibfield  {journal} {\bibinfo
  {journal} {Phys. Rev. Lett.}\ }\textbf {\bibinfo {volume} {118}},\ \bibinfo
  {pages} {142001} (\bibinfo {year} {2017})},\ \Eprint
  {http://arxiv.org/abs/1607.05214} {arXiv:1607.05214 [hep-lat]} \BibitemShut
  {NoStop}%
\bibitem [{\citenamefont {Francis}\ \emph {et~al.}(2019)\citenamefont
  {Francis}, \citenamefont {Hudspith}, \citenamefont {Lewis},\ and\
  \citenamefont {Maltman}}]{Francis:2018jyb}%
  \BibitemOpen
  \bibfield  {author} {\bibinfo {author} {\bibfnamefont {A.}~\bibnamefont
  {Francis}}, \bibinfo {author} {\bibfnamefont {R.~J.}\ \bibnamefont
  {Hudspith}}, \bibinfo {author} {\bibfnamefont {R.}~\bibnamefont {Lewis}}, \
  and\ \bibinfo {author} {\bibfnamefont {K.}~\bibnamefont {Maltman}},\ }\href
  {\doibase 10.1103/PhysRevD.99.054505} {\bibfield  {journal} {\bibinfo
  {journal} {Phys. Rev.}\ }\textbf {\bibinfo {volume} {D99}},\ \bibinfo {pages}
  {054505} (\bibinfo {year} {2019})},\ \Eprint
  {http://arxiv.org/abs/1810.10550} {arXiv:1810.10550 [hep-lat]} \BibitemShut
  {NoStop}%
\bibitem [{\citenamefont {Junnarkar}\ \emph {et~al.}(2019)\citenamefont
  {Junnarkar}, \citenamefont {Mathur},\ and\ \citenamefont
  {Padmanath}}]{Junnarkar:2018twb}%
  \BibitemOpen
  \bibfield  {author} {\bibinfo {author} {\bibfnamefont {P.}~\bibnamefont
  {Junnarkar}}, \bibinfo {author} {\bibfnamefont {N.}~\bibnamefont {Mathur}}, \
  and\ \bibinfo {author} {\bibfnamefont {M.}~\bibnamefont {Padmanath}},\ }\href
  {\doibase 10.1103/PhysRevD.99.034507} {\bibfield  {journal} {\bibinfo
  {journal} {Phys. Rev.}\ }\textbf {\bibinfo {volume} {D99}},\ \bibinfo {pages}
  {034507} (\bibinfo {year} {2019})},\ \Eprint
  {http://arxiv.org/abs/1810.12285} {arXiv:1810.12285 [hep-lat]} \BibitemShut
  {NoStop}%
\bibitem [{\citenamefont {Leskovec}\ \emph {et~al.}(2019)\citenamefont
  {Leskovec}, \citenamefont {Meinel}, \citenamefont {Pflaumer},\ and\
  \citenamefont {Wagner}}]{Leskovec:2019ioa}%
  \BibitemOpen
  \bibfield  {author} {\bibinfo {author} {\bibfnamefont {L.}~\bibnamefont
  {Leskovec}}, \bibinfo {author} {\bibfnamefont {S.}~\bibnamefont {Meinel}},
  \bibinfo {author} {\bibfnamefont {M.}~\bibnamefont {Pflaumer}}, \ and\
  \bibinfo {author} {\bibfnamefont {M.}~\bibnamefont {Wagner}},\ }\href
  {\doibase 10.1103/PhysRevD.100.014503} {\bibfield  {journal} {\bibinfo
  {journal} {Phys. Rev.}\ }\textbf {\bibinfo {volume} {D100}},\ \bibinfo
  {pages} {014503} (\bibinfo {year} {2019})},\ \Eprint
  {http://arxiv.org/abs/1904.04197} {arXiv:1904.04197 [hep-lat]} \BibitemShut
  {NoStop}%
\bibitem [{\citenamefont {Bicudo}\ \emph
  {et~al.}(2017{\natexlab{b}})\citenamefont {Bicudo}, \citenamefont {Cardoso},
  \citenamefont {Peters}, \citenamefont {Pflaumer},\ and\ \citenamefont
  {Wagner}}]{Bicudo:2017szl}%
  \BibitemOpen
  \bibfield  {author} {\bibinfo {author} {\bibfnamefont {P.}~\bibnamefont
  {Bicudo}}, \bibinfo {author} {\bibfnamefont {M.}~\bibnamefont {Cardoso}},
  \bibinfo {author} {\bibfnamefont {A.}~\bibnamefont {Peters}}, \bibinfo
  {author} {\bibfnamefont {M.}~\bibnamefont {Pflaumer}}, \ and\ \bibinfo
  {author} {\bibfnamefont {M.}~\bibnamefont {Wagner}},\ }\href {\doibase
  10.1103/PhysRevD.96.054510} {\bibfield  {journal} {\bibinfo  {journal} {Phys.
  Rev.}\ }\textbf {\bibinfo {volume} {D96}},\ \bibinfo {pages} {054510}
  (\bibinfo {year} {2017}{\natexlab{b}})},\ \Eprint
  {http://arxiv.org/abs/1704.02383} {arXiv:1704.02383 [hep-lat]} \BibitemShut
  {NoStop}%
\bibitem [{\citenamefont {Peters}\ \emph {et~al.}(2018)\citenamefont {Peters},
  \citenamefont {Bicudo},\ and\ \citenamefont {Wagner}}]{Peters:2017hon}%
  \BibitemOpen
  \bibfield  {author} {\bibinfo {author} {\bibfnamefont {A.}~\bibnamefont
  {Peters}}, \bibinfo {author} {\bibfnamefont {P.}~\bibnamefont {Bicudo}}, \
  and\ \bibinfo {author} {\bibfnamefont {M.}~\bibnamefont {Wagner}},\ }\href
  {\doibase 10.1051/epjconf/201817514018} {\bibfield  {journal} {\bibinfo
  {journal} {EPJ Web Conf.}\ }\textbf {\bibinfo {volume} {175}},\ \bibinfo
  {pages} {14018} (\bibinfo {year} {2018})},\ \Eprint
  {http://arxiv.org/abs/1709.03306} {arXiv:1709.03306 [hep-lat]} \BibitemShut
  {NoStop}%
\bibitem [{\citenamefont {Prelovsek}\ \emph {et~al.}(2019)\citenamefont
  {Prelovsek}, \citenamefont {Bahtiyar},\ and\ \citenamefont
  {Petkovic}}]{Prelovsek:2019yae}%
  \BibitemOpen
  \bibfield  {author} {\bibinfo {author} {\bibfnamefont {S.}~\bibnamefont
  {Prelovsek}}, \bibinfo {author} {\bibfnamefont {H.}~\bibnamefont {Bahtiyar}},
  \ and\ \bibinfo {author} {\bibfnamefont {J.}~\bibnamefont {Petkovic}}\
  }(\bibinfo {year} {2019})\ \Eprint {http://arxiv.org/abs/1909.02356}
  {arXiv:1909.02356 [hep-lat]} \BibitemShut {NoStop}%
\bibitem [{\citenamefont {Bali}\ \emph {et~al.}(2005)\citenamefont {Bali},
  \citenamefont {Neff}, \citenamefont {Duessel}, \citenamefont {Lippert},\ and\
  \citenamefont {Schilling}}]{Bali:2005fu}%
  \BibitemOpen
  \bibfield  {author} {\bibinfo {author} {\bibfnamefont {G.~S.}\ \bibnamefont
  {Bali}}, \bibinfo {author} {\bibfnamefont {H.}~\bibnamefont {Neff}}, \bibinfo
  {author} {\bibfnamefont {T.}~\bibnamefont {Duessel}}, \bibinfo {author}
  {\bibfnamefont {T.}~\bibnamefont {Lippert}}, \ and\ \bibinfo {author}
  {\bibfnamefont {K.}~\bibnamefont {Schilling}} (\bibinfo {collaboration}
  {SESAM}),\ }\href {\doibase 10.1103/PhysRevD.71.114513} {\bibfield  {journal}
  {\bibinfo  {journal} {Phys. Rev.}\ }\textbf {\bibinfo {volume} {D71}},\
  \bibinfo {pages} {114513} (\bibinfo {year} {2005})},\ \Eprint
  {http://arxiv.org/abs/hep-lat/0505012} {arXiv:hep-lat/0505012 [hep-lat]}
  \BibitemShut {NoStop}%
\bibitem [{\citenamefont {Okubo}(1963)}]{Okubo:1963fa}%
  \BibitemOpen
  \bibfield  {author} {\bibinfo {author} {\bibfnamefont {S.}~\bibnamefont
  {Okubo}},\ }\href {\doibase 10.1016/S0375-9601(63)92548-9} {\bibfield
  {journal} {\bibinfo  {journal} {Phys. Lett.}\ }\textbf {\bibinfo {volume}
  {5}},\ \bibinfo {pages} {165} (\bibinfo {year} {1963})}\BibitemShut {NoStop}%
\bibitem [{\citenamefont {Zweig}(1964)}]{Zweig:1981pd}%
  \BibitemOpen
  \bibfield  {author} {\bibinfo {author} {\bibfnamefont {G.}~\bibnamefont
  {Zweig}},\ }\href@noop {} {\bibfield  {journal} {\bibinfo  {journal} {report
  number CERN-TH-401}\ } (\bibinfo {year} {1964})}\BibitemShut {NoStop}%
\bibitem [{\citenamefont {Iizuka}(1966)}]{Iizuka:1966fk}%
  \BibitemOpen
  \bibfield  {author} {\bibinfo {author} {\bibfnamefont {J.}~\bibnamefont
  {Iizuka}},\ }\href {\doibase 10.1143/PTPS.37.21} {\bibfield  {journal}
  {\bibinfo  {journal} {Prog. Theor. Phys. Suppl.}\ }\textbf {\bibinfo {volume}
  {37}},\ \bibinfo {pages} {21} (\bibinfo {year} {1966})}\BibitemShut {NoStop}%
\bibitem [{\citenamefont {Jansen}\ \emph
  {et~al.}(2008{\natexlab{a}})\citenamefont {Jansen}, \citenamefont {Michael},
  \citenamefont {Shindler},\ and\ \citenamefont {Wagner}}]{Jansen:2008ht}%
  \BibitemOpen
  \bibfield  {author} {\bibinfo {author} {\bibfnamefont {K.}~\bibnamefont
  {Jansen}}, \bibinfo {author} {\bibfnamefont {C.}~\bibnamefont {Michael}},
  \bibinfo {author} {\bibfnamefont {A.}~\bibnamefont {Shindler}}, \ and\
  \bibinfo {author} {\bibfnamefont {M.}~\bibnamefont {Wagner}} (\bibinfo
  {collaboration} {ETM}),\ }\href {\doibase 10.22323/1.066.0122} {\bibfield
  {journal} {\bibinfo  {journal} {PoS}\ }\textbf {\bibinfo {volume}
  {LATTICE2008}},\ \bibinfo {pages} {122} (\bibinfo {year}
  {2008}{\natexlab{a}})},\ \Eprint {http://arxiv.org/abs/0808.2121}
  {arXiv:0808.2121 [hep-lat]} \BibitemShut {NoStop}%
\bibitem [{\citenamefont {Michael}\ \emph {et~al.}(2010)\citenamefont
  {Michael}, \citenamefont {Shindler},\ and\ \citenamefont
  {Wagner}}]{Michael:2010aa}%
  \BibitemOpen
  \bibfield  {author} {\bibinfo {author} {\bibfnamefont {C.}~\bibnamefont
  {Michael}}, \bibinfo {author} {\bibfnamefont {A.}~\bibnamefont {Shindler}}, \
  and\ \bibinfo {author} {\bibfnamefont {M.}~\bibnamefont {Wagner}} (\bibinfo
  {collaboration} {ETM}),\ }\href {\doibase 10.1007/JHEP08(2010)009} {\bibfield
   {journal} {\bibinfo  {journal} {JHEP}\ }\textbf {\bibinfo {volume} {08}},\
  \bibinfo {pages} {009} (\bibinfo {year} {2010})},\ \Eprint
  {http://arxiv.org/abs/1004.4235} {arXiv:1004.4235 [hep-lat]} \BibitemShut
  {NoStop}%
\bibitem [{\citenamefont {Juge}\ \emph {et~al.}(2003)\citenamefont {Juge},
  \citenamefont {Kuti},\ and\ \citenamefont {Morningstar}}]{Juge:2002br}%
  \BibitemOpen
  \bibfield  {author} {\bibinfo {author} {\bibfnamefont {K.~J.}\ \bibnamefont
  {Juge}}, \bibinfo {author} {\bibfnamefont {J.}~\bibnamefont {Kuti}}, \ and\
  \bibinfo {author} {\bibfnamefont {C.}~\bibnamefont {Morningstar}},\ }\href
  {\doibase 10.1103/PhysRevLett.90.161601} {\bibfield  {journal} {\bibinfo
  {journal} {Phys. Rev. Lett.}\ }\textbf {\bibinfo {volume} {90}},\ \bibinfo
  {pages} {161601} (\bibinfo {year} {2003})},\ \Eprint
  {http://arxiv.org/abs/hep-lat/0207004} {arXiv:hep-lat/0207004 [hep-lat]}
  \BibitemShut {NoStop}%
\bibitem [{\citenamefont {Bicudo}\ \emph {et~al.}(2018)\citenamefont {Bicudo},
  \citenamefont {Cardoso},\ and\ \citenamefont {Cardoso}}]{Bicudo:2018jbb}%
  \BibitemOpen
  \bibfield  {author} {\bibinfo {author} {\bibfnamefont {P.}~\bibnamefont
  {Bicudo}}, \bibinfo {author} {\bibfnamefont {N.}~\bibnamefont {Cardoso}}, \
  and\ \bibinfo {author} {\bibfnamefont {M.}~\bibnamefont {Cardoso}},\ }\href
  {\doibase 10.1103/PhysRevD.98.114507} {\bibfield  {journal} {\bibinfo
  {journal} {Phys. Rev.}\ }\textbf {\bibinfo {volume} {D98}},\ \bibinfo {pages}
  {114507} (\bibinfo {year} {2018})},\ \Eprint
  {http://arxiv.org/abs/1808.08815} {arXiv:1808.08815 [hep-lat]} \BibitemShut
  {NoStop}%
\bibitem [{\citenamefont {Jansen}\ \emph
  {et~al.}(2008{\natexlab{b}})\citenamefont {Jansen}, \citenamefont {Michael},
  \citenamefont {Shindler},\ and\ \citenamefont {Wagner}}]{Jansen:2008si}%
  \BibitemOpen
  \bibfield  {author} {\bibinfo {author} {\bibfnamefont {K.}~\bibnamefont
  {Jansen}}, \bibinfo {author} {\bibfnamefont {C.}~\bibnamefont {Michael}},
  \bibinfo {author} {\bibfnamefont {A.}~\bibnamefont {Shindler}}, \ and\
  \bibinfo {author} {\bibfnamefont {M.}~\bibnamefont {Wagner}} (\bibinfo
  {collaboration} {ETM}),\ }\href {\doibase 10.1088/1126-6708/2008/12/058}
  {\bibfield  {journal} {\bibinfo  {journal} {JHEP}\ }\textbf {\bibinfo
  {volume} {12}},\ \bibinfo {pages} {058} (\bibinfo {year}
  {2008}{\natexlab{b}})},\ \Eprint {http://arxiv.org/abs/0810.1843}
  {arXiv:0810.1843 [hep-lat]} \BibitemShut {NoStop}%
\bibitem [{\citenamefont {Patrignani}\ \emph {et~al.}(2016)\citenamefont
  {Patrignani} \emph {et~al.}}]{Patrignani:2016xqp}%
  \BibitemOpen
  \bibfield  {author} {\bibinfo {author} {\bibfnamefont {C.}~\bibnamefont
  {Patrignani}} \emph {et~al.} (\bibinfo {collaboration} {Particle Data
  Group}),\ }\href {\doibase 10.1088/1674-1137/40/10/100001} {\bibfield
  {journal} {\bibinfo  {journal} {Chin. Phys.}\ }\textbf {\bibinfo {volume}
  {C40}},\ \bibinfo {pages} {100001} (\bibinfo {year} {2016})}\BibitemShut
  {NoStop}%
\bibitem [{\citenamefont {Bulava}\ \emph {et~al.}(2019)\citenamefont {Bulava},
  \citenamefont {Hörz}, \citenamefont {Knechtli}, \citenamefont {Koch},
  \citenamefont {Moir}, \citenamefont {Morningstar},\ and\ \citenamefont
  {Peardon}}]{Bulava:2019iut}%
  \BibitemOpen
  \bibfield  {author} {\bibinfo {author} {\bibfnamefont {J.}~\bibnamefont
  {Bulava}}, \bibinfo {author} {\bibfnamefont {B.}~\bibnamefont {Hörz}},
  \bibinfo {author} {\bibfnamefont {F.}~\bibnamefont {Knechtli}}, \bibinfo
  {author} {\bibfnamefont {V.}~\bibnamefont {Koch}}, \bibinfo {author}
  {\bibfnamefont {G.}~\bibnamefont {Moir}}, \bibinfo {author} {\bibfnamefont
  {C.}~\bibnamefont {Morningstar}}, \ and\ \bibinfo {author} {\bibfnamefont
  {M.}~\bibnamefont {Peardon}},\ }\href {\doibase
  10.1016/j.physletb.2019.05.018} {\bibfield  {journal} {\bibinfo  {journal}
  {Phys. Lett.}\ }\textbf {\bibinfo {volume} {B793}},\ \bibinfo {pages} {493}
  (\bibinfo {year} {2019})},\ \Eprint {http://arxiv.org/abs/1902.04006}
  {arXiv:1902.04006 [hep-lat]} \BibitemShut {NoStop}%
\bibitem [{Note1()}]{Note1}%
  \BibitemOpen
  \bibinfo {note} {For $r$ larger than the size of a static-light meson, i.e.\
  $r \protect \raisebox {-0.5ex}{$\protect \tmspace +\thinmuskip
  {.1667em}\mathrel {\mathop {\scriptstyle \sim }\limits ^{>}}\protect \tmspace
  +\thinmuskip {.1667em}$}0.5 \protect \tmspace +\thinmuskip {.1667em} \protect
  \textrm {fm} \protect \ldots 1.0 \protect \tmspace +\thinmuskip {.1667em}
  \protect \textrm {fm}$, this assumption is exactly fulfilled, because the
  overlaps $a_{\protect \mathaccentV {bar}016{M} M}^{\Lambda _\eta ^{(\epsilon
  )}}(r)$ are then equal to the squares of the overlaps of a $P = -$
  static-light meson with the corresponding single-meson creation operator,
  which do not depend on the static spin orientation.}\BibitemShut {Stop}%
\bibitem [{\citenamefont {Sommer}(1994)}]{Sommer:1993ce}%
  \BibitemOpen
  \bibfield  {author} {\bibinfo {author} {\bibfnamefont {R.}~\bibnamefont
  {Sommer}},\ }\href {\doibase 10.1016/0550-3213(94)90473-1} {\bibfield
  {journal} {\bibinfo  {journal} {Nucl. Phys.}\ }\textbf {\bibinfo {volume}
  {B411}},\ \bibinfo {pages} {839} (\bibinfo {year} {1994})},\ \Eprint
  {http://arxiv.org/abs/hep-lat/9310022} {arXiv:hep-lat/9310022 [hep-lat]}
  \BibitemShut {NoStop}%
\bibitem [{\citenamefont {Godfrey}\ and\ \citenamefont
  {Isgur}(1985)}]{Godfrey:1985xj}%
  \BibitemOpen
  \bibfield  {author} {\bibinfo {author} {\bibfnamefont {S.}~\bibnamefont
  {Godfrey}}\ and\ \bibinfo {author} {\bibfnamefont {N.}~\bibnamefont
  {Isgur}},\ }\href {\doibase 10.1103/PhysRevD.32.189} {\bibfield  {journal}
  {\bibinfo  {journal} {Phys. Rev.}\ }\textbf {\bibinfo {volume} {D32}},\
  \bibinfo {pages} {189} (\bibinfo {year} {1985})}\BibitemShut {NoStop}%
\bibitem [{\citenamefont {NVIDIA}(2019{\natexlab{a}})}]{CUDA1}%
  \BibitemOpen
  \bibfield  {author} {\bibinfo {author} {\bibnamefont {NVIDIA}},\ }\href@noop
  {} {\enquote {\bibinfo {title} {{CUDA}},}\ }\bibinfo {howpublished}
  {\url{https://developer.nvidia.com/cuda-zone}} (\bibinfo {year}
  {2019}{\natexlab{a}}),\ \bibinfo {note} {[Online; accessed
  03-May-2019]}\BibitemShut {NoStop}%
\bibitem [{\citenamefont {NVIDIA}(2019{\natexlab{b}})}]{CUDA2}%
  \BibitemOpen
  \bibfield  {author} {\bibinfo {author} {\bibnamefont {NVIDIA}},\ }\href@noop
  {} {\enquote {\bibinfo {title} {{cuSOLVER Library User Guide, v10.1.105}},}\
  }\bibinfo {howpublished} {\url{http://docs.nvidia.com/cuda/cusolver/}}
  (\bibinfo {year} {2019}{\natexlab{b}}),\ \bibinfo {note} {[Online; accessed
  03-May-2019]}\BibitemShut {NoStop}%
\bibitem [{\citenamefont {Braaten}\ \emph {et~al.}(2001)\citenamefont
  {Braaten}, \citenamefont {Fleming},\ and\ \citenamefont
  {Leibovich}}]{Braaten:2000cm}%
  \BibitemOpen
  \bibfield  {author} {\bibinfo {author} {\bibfnamefont {E.}~\bibnamefont
  {Braaten}}, \bibinfo {author} {\bibfnamefont {S.}~\bibnamefont {Fleming}}, \
  and\ \bibinfo {author} {\bibfnamefont {A.~K.}\ \bibnamefont {Leibovich}},\
  }\href {\doibase 10.1103/PhysRevD.63.094006} {\bibfield  {journal} {\bibinfo
  {journal} {Phys. Rev.}\ }\textbf {\bibinfo {volume} {D63}},\ \bibinfo {pages}
  {094006} (\bibinfo {year} {2001})},\ \Eprint
  {http://arxiv.org/abs/hep-ph/0008091} {arXiv:hep-ph/0008091 [hep-ph]}
  \BibitemShut {NoStop}%
\bibitem [{\citenamefont {Bodwin}\ and\ \citenamefont
  {Chen}(2001)}]{Bodwin:2001pt}%
  \BibitemOpen
  \bibfield  {author} {\bibinfo {author} {\bibfnamefont {G.~T.}\ \bibnamefont
  {Bodwin}}\ and\ \bibinfo {author} {\bibfnamefont {Y.-Q.}\ \bibnamefont
  {Chen}},\ }\href {\doibase 10.1103/PhysRevD.64.114008} {\bibfield  {journal}
  {\bibinfo  {journal} {Phys. Rev.}\ }\textbf {\bibinfo {volume} {D64}},\
  \bibinfo {pages} {114008} (\bibinfo {year} {2001})},\ \Eprint
  {http://arxiv.org/abs/hep-ph/0106095} {arXiv:hep-ph/0106095 [hep-ph]}
  \BibitemShut {NoStop}%
\bibitem [{\citenamefont {Maltoni}\ and\ \citenamefont
  {Polosa}(2004)}]{Maltoni:2004hv}%
  \BibitemOpen
  \bibfield  {author} {\bibinfo {author} {\bibfnamefont {F.}~\bibnamefont
  {Maltoni}}\ and\ \bibinfo {author} {\bibfnamefont {A.~D.}\ \bibnamefont
  {Polosa}},\ }\href {\doibase 10.1103/PhysRevD.70.054014} {\bibfield
  {journal} {\bibinfo  {journal} {Phys. Rev.}\ }\textbf {\bibinfo {volume}
  {D70}},\ \bibinfo {pages} {054014} (\bibinfo {year} {2004})},\ \Eprint
  {http://arxiv.org/abs/hep-ph/0405082} {arXiv:hep-ph/0405082 [hep-ph]}
  \BibitemShut {NoStop}%
\bibitem [{\citenamefont {Chen}\ \emph {et~al.}(2019)\citenamefont {Chen},
  \citenamefont {Zhang},\ and\ \citenamefont {He}}]{Chen:2019uzm}%
  \BibitemOpen
  \bibfield  {author} {\bibinfo {author} {\bibfnamefont {B.}~\bibnamefont
  {Chen}}, \bibinfo {author} {\bibfnamefont {A.}~\bibnamefont {Zhang}}, \ and\
  \bibinfo {author} {\bibfnamefont {J.}~\bibnamefont {He}},\ }\href@noop {} {\
  (\bibinfo {year} {2019})},\ \Eprint {http://arxiv.org/abs/1910.06065}
  {arXiv:1910.06065 [hep-ph]} \BibitemShut {NoStop}%
\bibitem [{\citenamefont {Choi}\ \emph {et~al.}(2003)\citenamefont {Choi} \emph
  {et~al.}}]{Choi:2003ue}%
  \BibitemOpen
  \bibfield  {author} {\bibinfo {author} {\bibfnamefont {S.~K.}\ \bibnamefont
  {Choi}} \emph {et~al.} (\bibinfo {collaboration} {Belle}),\ }\href {\doibase
  10.1103/PhysRevLett.91.262001} {\bibfield  {journal} {\bibinfo  {journal}
  {Phys. Rev. Lett.}\ }\textbf {\bibinfo {volume} {91}},\ \bibinfo {pages}
  {262001} (\bibinfo {year} {2003})},\ \Eprint
  {http://arxiv.org/abs/hep-ex/0309032} {arXiv:hep-ex/0309032 [hep-ex]}
  \BibitemShut {NoStop}%
\bibitem [{\citenamefont {Bali}\ \emph {et~al.}(1997)\citenamefont {Bali},
  \citenamefont {Schilling},\ and\ \citenamefont {Wachter}}]{Bali:1997am}%
  \BibitemOpen
  \bibfield  {author} {\bibinfo {author} {\bibfnamefont {G.~S.}\ \bibnamefont
  {Bali}}, \bibinfo {author} {\bibfnamefont {K.}~\bibnamefont {Schilling}}, \
  and\ \bibinfo {author} {\bibfnamefont {A.}~\bibnamefont {Wachter}},\ }\href
  {\doibase 10.1103/PhysRevD.56.2566} {\bibfield  {journal} {\bibinfo
  {journal} {Phys. Rev.}\ }\textbf {\bibinfo {volume} {D56}},\ \bibinfo {pages}
  {2566} (\bibinfo {year} {1997})},\ \Eprint
  {http://arxiv.org/abs/hep-lat/9703019} {arXiv:hep-lat/9703019 [hep-lat]}
  \BibitemShut {NoStop}%
\bibitem [{\citenamefont {Brambilla}\ \emph {et~al.}(2001)\citenamefont
  {Brambilla}, \citenamefont {Pineda}, \citenamefont {Soto},\ and\
  \citenamefont {Vairo}}]{Brambilla:2000gk}%
  \BibitemOpen
  \bibfield  {author} {\bibinfo {author} {\bibfnamefont {N.}~\bibnamefont
  {Brambilla}}, \bibinfo {author} {\bibfnamefont {A.}~\bibnamefont {Pineda}},
  \bibinfo {author} {\bibfnamefont {J.}~\bibnamefont {Soto}}, \ and\ \bibinfo
  {author} {\bibfnamefont {A.}~\bibnamefont {Vairo}},\ }\href {\doibase
  10.1103/PhysRevD.63.014023} {\bibfield  {journal} {\bibinfo  {journal} {Phys.
  Rev.}\ }\textbf {\bibinfo {volume} {D63}},\ \bibinfo {pages} {014023}
  (\bibinfo {year} {2001})},\ \Eprint {http://arxiv.org/abs/hep-ph/0002250}
  {arXiv:hep-ph/0002250 [hep-ph]} \BibitemShut {NoStop}%
\bibitem [{\citenamefont {Pineda}\ and\ \citenamefont
  {Vairo}(2001)}]{Pineda:2000sz}%
  \BibitemOpen
  \bibfield  {author} {\bibinfo {author} {\bibfnamefont {A.}~\bibnamefont
  {Pineda}}\ and\ \bibinfo {author} {\bibfnamefont {A.}~\bibnamefont {Vairo}},\
  }\href {\doibase 10.1103/PhysRevD.64.039902, 10.1103/PhysRevD.63.054007}
  {\bibfield  {journal} {\bibinfo  {journal} {Phys. Rev.}\ }\textbf {\bibinfo
  {volume} {D63}},\ \bibinfo {pages} {054007} (\bibinfo {year} {2001})},\
  \bibinfo {note} {[Erratum: Phys.\ Rev.\ D64, 039902 (2001)]},\ \Eprint
  {http://arxiv.org/abs/hep-ph/0009145} {arXiv:hep-ph/0009145 [hep-ph]}
  \BibitemShut {NoStop}%
\bibitem [{\citenamefont {Brambilla}\ \emph {et~al.}(2005)\citenamefont
  {Brambilla}, \citenamefont {Pineda}, \citenamefont {Soto},\ and\
  \citenamefont {Vairo}}]{Brambilla:2004jw}%
  \BibitemOpen
  \bibfield  {author} {\bibinfo {author} {\bibfnamefont {N.}~\bibnamefont
  {Brambilla}}, \bibinfo {author} {\bibfnamefont {A.}~\bibnamefont {Pineda}},
  \bibinfo {author} {\bibfnamefont {J.}~\bibnamefont {Soto}}, \ and\ \bibinfo
  {author} {\bibfnamefont {A.}~\bibnamefont {Vairo}},\ }\href {\doibase
  10.1103/RevModPhys.77.1423} {\bibfield  {journal} {\bibinfo  {journal} {Rev.
  Mod. Phys.}\ }\textbf {\bibinfo {volume} {77}},\ \bibinfo {pages} {1423}
  (\bibinfo {year} {2005})},\ \Eprint {http://arxiv.org/abs/hep-ph/0410047}
  {arXiv:hep-ph/0410047 [hep-ph]} \BibitemShut {NoStop}%
\bibitem [{\citenamefont {Koma}\ \emph {et~al.}(2006)\citenamefont {Koma},
  \citenamefont {Koma},\ and\ \citenamefont {Wittig}}]{Koma:2006si}%
  \BibitemOpen
  \bibfield  {author} {\bibinfo {author} {\bibfnamefont {Y.}~\bibnamefont
  {Koma}}, \bibinfo {author} {\bibfnamefont {M.}~\bibnamefont {Koma}}, \ and\
  \bibinfo {author} {\bibfnamefont {H.}~\bibnamefont {Wittig}},\ }\href
  {\doibase 10.1103/PhysRevLett.97.122003} {\bibfield  {journal} {\bibinfo
  {journal} {Phys. Rev. Lett.}\ }\textbf {\bibinfo {volume} {97}},\ \bibinfo
  {pages} {122003} (\bibinfo {year} {2006})},\ \Eprint
  {http://arxiv.org/abs/hep-lat/0607009} {arXiv:hep-lat/0607009 [hep-lat]}
  \BibitemShut {NoStop}%
\bibitem [{\citenamefont {Koma}\ and\ \citenamefont
  {Koma}(2012)}]{Koma:2012bc}%
  \BibitemOpen
  \bibfield  {author} {\bibinfo {author} {\bibfnamefont {Y.}~\bibnamefont
  {Koma}}\ and\ \bibinfo {author} {\bibfnamefont {M.}~\bibnamefont {Koma}},\
  }\href {\doibase 10.22323/1.164.0140} {\bibfield  {journal} {\bibinfo
  {journal} {PoS}\ }\textbf {\bibinfo {volume} {LATTICE2012}},\ \bibinfo
  {pages} {140} (\bibinfo {year} {2012})},\ \Eprint
  {http://arxiv.org/abs/1211.6795} {arXiv:1211.6795 [hep-lat]} \BibitemShut
  {NoStop}%
\bibitem [{\citenamefont {Brambilla}\ \emph
  {et~al.}(2019{\natexlab{a}})\citenamefont {Brambilla}, \citenamefont {Lai},
  \citenamefont {Segovia}, \citenamefont {Tarrús~Castella},\ and\
  \citenamefont {Vairo}}]{Brambilla:2018pyn}%
  \BibitemOpen
  \bibfield  {author} {\bibinfo {author} {\bibfnamefont {N.}~\bibnamefont
  {Brambilla}}, \bibinfo {author} {\bibfnamefont {W.~K.}\ \bibnamefont {Lai}},
  \bibinfo {author} {\bibfnamefont {J.}~\bibnamefont {Segovia}}, \bibinfo
  {author} {\bibfnamefont {J.}~\bibnamefont {Tarrús~Castella}}, \ and\
  \bibinfo {author} {\bibfnamefont {A.}~\bibnamefont {Vairo}},\ }\href
  {\doibase 10.1103/PhysRevD.99.014017} {\bibfield  {journal} {\bibinfo
  {journal} {Phys. Rev.}\ }\textbf {\bibinfo {volume} {D99}},\ \bibinfo {pages}
  {014017} (\bibinfo {year} {2019}{\natexlab{a}})},\ \Eprint
  {http://arxiv.org/abs/1805.07713} {arXiv:1805.07713 [hep-ph]} \BibitemShut
  {NoStop}%
\bibitem [{\citenamefont {Brambilla}\ \emph
  {et~al.}(2019{\natexlab{b}})\citenamefont {Brambilla}, \citenamefont {Lai},
  \citenamefont {Segovia},\ and\ \citenamefont
  {Tarrus~Castella}}]{Brambilla:2019jfi}%
  \BibitemOpen
  \bibfield  {author} {\bibinfo {author} {\bibfnamefont {N.}~\bibnamefont
  {Brambilla}}, \bibinfo {author} {\bibfnamefont {W.~K.}\ \bibnamefont {Lai}},
  \bibinfo {author} {\bibfnamefont {J.}~\bibnamefont {Segovia}}, \ and\
  \bibinfo {author} {\bibfnamefont {J.}~\bibnamefont {Tarrus~Castella}},\
  }\href@noop {} {\  (\bibinfo {year} {2019}{\natexlab{b}})},\ \Eprint
  {http://arxiv.org/abs/1908.11699} {arXiv:1908.11699 [hep-ph]} \BibitemShut
  {NoStop}%
\end{thebibliography}%

\end{document}